\documentclass[smallextended]{svjour3}
\usepackage{import}
\makeatletter
\let\cl@chapter\undefined
\makeatother

\usepackage{amsmath,amssymb,mathtools}
\usepackage{graphicx}
\usepackage{verbatim}
\usepackage[british]{babel}
\usepackage{longtable,booktabs}
\usepackage[labelfont=bf]{caption}
\usepackage{pdflscape}
\usepackage{fancyhdr}
\usepackage{etoolbox}
\usepackage{bm}
\usepackage[backref=section]{hyperref}   
\hypersetup{
    colorlinks=true,
    linkcolor=blue,
    filecolor=magenta,      
    urlcolor=cyan,
    pdfpagemode=FullScreen,
    citecolor = blue
    }
\usepackage[capitalize]{cleveref}        
\usepackage[square,numbers,sort&compress]{natbib}  
\usepackage{siunitx}                     
\usepackage{textcomp}
\usepackage{calc}
\usepackage{xcolor}
\usepackage{lipsum,tabularx}
\usepackage[usestackEOL]{stackengine}
\edef\tmp{\the\baselineskip}
\setstackgap{L}{\tmp}
\usepackage{makecell}
\setcellgapes{4pt}

\usepackage{subfiles}

\usepackage{titlesec}
\titleformat{\section}{\normalfont\Large\bfseries}{\thesection}{1em}{}
\titleformat{\subsection}{\normalfont\large\bfseries}{\thesubsection}{1em}{}
\titleformat{\subsubsection}{\normalfont\normalsize\bfseries}{\thesubsubsection}{1em}{}
\titleformat{\paragraph}{\normalfont\itshape}{}{}{}

\renewcommand{\arraystretch}{1.5}

\newcommand{\PBH}{\text{\tiny PBH}}

\newcommand{\be}{\begin{equation}}
\newcommand{\ee}{\end{equation}}
\newcommand{\bea}{\begin{eqnarray}}
\newcommand{\eea}{\end{eqnarray}}

\newcommand{\reply}[1]{\textcolor{black}{#1}}

\renewcommand*{\backrefalt}[4]{%
    \ifcase #1%
     \or (Section:~#2)%
     \else (Sections:~#2)%
    \fi%
    }

\journalname{Living Reviews in Relativity}

\graphicspath{{images/}}

\begin{document}

\fancyhead{}
\fancyfoot{}
\fancyhead[RE]{\the\authorrunning, \textit{\the\titlerunning}}
\fancyhead[LE,RO]{\thepage}

\newif\ifinappendix            
\pretocmd{\appendix}{\inappendixtrue}{}{}

\newif\ifunnumbered            

\fancyhead[LO]{%
  \ifunnumbered
    \nouppercase{\leftmark}
  \else
    \ifinappendix
      Appendix \thesection: \nouppercase{\leftmark}
    \else
      Section \thesection: \nouppercase{\leftmark}
    \fi
  \fi
}

\renewcommand{\sectionmark}[1]{\markboth{#1}{}}

\pagestyle{fancy}
\unnumberedtrue     
\markboth{}{}       
\title{Challenges and Opportunities \\ of Gravitational Wave Searches above 10~kHz}

\titlerunning{Gravitational Wave Searches above 10~kHz}


\author{\parbox{\textwidth}{\raggedright
\mbox{Nancy Aggarwal}$^{a}$ \and  
\mbox{Odylio D.\ Aguiar}$^b$ \and 
\mbox{Diego Blas}$^{c,d}$ \and
\mbox{Andreas Bauswein}$^e$ \and 
\mbox{Giancarlo Cella}$^f$ \and 
\mbox{Sebastian Clesse}$^g$ \and 
\mbox{Adrian Michael Cruise}$^h$ \and 
\mbox{Valerie Domcke}$^{i,*}$ \and
\mbox{Sebastian Ellis}$^{j,*}$ \and
\mbox{Daniel G.\ Figueroa}$^k$ \and 
\mbox{Gabriele Franciolini}$^{i,*}$ \and
\mbox{Camilo Garcia-Cely}$^k$ \and
\mbox{Andrew Geraci}$^a$ \and 
\mbox{Maxim Goryachev}$^l$ \and 
\mbox{Hartmut Grote}$^m$ \and 
\mbox{Mark Hindmarsh}$^{n,o}$ \and 
\mbox{Asuka Ito}$^{p,q}$ \and
\mbox{Joachim Kopp}$^{i,r,*}$ \and
\mbox{Sung Mook Lee}$^{i,*}$ \and
\mbox{Killian Martineau}$^s$  \and
\mbox{Jamie McDonald}$^t$ \and
\mbox{Francesco Muia}$^u$ \and
\mbox{Nikhil Mukund}$^v$ \and 
\mbox{David Ottaway}$^w$ \and 
\mbox{Marco Peloso}$^{x,y}$ \and 
\mbox{Krisztian Peters}$^z$ \and
\mbox{Fernando Quevedo}$^{u,\alpha}$ \and
\mbox{Angelo Ricciardone}$^{f,\beta}$ \and
\mbox{Andreas Ringwald}$^z$ \and
\mbox{Jessica Steinlechner}$^{\gamma,\delta,\epsilon}$ \and
\mbox{Sebastian Steinlechner}$^{\gamma,\delta}$ \and
\mbox{Sichun Sun}$^{\zeta}$ \and
\mbox{Carlos Tamarit}$^r$ \and
\mbox{Michael E.\ Tobar}$^l$ \and 
\mbox{Francisco Torrenti}$^\eta$ \and 
\mbox{Caner \"Unal}$^{\theta,\lambda}$ \and
\mbox{Graham White}$^\mu$
}}

\authorrunning{N.\ Aggarwal et al.} 

\institute{Corresponding authors: Valerie Domcke (\email{valerie.domcke@cern.ch}) \and Sebastian Ellis (\email{Sebastian.Ellis@unige.ch}) \and Gabriele Franciolini (\email{gabriele.franciolini@cern.ch}) \and Joachim Kopp (\email{jkopp@uni-mainz.de}) \and Sung Mook Lee (\email{sungmook.lee@cern.ch})}

\date{}
\maketitle

\begin{abstract}
The first direct measurement of gravitational waves by the LIGO and Virgo collaborations has opened up new avenues to explore our Universe. This white paper outlines the challenges and gains expected in gravitational-wave searches at frequencies above the LIGO/Virgo band.
The scarcity of possible astrophysical sources in most of this frequency range provides a unique opportunity to discover physics beyond the Standard Model operating both in the early and late Universe, and we highlight some of the most promising of these sources. We review several detector concepts that have been proposed to take up this challenge, and compare their expected sensitivity with the signal strength predicted in various models. This report is the summary of a series of workshops on the topic of high-frequency gravitational wave detection, held in 2019 (ICTP, Trieste, Italy), 2021 (online) and 2023 (CERN, Geneva, Switzerland).
\end{abstract}

\newpage

\noindent
{\small
$^a${Department of Physics and Astronomy, Northwestern University, Evanston, IL, USA} \\
$^b${Instituto Nacional de Pesquisas Espaciais (INPE), S\~ao Jos\'e dos Campos, S\~ao Paulo, Brazil} \\
$^c${IFAE, The Barcelona Institute of Science and Technology, Bellaterra (Barcelona), Spain}\\
$^d${Instituci\'{o} Catalana de Recerca i Estudis Avan\c{c}ats (ICREA), Barcelona, Spain}\\
$^e${GSI Helmholtzzentrum f\"ur Schwerionenforschung, Darmstadt, Germany} \\
$^f${Istituto Nazionale di Fisica Nucleare, Sezione di Pisa, Italy} \\
$^g${Service de Physique Th\'eorique, Universit\'e  Libre de Bruxelles, Brussels, Belgium} \\
$^h${School of Physics and Astronomy, University of Birmingham, Edgbaston, Birmingham, UK} \\
$^i${Theoretical Physics Department, CERN, Geneva, Switzerland} \\
$^j${D\'epartement de Physique Th\'eorique, Universit\'e de Gen\`eve, Geneva, Switzerland} \\
$^k${Instituto de F\'isica Corpuscular (IFIC), CSIC-Universitat de Val\`encia, Spain}\\
$^l${Department of Physics, University of Western Australia, Crawley, WA, Australia} \\
$^m${Cardiff University, Cardiff, UK} \\
$^n${Department of Physics and Helsinki Institute of Physics, University of Helsinki, Finland} \\
$^o${Department of Physics and Astronomy, University of Sussex, Brighton, UK} \\
$^p${Department of Physics, Kobe University, Kobe, Japan}\\
$^q${QUP, KEK, Tsukuba, Japan}\\
$^r${PRISMA+ Cluster of Excellence, Johannes Gutenberg University, Mainz, Germany} \\
$^s${LPSC, Universit\'{e} Grenoble--Alpes, CNRS-IN2P3, Grenoble, France}\\
$^t${Department of Physics and Astronomy, University of Manchester, UK} \\
$^u${DAMTP, Centre for Mathematical Sciences, Cambridge, UK} \\
$^v${Massachusetts Institute of Technology, Cambridge, MA, USA} \\
$^w${ARC Centre of Excellence OzGrav, The University of Adelaide, Australia} \\
$^x${Dipartimento di Fisica e Astronomia `Galileo Galilei' Universit\`a di Padova, Italy} \\
$^y${INFN, Sezione di Padova, Italy} \\
$^z${Deutsches Elektronen-Synchrotron DESY, Hamburg, Germany} \\
$^\alpha${New York University, Abu Dhabi, United Arab Emirates} \\
$^\beta${Dipartimento di Fisica `Enrico Fermi', Universit\`a di Pisa, Italy}\\
$^\gamma${Maastricht University, The Netherlands} \\
$^\delta${Nikhef, Amsterdam, The Netherlands} \\
$^\epsilon${SUPA, School of Physics and Astronomy, University of Glasgow, Scotland, UK} \\
$^\zeta${School of Physics, Beijing Institute of Technology, Beijing, People's Republic of China} \\
$^\eta${Departament de Física Quàntica i Astrofísıca \& ICCUB, Universitat de Barcelona, Spain} \\
$^\theta${CEICO, Institute of Physics of the Czech Academy of Sciences, Prague, Czechia} \\
$^\lambda${Department of Physics, Ben-Gurion University of the Negev, Be'er Sheva, Israel} \\
$^\mu${School of Physics and Astronomy, University of Southampton, Southampton, UK}
}

\vskip 1cm

\renewcommand*{\thefootnote}{\arabic{footnote}}
\setcounter{footnote}{0}

\clearpage
\markboth{}{}
\setcounter{tocdepth}{3}
\tableofcontents


\section{Introduction}
\label{sec:intro}
\unnumberedfalse

Over centuries, the main tool for observing the Universe has been electromagnetic waves, covering more than 20 orders of magnitude in frequency, from radio waves to gamma rays. The recent onset of gravitational wave astronomy has opened up a totally new window to observe our Universe \citep{Abbott:2016blz}.
As for electromagnetic observations, we may expect that at every scale in gravitational wave frequency, there should be interesting and unique physics to be discovered.  Current and planned projects such as pulsar timing arrays and ground- or space-based interferometers will explore gravitational waves in the well-motivated frequency range between nHz and kHz. However, both from the experimental and from the theoretical point of view it is worth considering also gravitational waves at much higher frequencies, such as the MHz and GHz bands.

A strong theoretical motivation for exploring frequencies above kHz is that there are hardly any known astrophysical objects small and dense enough to potentially emit at frequencies beyond 10\,kHz with a sizeable amplitude. Any discovery of gravitational waves at higher frequencies would thus indicate new physics beyond the Standard Model of particle physics, linked for instance to exotic astrophysical objects (such as primordial black holes or boson stars) or to cosmological events in the early Universe such as phase transitions, preheating after inflation, oscillons, cosmic strings, thermal fluctuations after reheating, etc., see \cite{Caprini:2018mtu} for a review. This should be seen in contrast to GW astronomy at lower frequencies, where, as detector sensitives increase, astrophysical gravitational wave foregrounds are posing an increasing challenge to searches for fainter signals from new physics.
In cosmology, gravitational waves may be the only way to observe certain epochs. In particular, before the recombination of electrons and nuclei into neutral atoms and the associated emission of the cosmic microwave background radiation, electromagnetic waves cannot propagate freely, so no electromagnetic signal can reach us from these early epochs. Gravitational waves, on the other hand, decouple essentially immediately after being produced thanks to the weakness of gravity. They travel undisturbed through the Universe, forming a stochastic background that could eventually be detected. Even though it may not be easy to unambiguously determine the specific cosmological source of a gravitational-wave signal, indications of its cosmological nature can be gained from properties such as isotropy and stationarity, in analogy to the original discovery of the cosmic microwave background.

The frequency of a cosmological gravitational wave signal is related to the epoch at which it is emitted: causality restricts the wavelength to be smaller than the cosmological horizon size at the time of gravitational wave production. This roughly implies that signals at frequencies above the range of the existing laser interferometers LIGO \citep{aLIGO2015,PhysRevLett.116.131103,PhysRevD.102.062003,PhysRevLett.123.231107}, Virgo \citep{AdvVirgo,PhysRevLett.123.231108}, and KAGRA \citep{Akutsu:2018axf,PhysRevD.88.043007} correspond to gravitational waves produced at temperatures $\gtrsim 10^{10}$\,GeV.\footnote{Cosmological events occurring at lower temperatures can also source such high-frequencies gravitational waves if the typical scale of the source is hierarchically smaller than the horizon scale.}
(Here, we have assumed radiation domination all the way to matter--radiation equality, as is the case in standard cosmology.) In particular, GHz frequencies correspond to the horizon size at the highest energies conceivable in particle physics (such as the Grand Unification or string scale) and phenomena like phase transitions and preheating after inflation would naturally produce gravitational waves with frequencies in the range from around 10\,kHz (the upper end of the LIGO detection band) to GHz.
Astrophysical sources such as mergers of compact binaries can generate gravitational waves at even higher frequencies. We moreover stress that essentially all detector concepts discussed in this review are probing uncharted territory. Even in regions of parameter space where no signals are expected or even envisaged, one may find unexpected surprises once one starts experimentally probing these regions for the first time.

Several proposals have been put forward for pushing the high-frequency end of interferometric detectors into the high-frequency region. Detectors for the MHz, GHz and THz frequency bands, however, require radically different experimental approaches. Over the years, there have been isolated attempts to search for such gravitational waves of very high frequencies, but interest in the field has increased significantly in recent years, with many new proposals and numerous emerging R\&D efforts. The current status of many of these ideas must be regarded as highly preliminary, driven by theoretical work rather than serious discussion of experimental noise sources, while others are already at the prototyping stage or setting first limits. The published concepts span a wide range of technologies with no real consensus yet as to where the community effort should be concentrated.
Concrete R\&D efforts are crucial to evaluate the suitability and potential of different technologies and are thus key to progress even when the sensitivity of prototypes falls short of the expected signals by several orders of magnitude.
In addition to the selection of suitable technological pathways towards a serious attempt at a detection at high frequencies, there needs to be an identification of the most realistic sources and thereby the waveforms and spectra for which such detectors should be optimised. This process demands a close collaboration of theorists and experimentalists.

The goal of this report is to summarise and start a dialogue among the various communities involved regarding the importance and feasibility of searches for high-frequency gravitational waves. We are aware that this may be a long-term goal but, keeping in mind that the strain sensitivity of the first historical gravitational wave detectors was eight orders of magnitude less than achieved in the current generation, we are convinced that the physics motivation is strong enough to start a systematic study of the different sources of high-frequency gravitational waves and their potential detectability already now. The origin of this initiative was a workshop organised at ICTP in October 2019 called ``Challenges and Opportunities of High-Frequency Gravitational Wave Detection'', where members of the theoretical and experimental communities interested on high-frequency gravitational waves got together to explore the motivations and challenges towards this search. Follow up workshops were held hosted by CERN in October 2021 and December 2023, and a 4th edition is planned for July 2025 in Mainz.\footnote{Slides and recordings of the contributions to these workshops can be found at: \url{http://indico.ictp.it/event/9006/} (2019), \url{https://indico.cern.ch/event/1074510/} (2021) and \url{https://indico.cern.ch/event/1257532/} (2023).} This series of workshops and the present white paper set the stage for the launch of the Ultra-High-Frequency Gravitational Wave (UHF-GW) initiative\footnote{\url{http://www.ctc.cam.ac.uk/activities/UHF-GW.php}.}, whose goals include supporting the R\&D and prototyping phase of experimental projects, stimulating technological advancements that may lead to new detection schemes, and fostering a vibrant theoretical community.

The remainder of this report is organized as follows: \cref{sec:Notation} introduces basic concepts and notation for the subsequent discussion of high-frequency gravitational-wave sources and detectors. \Cref{sec:overview} provides an executive summary of the sensitivities of different detector concepts discussed in this report and illustrates their reach to some exemplary categories of gravitational wave signals.  A more detailed discussion of sources then follows in \cref{sec:th}, while detectors are discussed in detail in \cref{sec:exp}.  We conclude in \cref{sec:conclusion}.
For a summary of the various detector concepts as well as the corresponding frequency ranges and sensitivities see \cref{fig:sens_HF,fig:sens_VHF,fig:sharkfin,fig:PLS} in \cref{sec:overview}, as well as \cref{tab:SummarySensitivity,tab:signal_sensing_resonator} in \cref{sec:SummarySensitivities} . For a summary of sources see \cref{tab:summary-coherent,tab:summary-stochastic} in \cref{sec:SummaryTable}.


\section{Basic Concepts and Notation}
\label{sec:Notation}

We introduce here the main concepts, and set out the notation, that will be used in this report \reply{in order to appropriately characterise GW sources and the ability of detectors to measure them. We start by discussing sources in a general way, and follow up with a similarly general discussion of detectors.}

\subsection{Acronyms and Conventions}

We will frequently use the following acronyms
\begin{center}
\begin{tabular}{ll}
    BBN     & Big Bang Nucleosynthesis \\
    BH      & black hole \\
    CMB     & cosmic microwave background \\
    FOPT    & first-order phase transition \\
    GW      & gravitational wave \\
    ISCO    & innermost stable circular orbit (of a black hole) \\
    LVK     & LIGO--VIRGO--KAGRA \\
    QCD     & quantum chromodynamics \\
    SGWB    & stochastic gravitational wave background \\
    SMBH    & supermassive black hole \\
    SNR     & signal-to-noise ratio \\
    UHF-GWs & ultra-high frequency gravitational waves
\end{tabular}
\end{center}

We will frequently encounter Fourier transforms, which, for a time-dependent quantity $q(t)$, we write as
\begin{align}
    q(t) = \int_{-\infty}^{\infty} \! df\, e^{2\pi i f t} q(f) \,,
    \qquad
    q(f) = \int_{-\infty}^{\infty} dt\, e^{-2 \pi i f t} q(t) \,,
    \label{eq:FTconv}
\end{align}
where $f$ is the frequency.  Even though $q(t)$ are typically real-valued data, $q(f)$ will in general be complex. A related quantity we will frequently encounter is the two-sided power spectral density (PSD), which we denote $S_q(f)$.%
\footnote{\label{fn:one-sided} The GW literature often uses the \emph{one-sided} PSD, $S_q(f)^{(1)}$, which is related to the two-sided PSD according to $S_q(f)^{(1)} = 2 S_q(f)$. We admit to adding to this confusion by switching from one-sided PSDs (used in the first version of this review) to two-sided PSDs here. This minimal change allows us to improve consistency in our notation while keeping key conversion formulas between $\Omega_\text{GW}$ (normalized energy density), $S_h$ (strain-equivalent noise PSD) and $h_c$ (characteristic strain) formally identical.}
It is defined as
\begin{align}
    \langle q(f) \, q^*(f') \rangle \equiv S_q(f) \, \delta(f-f') \,,
    \label{eq:Sq}
\end{align}
and is related to the auto-correlation function $R_q(\tau) = \langle q^*(t) \, q(t-\tau)\rangle$ by
\begin{align}
    R_q(\tau) = \int_{-\infty}^{\infty} \! df \, e^{i 2\pi\tau} \, S_q(f) \,.
\end{align}

Gravitational waves can be conveniently described in either the ``transverse-traceless'' (TT) gauge, or in the local inertial frame (LIF) of the detector, usually called the ``proper detector frame'' (PDF) \citep{Rakhmanov:2004eh, Rakhmanov:2014noa, Maggiore:1900zz}. The PDF is typically constructed with respect to the center of mass of the detector.

It is important to keep in mind that the choice of gauge, while often having a big impact on the complexity of calculations, does not affect the physics. When characterizing GW sources, one often works in TT gauge. Meanwhile when discussing detectors, the choice of gauge often depends on whether the detector components can be considered as freely falling or not. If they are freely falling, then the TT gauge description is often most useful. If they are not, the PDF is usually favored, especially if the GW wavelength is much larger than the size of the detector, $\omega_g L \ll 1$. Of course, general relativity requires that the final result be gauge/frame-independent, so that calculations in both approaches must agree. In practice, verifying this frame-independence for a given experimental setup can be cumbersome, although significant efforts have been made to show the equivalence for HFGW detectors (see, e.g., Ref.~\cite{Ratzinger:2024spd}). The metric perturbation for a gravitational wave in TT gauge can be written as 
\begin{align}
    h_{ij}^{\rm TT}(\mathbf{x},t) &=
        \sum_{a=+,\times} \int_{- \infty}^{+ \infty}
         h_a(f,\hat{\mathbf{k}})\,e_{ij}^a(\hat{\mathbf{k}})
            \exp(-2 \pi i( f t - \hat{\mathbf{k}}\cdot\mathbf{x})) \,,
    \label{eq:hFT}
\end{align}
where the polarization tensors $e_{ij}^a(\hat{\mathbf{k}})$ are defined as
\begin{align}
    e_{ij}^+(\hat{\mathbf{k}}) =
        \frac{1}{\sqrt{2}} ( \hat{\mathbf{u}}_i \hat{\mathbf{u}}_j
                           - \hat{\mathbf{v}}_i \hat{\mathbf{v}}_j )
    \,, \qquad
    e_{ij}^\times(\hat{\mathbf{k}}) =
        \frac{1}{\sqrt{2}} ( \hat{\mathbf{u}}_i \hat{\mathbf{v}}_j
                           + \hat{\mathbf{v}}_i \hat{\mathbf{u}}_j ) \,.
\end{align}
Here, the unit vectors $\hat{\mathbf{u}}$, $\hat{\mathbf{v}}$ are orthogonal to the direction of propagation of the GW $\hat{\mathbf{k}} = \mathbf{k}/|\mathbf{k}|$ and to each other. When traced over spatial and polarization indices, the polarization tensors satisfy the completeness relation $e_{ij}^a e_{ij}^a = 2$.\footnote{Note that the literature is split between this convention and $e_{ij}^a e_{ij}^a = 4$. The latter convention is obtained by removing the factor $1/\sqrt{2}$ in our definition of the polarization tensors.}

\reply{We will work in natural units, $c = \hbar = G = 1$, though we will occasionally reintroduce when displaying explicit dimensions helps clarify the physical meaning of the expressions.}

\subsection{Characterizing Sources}
\label{sec:characterizing-sources}

Sources of HFGWs can be classified into three broad categories: stochastic, transient, and persistent. In the case of the two latter categories, we assume that the signal is resolvable, either through its spatial origin, time-dependence, or both. A precise discussion of the physical origin of these three categories will follow in subsequent sections. Nevertheless, it is useful to recall that cosmological mechanisms will typically generate stochastic GWs; inspirals and mergers of compact objects can lead to resolvable transient GWs; processes such as decays or annihilation of axions in superradiant clouds can lead to resolvable persistent and coherent GWs. Below we introduce these categories in turn, introducing the notation required to quantify the GW strength at each stage.

\subsubsection{Stochastic Gravitational Waves}
\label{subs:stochastic}

Stochastic gravitational waves can be produced by various processes, including for instance phase transitions in the early Universe, the dynamics of inflation, subsequent (p)reheating, or fluctuations in the thermal plasma. They are often characterized by their spectral energy density,
\begin{align}
    \Omega_{\rm GW}(f) = \frac{1}{\rho_c} \frac{d\rho_g}{d\ln f} \,,
    \label{eq:Omega_g}
\end{align}
which normalizes the GW energy density per log-frequency interval, $\frac{d\rho_g}{d\ln f}$, to the critical energy density of the Universe, $\rho_c = 3 H_0^2/(8\pi G)$, where $H_0$ is the Hubble parameter today, and $G$ is Newton's constant. The total energy density in GWs, $\rho_g$, is related to the metric perturbation according to
\begin{align}
    \rho_g = \frac{1}{32\pi G} \langle\dot{h}_{\mu\nu} \dot{h}^{\mu\nu} \rangle
           = \frac{1}{32\pi G} \langle\dot{h}_+^2 + \dot{h}_\times^2\rangle \,,
\end{align}
where the first equality is exact and can be computed in the transverse-traceless (TT) gauge, resulting in the second equality. Since $\rho_g$ is a Lorentz scalar, it is frame-invariant. The averaging $\langle \ldots \rangle$ is over time.

This definition of the GW energy density lends itself to being related to the two-sided power spectral density $S_h(f)$ (cf.\ \cref{eq:Sq}) \citep{Allen:1999stochastic, Maggiore:1900zz, Thrane:2013_sen, Moore:2014sen}:\footnote{Note the different conventions in the literature, which we discuss below.}
\begin{align}
    \langle h_{a}(f,\phi,\theta) \, h^*_{a'}(f',\phi',\theta') \rangle
        \equiv \frac{1}{4\pi} S^a_h(f) \, \delta(f-f') \, \delta(\phi-\phi') \,
               \delta(\cos\theta - \cos\theta') \, \delta_{aa'} \ ,
    \label{eq:twopoint}
\end{align}
where $\phi$, $\theta$ are angles on the celestial sphere and $a = +, \times$ is the polarization. This expression is only valid for an isotropic stochastic background. If the background is anisotropic, $S^a_h(f)$ retains a dependence on $\phi$ and $\theta$, i.e., $S^a_h(f) \to S^a_h(f,\phi,\theta)$. The total power spectral density $S_h(f) = \tfrac12 \sum_a S_h^a(f)$ is related to the relative energy density in GWs by
\begin{align}
    \Omega_{\rm GW}(f) = \frac{4\pi^2}{3 H_0^2} |f|^3 \, S_h(|f|) \ .
    \label{eq:OmegaShRelation}
\end{align}
The GW power spectral density is therefore a useful proxy for the relative energy density in stochastic GWs. As already emphasized above in \cref{fn:one-sided}, conventions for $S_h(f)$ differ in the literature. We follow Ref.~\citep{Allen:1999stochastic} in using two-sided PSDs, but follow Ref.~\citep{Maggiore:1900zz} in normalizing by a factor $4\pi$ so that the integration over solid angle yields $1$, such that \cref{eq:hFT,eq:twopoint} yield
\begin{align}
 \langle h_a(t,\mathbf{x})^2 \rangle = \int_{- \infty}^{+ \infty} df S_h^a \,.
\end{align}
Taking into account also the different choices for the normalization of the polarization tensors (with our convention being $e^a_{ij} e^{a}_{ij} = 2$), we obtain
\begin{align}
    S_h(f) = 4\pi S_h^\text{Allen--Romano}(f)
           = \tfrac{1}{2} S_h^\text{Maggiore}(f)
           = S_h^\text{Thrane-Romano,~Moore}(f) \,.
\end{align}

Another useful quantity that is often used in the literature to characterize the amplitude of stochastic GWs is the `characteristic strain' $h_{c,\text{sto}}$. It is defined with respect to the GW power spectral density according to
\begin{align}
    h_{c,\text{sto}} \equiv \sqrt{f S_h(f)} \,,
    \label{eq:hcStoch}
\end{align}
and is therefore dimensionless. Using \cref{eq:OmegaShRelation}, we can also relate the characteristic strain to the relative GW energy density,
\begin{align}
    h_{c,\text{sto}} = \frac{H_0}{2\pi f} \big(3 \Omega_{\rm GW}(|f|) \big)^{1/2} \ .
    \label{eq:hcStoch_Omegag}
\end{align}

\subsubsection{Transient Gravitational Wave Sources}
\label{subs:transient}

Transient sources such as primordial black hole mergers (see \cref{sec:PBHmergers} or GW bursts (e.g. from hyperbolic encounters of compact objects or from cosmic string cusps) lead to signals with a short duration compared with the experimental measurement time. Nevertheless, such signals can still be characterized in terms of their PSD
\begin{align}
    \langle h(f) h^*(f')\rangle \equiv S_h(f) \delta (f-f') \,,
    \label{eq:ShTransient}
\end{align}
with $h(f)$ being the Fourier transform of the GW strain amplitude, $h(t)$, as defined in \cref{eq:FTconv}. The frequency dependence of $h(f)$ is dictated by the source properties, while the overall amplitude is inversely proportional to the source distance.  The latter allows \reply{one} to express detector capabilities in terms of a ``distance reach'' (see e.g.\ \cref{sec:PBHmergers}). Of importance for data analysis is that for given assumptions on the properties of the source, the frequency of the GW signal and its phase are known. Depending on the specifics of the detector, this can allow for a matched filtering analysis that improves the sensitivity to such sources. As a result, rather than using the strain PSD to characterise the signal-to-noise ratio for BH mergers, it is best to use $h(f)$, as we discuss below.

\subsubsection{Persistent Coherent Gravitational Wave Sources}

Various sources can lead to GWs that are monochromatic or at least coherent over a long timescale. An example is black hole superradiance, discussed in \cref{sec:Superradiance}. Such sources can also be treated in Fourier space, where their PSD takes on a particularly simple form, namely
\begin{align}
    S_h(f) = \frac{1}{2} h_0^2 \big[ \delta(f-f_g) + \delta(f+f_g) \big] \,.
\end{align}
The second Delta function, which accounts for negative frequencies, appears due to our choice of working with two-sided PSDs. In a scenario where the signal is very coherent, but not perfectly monochromatic, it can be useful to instead assume that the signal has a fixed bandwidth $\Delta f_g \ll f_g$, whereupon we can write the PSD in a simple form by replacing the delta functions with a broader peaked distribution with a width $\Delta f_g = f_g/Q_g$, where we have defined $Q_g$ as the quality factor of the signal.

\subsection{Characterizing Detectors}

It is inherently difficult to compare detection technologies and approaches, as they each have very different noise sources and amplitudes, bandwidths, antenna patterns, analysis strategy, etc. Nevertheless, certain quantities lend themselves to comparing detectors. In particular, the noise-equivalent strain power spectral density, $S_h^{\rm noise}(f)$ gives a measure of the noise in the detector as well as its response to a signal of generic spectral density $S_h(f)$. In simple terms, $S_h^{\rm noise}(f)$ can be viewed as the detector noise folded with the inverse of the detector response function.

In more detail, let us consider what a detector measures in its data stream. A detector taking data in the frequency domain can be viewed as recording a quantity $d(f) = n(f) + s(f)$, where $n(f)$ is the noise in the detector and $s(f)$ is the signal (if present).\footnote{In the interferometer literature, $d(f)$ is often normalized such that $s(f) = h(f)$, meaning that $n(f)$ carries information about how the strain is imprinted on the data measured by the detector.} The quantity $s(f)$ is itself a convolution of the GW signal $h(f)$ and the detector response, often characterized by its ``transfer function'' $T_h(f)$, such that $s(f) \equiv T_h(f) \, h(f)$.\footnote{For example, an experiment whose observable is a voltage has a dimensionful transfer function that encodes how the dimensionless strain signal is converted into a pure-signal voltage measurable at the output.} We add a subscript $h$ to this transfer function to distinguish it from the possibly different detector transfer function for noise, $T_n(f)$, defined such that $n(f) = T_n(f) \, \bar{n}(f)$, with $\bar{n}(f)$ the raw noise in the detector. The quantities $s(f)$ and $n(f)$ can each be characterized by two-sided PSDs, $S_{\rm sig}(f) \equiv |T_h(f)|^2 \, S_h(f)$ and $S_{\rm noise}(f) \equiv |T_n(f)|^2 \, S_{\bar{n}}(f)$, respectively. If a detector has multiple noise sources, as most of them do, each noise source must be calibrated separately to the readout channel and added in quadrature.  
These quantities allow us to finally define the noise-equivalent strain PSD as
\begin{align}
    S_h^{\rm noise}(f)
        \equiv S_{h}(f) \frac{S_{\rm noise}(f)}{S_{\rm sig}(f)}
        =      S_{\bar{n}}(f)\frac{|T_n(f)|^2}{|T_h(f)|^2}\ .
    \label{eq:shnoise}
\end{align}
The interpretation of this quantity is that the detector is sensitive to a given signal power spectral density $S_{\rm sig}(f)$, which is a combination of the intrinsic properties of the GW, $S_h(f)$, and the response of the detector to this input. Evaluating a detector's sensitivity to an unknown GW input therefore reduces to computing the quantities $S_{\bar{n}}(f)$, $T_n(f)$, and $T_h(f)$.

Numerically, the sensitivity of a detector is quantified by the signal-to-noise ratio (SNR). The data output is typically fed through a filter $F(t)$, which is usually implemented in software. The recorded data is therefore the convolution $d'(f) \sim F^*(f) \, d(f)$. The SNR is maximized by finding the optimal filter. The variance $\sigma_d^2$ of $\bar{d}(f)$ in the absence of signal ($s(f) = 0$) sets the noise level in the detector. We must now distinguish between recorded data that depends linearly on $s(f)$, $n(f)$, and data that depends quadratically on these quantities. In the case of linear data, $d'(f)_{s=0} \sim F^*(f) \, n(f)$ implies 
\begin{align}
    \left(\sigma_d^2\right)^{\rm lin}
        &= \big\langle (d'_{s=0})^2 \big\rangle
         - \big\langle d'_{s=0} \big\rangle^2 \notag\\
        &\simeq \int \! df \, |F(f)|^2 S_{\rm noise}(f) \,.
\end{align}
Meanwhile, if the data is quadratic in $s(f)$ and $n(f)$, we have $d'(t)_{s=0} \sim F^2(t) \, n^2(t)$, which in turn implies
\begin{align}
    \left(\sigma_d^2\right)^{\rm quad}
        &= \big\langle (d'_{s=0})^2 \big\rangle
         - \big\langle d'_{s=0} \big\rangle^2 \notag\\
        &\simeq \frac{1}{\Delta t} \int \! df |F(f)|^4 \, S^2_{\rm noise}(f) \,.
\end{align} 
Here, $\Delta t = \min[t_{\rm int}, \tau]$ is the smaller of the experimental integration time $t_{\rm int}$ or the signal duration $\tau$. We are implicitly assuming here that the integration time is the longest timescale in the experiment.  This is true for stochastic backgrounds, for example, but not for short transient sources.

For the signal in the absence of noise, we can define the signal power as
\begin{align}
    P_{\rm sig} \simeq
    \begin{dcases}
        \int \! df \, F^*(f) \, s(f)             & \text{linear} \,, \\
        \int \! df \, |F(f)|^2 \, S_{\rm sig}(f) & \text{quadratic} \,.
    \end{dcases}
\end{align}
For a linear signal, $P_{\rm sig}$ can be interpreted as being equivalent to the time-average of the data stream $d'(t)$, since $\langle n(t) \rangle = 0$, so the only possible contribution comes from $s(t)$. This further implies that if $\langle s(t) \rangle = 0$ as well, the signal must be auto-correlated with itself to be observable, rendering the measurement quadratic. For a quadratic signal, $P_{\rm sig}$ should be thought of as the power in excess of the mean noise power.

The SNR is then straightforwardly given by
\begin{align}
    \text{SNR}
        &= \frac{P_{\rm sig}}{\sqrt{\sigma_d^2(f)}} \\
        &= \begin{dcases}
               \frac{\int \! df \, F^*(f) \, s(f)}
                    {(\int \! df \, |F(f)|^2 S_{\rm noise}(f))^{1/2}}
               &\text{linear} \,, \\
               \frac{\int \! df \, |F(f)|^2 \, S_{\rm sig}(f)}
                    {\big( \frac{1}{\Delta t}
                         \int \! df \, |F(f)|^4 \, S^2_{\rm noise}(f) \big)^{1/2}}
               &\text{quadratic} \,.
           \end{dcases}
\end{align}
In the linear case, the optimal filter is $F(f) = K s(f)/S_{\rm noise}(f)$, where $K$ is an arbitrary constant. Meanwhile in the quadratic case, the optimal filter is $|F(f)|^2 = K' S_{\rm sig}(f)/S_{\rm noise}^2(f)$, with $K'$ another arbitrary constant.

The end result is that the optimal SNR for a generic signal is~\citep{Maggiore:1900zz}
\begin{align}
    \text{SNR}^{\rm lin} &= \bigg[2\,\Delta t \int_0^{\infty} \! df \,
        \frac{S_h(f)}{S_h^{\rm noise}(f)} \bigg]^{1/2} \,,
    \label{eq:LinSNR} \\
    \text{SNR}^{\rm quad} &= \bigg[2\,\Delta t \int_0^{\infty} \! df \, 
        \bigg(\frac{S_h(f)}{S_h^{\rm noise}(f)} \bigg)^2 \bigg]^{1/2} \,,
    \label{eq:QuadSNR}
\end{align}
for a detector sensitive to an observable linear in the GW strain $h$ in the first line, and for a detector sensitive to an observable that is quadratic in $h$ in the second line.\footnote{Note that the linear SNR is often written in the literature without the factor $t_{\rm int}$, and in terms of $|h(f)|^2$ instead of $S_h(f)$. To recover the form above, we can use that $|h(f)|^2 \sim S_h(f) \delta(f-f)$, and that $\delta(0)$ can only be resolved at the level of $t_{\rm int}$.} To obtain an expression in terms of $S_h(f)$ and $S_h^{\rm noise}(f)$, we have used \cref{eq:shnoise}. We see that the difference between a linearly sensitive detector and a quadratically sensitive detector is the relative scaling with $S_h(f)$ and the integration time $\Delta t$, that is, the degree to which a longer integration time can compensate for a smaller signal while keeping the SNR fixed. In order to compare the ability of a given detector to establish an exclusion limit or make a discovery, care must be taken in establishing what the appropriate threshold value is for the SNR. For this purpose, it is often useful to relate the SNR to the test statistic given a likelihood function~\citep{Cowan:2010js}. Below, we consider the resulting sensitivity of detectors to various types of GW sources in terms of SNR.

In practice, one often works with binned data, in which case the integral over frequencies in \cref{eq:LinSNR,eq:QuadSNR} reduces to a sum over bins in frequency-space, where each bin has a size $\delta f = 1/t_{\rm FFT}$ that comes from the ability to resolve a signal in the frequency domain. The quantity $t_{\rm FFT}$ is the timescale of the fast Fourier transform used in the data analysis. The frequency integral or sum should be limited to the frequency range over which the detector or signal has support, $\Delta f = \text{min}[\Delta f_{\rm det},~\Delta f_g]$ which effectively limits the bandwidth.

\subsubsection{Detector Sensitivity to Stochastic GWs}
\label{subs:sbsensitivity}

Stochastic GWs are by nature signals for which we lack phase information. Searching for them therefore requires a different strategy from that used to search for, e.g., inspirals where a waveform can be matched to the signal. For $N$ detectors sensitive to an observable linear in the GW strain, the signal can be cross-correlated between detectors, leading to an SNR which is similar to that of an observable quadratic in the strain. In particular~\citep{Maggiore:1900zz}
\begin{align}
    \text{SNR} \simeq \bigg[ N (N-1) \, t_{\rm int}
        \int_0^{\infty} \! df \, \Gamma(f)^2 \,
        \bigg( \frac{S_h(f)}{S_h^{\rm noise}(f)} \bigg)^2 \bigg]^{1/2} \,,
	\label{eq:SNR_SB}
\end{align}
where the function $\Gamma(f)$ is the ``overlap reduction function'', which captures the fact that the pairs of detectors may exhibit different responses to GW signals due to, e.g., different orientations, locations, etc.~\citep{Maggiore:1900zz}. Here, we have assumed for simplicity that $\Gamma(f)$ is the same for all detector pairings.

For observables quadratic in the strain, a single detector searching for a stochastic background will have an SNR given by \cref{eq:QuadSNR}. This is identical to \cref{eq:SNR_SB} without the combinatorial prefactor $N (N-1)$ and without the overlap reduction function. Combining multiple quadratic-in-strain detectors assuming the signal (but not the noise) to be correlated across detectors and taking the signal and noise to be independent so that they can be added in quadrature, the SNR scales  as $\sqrt{N}$.

For both types of detector, we observe that the SNR can be improved by increasing the integration time. If we approximate $S_h \sim h^2/\Delta f$, the sensitivity scales as
\begin{align}
    \text{SNR} \propto t_{\rm int}^{1/2} \,.
    \label{eq:SNR-stochastic-approx}
\end{align}
For cosmological GW backgrounds, we can express $S_h(f)$ in \cref{eq:SNR_SB} in terms of $\Omega_\text{GW}(f)$ using \cref{eq:OmegaShRelation}, which leads to
\begin{align}
    \text{SNR} \simeq \frac{3 H_0^2}{4\pi^2}
        \bigg[ N (N-1) \, t_{\rm int} \int_0^\infty \! df \,
        \bigg( \frac{\Gamma(f) \, \Omega_{\rm GW}(f)}
             {f^3 \, S_h^{\rm noise}(f)} \bigg)^2 \bigg]^{1/2} \,.
    \label{eq:SNRGWBdensity}
\end{align}
Given that typical cosmological sources emit over a fairly broad frequency range, the frequency integral is likely to yield a factor $\sim \min(\Delta f, f)$, such that the scaling is often $\text{SNR} \propto (t_{\rm int} \Delta f)^{1/2}$ (see, e.g., Chapter~7 of \cite{Maggiore:1900zz}).

\subsubsection{Detector Sensitivity to Transient GWs}
\label{sec:sensitivity-transients}

In \cref{subs:transient} above, we have argued that transient GWs can be characterized by the PSD of the GW signal, given by \cref{eq:ShTransient}. If the signal PSD and noise-equivalent strain PSD can be treated as being approximately flat in a band of width $\Delta f$ around the central frequency $f$, we can write the sensitivity as
\begin{align}
    S_h^{\rm lin}(f) &\gtrsim \text{SNR}^2 \times S_h^{\rm noise}(f)
        \bigg( \frac{1}{\Delta f\, \Delta t} \bigg)\,,
    \label{eq:S-sens-transient-lin} \\
    S_h^{\rm quad}(f) &\gtrsim \text{SNR}^2 \times S_h^{\rm noise}(f)
        \bigg( \frac{1}{\Delta f\, \Delta t} \bigg)^{1/2} \,.
    \label{eq:S-sens-transient-quad}
\end{align}
The time scale relevant in the denominator is either the signal duration, $\tau$, or the data taking time, $t_{\rm int}$, whichever is shorter. In the second line we observe that a quadratic-in-strain detector is necessarily limited to be less sensitive than a linear-in-strain detector unless the bandwidth saturates the maximum possible resolution, i.e.\ $\Delta f = 1/\Delta t$.

From the signal PSD, the dimensionless strain sensitivity can be obtained, though the exact relation depends on the type of source. For example, for a monochromatic burst of duration $\tau$, the strain is
\begin{align}
    h \sim \sqrt{S_h / \tau} \,.
\end{align}
This allows us to estimate the dimensionless strain sensitivity based on \cref{eq:S-sens-transient-lin,eq:S-sens-transient-quad}.

In addition, also the frequency-evolution of the signal must also be taken into account. For high-frequency GW sources, this can often be very fast, for instance $\dot{f} \propto f^{11/3}$ for inspiralling primordial black hole binaries. In the frequency domain, this can be accounted for by determining the total number of cycles $\mathcal{N}$ the signal spends inside a detector bandwidth. 

A further convenient way of parameterizing the sensitivity to transient sources is the so-called distance reach $d$ for a fixed SNR. If we schematically write $S_h(f) = S_h^0(f) / d^2$ and assume optimal filtering, $d$ is given by~\citep{Maggiore:1900zz}
\begin{align}
    d^k = \frac{2}{\text{SNR}} \bigg[ \Delta t
              \int_{f_{\rm min}}^{f_{\rm max}} \! df \,
              \bigg( \frac{S_h^0(f)}{S_h^{\rm noise}(f)} \bigg)^k \bigg]^{1/2} \,,
    \label{eq:reach}
\end{align}
with $k = 1, 2$ for linear and quadratic detectors, respectively.

\subsubsection{Detector Sensitivity to Persistent Coherent GWs}

For sufficiently persistent coherent GWs, we have argued previously that the signal PSD could be approximated by a Dirac delta-function in frequency space, or by a window function over some narrow width $\Delta f_g$. If the signal PSD is approximated as a delta-function and the detector response has a width $\Delta f_{\rm det} \leq 1/t_{\rm int}$, then the frequency resolution is given by $\delta f = 1/t_{\rm int}$.
We can then write the sensitivity to the GW strain as
\begin{align}
    h &\gtrsim \text{SNR} \times
        \bigg( \frac{S_h^{\rm noise}(f_g)}{t_{\rm int}} \bigg)^{1/2}
        & \text{(linear)} \,,
    \label{eq:h-sens-persistent-lin} \\
    h &\gtrsim \text{SNR}^{1/2} \times
        \bigg( \frac{S_h^{\rm noise}(f_g)}{t_{\rm int}} \bigg)^{1/2}
        & \text{(quadratic)} \,.
    \label{eq:h-sens-persistent-quad}
\end{align}
If the detector response is broad, $\Delta f_{\rm det} \geq 1/t_{\rm int}$, the sensitivity for linear-in-strain detectors is still given by \cref{eq:h-sens-persistent-lin}, but for quadratic-in-strain detectors it is modified to
\begin{align}
    h &\gtrsim \text{SNR}^{1/2} \times (S_h^{\rm noise}(f_g))^{1/2}
        \bigg( \frac{\Delta f_\text{det}}{t_{\rm int}} \bigg)^{1/4}
        & \text{(quadratic, $\Delta f_{\rm det} \geq 1/t_{\rm int}$)} \,.
\end{align}
owing to the fact that the integral $\int \! df S_h(f)^2 \sim h^4 \delta(0)$, and the ability to resolve $\delta(0)$ is limited by the detector response, i.e., $\delta(0) \sim 1/\Delta f_{\rm det}$.

\subsection{Note on Characteristic Strain for HFGWs}

The quantity ``characteristic strain'' is often used in the literature (see, e.g.,~\citep{Moore:2014sen}). It is particularly useful for inspiralling sources, since it is designed to include the effect of the frequency evolution of the signal, keeping track of how many cycles of a given signal can be seen within some detector bandwidth.

However, most definitions of $h_c$ in the literature start from the assumption of a matched filtering search for a signal of known frequency and phase, and a broadband detector such as an interferometer. As such, the definitions often seen in the literature on interferometers should not be directly applied to other signals/detectors. In this review, many detectors and signals are considered that do not have the same properties as the combination of BH inspirals at interferometers. Therefore, great care must be taken when considering the characteristic strain of the source, and mapping it onto a formula for the signal-to-noise ratio in a given detector.

\section{Overview of Detector Sensitivities and Possible Signals}
\label{sec:overview}

The goal of this section is to provide a brief overview of the different detector concepts discussed in this review together with their sensitivity to some exemplary GW signals. The latter will be discussed in more detail in 
\cref{sec:lateU} (astrophysical sources) and \cref{sec:earlyU} (cosmological source), while the detector concepts are the topic of \cref{sec:exp}. All details and references are given there. We caution that the figures below are indicative only and subject to a range of caveats.\footnote{%
In the first version of this review, we attempted to show different detector sensitivities together with the strengths of different types of signals in a single plot, using characteristic strain as a measure. We caution that a plot of that type contains many hidden variables (such as time scales associated with the the signal, the detector integration time, and the detector bandwidth), which may lead to misleading conclusions. In this updated version of the review, we therefore choose a different approach and compare different detector concepts only in terms of noise-equivalent strain (which contains no information on the GW source) or for specific source classes.}
\reply{The sensitivity curves shown in these figures are available at the HFGWPlotter webpage \citep{HFGWPlotter_Sh,HFGWPlotter_Omega}.}
\begin{figure}
    \includegraphics[width=\textwidth]{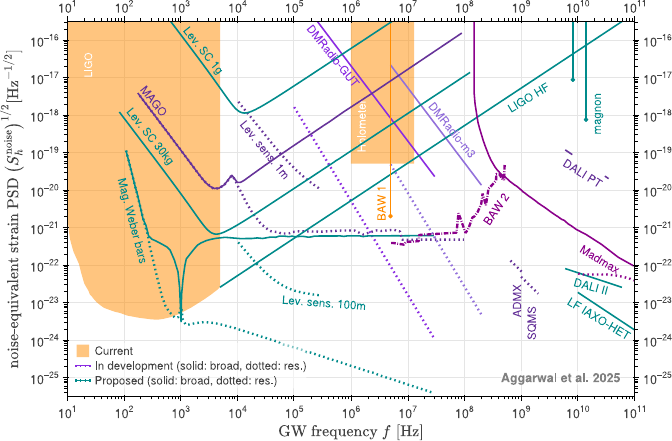}
    \caption[Strain sensitivities of gravitational wave detectors up to \SI{100}{GHz}]
    { Overview of achieved and projected strain sensitivities of high-frequency gravitational wave detectors up to \SI{100}{GHz}. Solid (dashed) lines indicate broadband (resonant) detectors. The color coding (see text for details) indicates the development stage ranging from published GW results (orange) to active R\&D efforts (purple) and proposed concepts (cyan). Details on the different proposals are given in \cref{sec:exp}.}
    \label{fig:sens_HF}
\end{figure}

\begin{figure}
    \includegraphics[width=\textwidth]{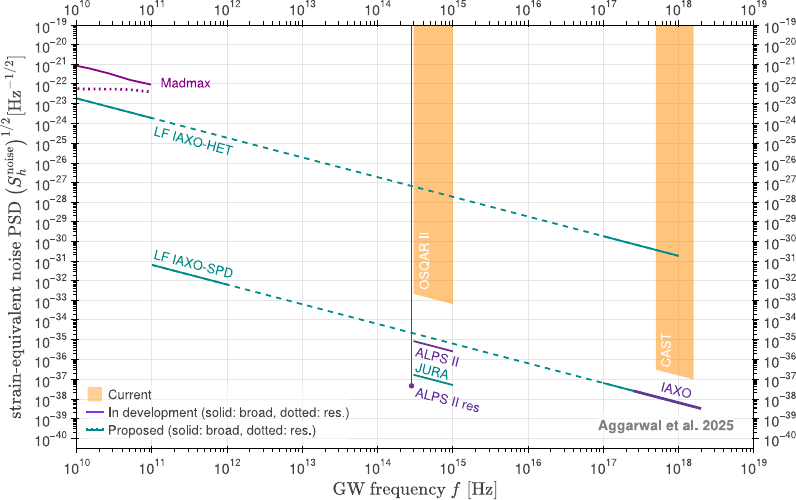}
    \caption[Strain sensitivities of gravitational wave detectors above \SI{10}{GHz}.]
    { Overview of projected strain sensitivities of gravitational wave detectors above \SI{100}{GHz}. The color coding is the same as in \cref{fig:sens_HF}, with orange, purple and cyan curves indicating published GW results, active R\&D efforts, and proposed concepts, respectively. Details on the different proposals are given in \cref{sec:exp}.}
    \label{fig:sens_VHF}
\end{figure}

\Cref{fig:sens_HF,fig:sens_VHF} provide on overview of the noise-equivalent strain sensitivities (see \cref{eq:shnoise}) of a range of ultra-high frequency gravitational wave detectors discussed in more detail in \cref{sec:exp}. Here and throughout this review, we show sensitivities of resonant detectors ($\Delta f_\text{det} \ll f$) as dashed curves and those of broadband detectors as solid curves. The color coding indicates the development stage of different experiments: orange curves correspond to detectors for which results on GW searches have been published. Purple indicates detector concepts under active research and development, which can either mean that a detector or detector prototype exists, or that there is a detailed technical proposal, funding is available for R\&D, and/or a collaborative effort is underway in the community supporting the proposal. This category includes concepts whose development is driven by physics goals other than GWs, for instance light dark matter searches. Finally, cyan curves indicate detector concepts which have been proposed but are, to our knowledge, not yet under active R\&D. This classification is necessarily somewhat subjective and will evolve over time; it should therefore be taken as indicative only.  For better visibility, we have split these summary plots into two frequency regimes, namely below \SI{100}{GHz} (\cref{fig:sens_HF}) and above (\cref{fig:sens_VHF}).

Given the sensitivity curves in \cref{fig:sens_HF,fig:sens_VHF}, the detectability of possible signals can be estimated by determining the corresponding signal-to-noise ratio as given in \cref{eq:QuadSNR,eq:LinSNR}. Various types of sources and signals will be comprehensively discussed in \cref{sec:th}; here, we focus on three exemplary cases: a monochromatic signal, a PBH binary inspiral, and a stochastic GW background.

For {\bf a persistent, monochromatic GW signal} (arising e.g.\ from black hole superradiance, see \cref{sec:Superradiance}) and a detector performing a linear measurement of the GW, the sensitivity to the GW amplitude can be estimated as (see \cref{eq:h-sens-persistent-lin})
\begin{align}
    h_0^\text{sens} \simeq (S_h^\text{noise}/t_\text{int})^{1/2} \,.
\end{align}

\begin{figure}
    \includegraphics[width = \textwidth]{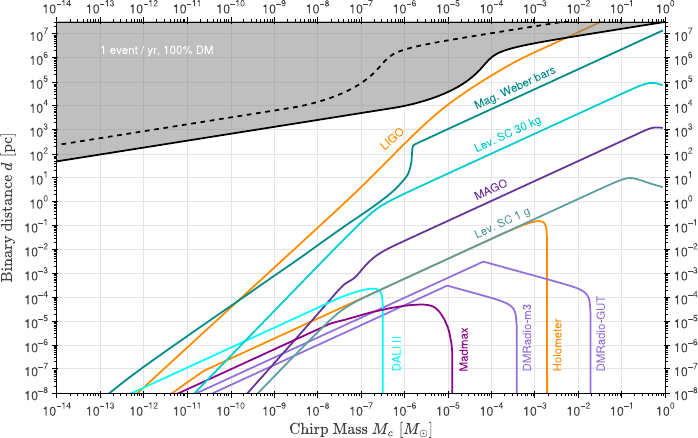}
    \caption[Distance reach of high-frequency GW detectors for PBH binaries.]
    {Distance reach of different broad-band high-frequency GW detectors for equal-mass PBH binaries with chirp mass $M_c$. The color code matches the one used in \cref{fig:sens_HF,fig:sens_VHF}, with orange, purple and cyan curves indicating published GW results, active R\&D efforts, and proposed concepts, respectively. The upper shaded region corresponds to distances within which $\geq 1$~event/yr is expected, assuming PBHs to account for all of the dark matter in the Universe (solid) or 0.1\% of it (dashed).}
    \label{fig:sharkfin}
\end{figure}

For {\bf mergers of primordial black holes} (see \cref{sec:PBHmergers}), \cref{fig:sharkfin} shows the astrophysical reach of a range of proposed broadband UHF GW detectors. This is obtained by integrating the GW waveform across the detector bandwidth using \eqref{eq:reachPBH} and assuming an SNR threshold of $10$. For simplicity, we have here assumed equal mass PBHs, circular orbits, no inclination angle, optimal sky position and we are working in the Newtonian approximation, integrating up to the innermost stable circular orbit, see \cref{sec:PBHmergers} for details. The `chirp' signal of PBH mergers, increasing rapidly in frequency and amplitude as the merger approaches, makes it challenging for resonant detectors to pick up a significant part of the signal strength, and hence these detectors are not shown in \cref{fig:sharkfin}.
Similarly, single photon detectors suffer from the short duration of these signals as it prevents them from reaching the energy threshold of a single photon per merger event. The corresponding line for LF-IAXO SPD is thus below the plot range shown. 

To estimate the {\bf sensitivity to stochastic GW backgrounds (SGWBs)}, we distinguish between broadband detectors (with a typical bandwidth of about a decade in frequency) and resonant detectors, which profit from a resonance with a large quality factor $Q$, but are sensitive only to a very narrow bandwidth, $\Delta f_\text{det} = f/Q$. In the latter case, coverage over a wider frequency range can often be achieved by a scanning strategy, amounting to tuning the detector to different frequencies over time.

\begin{figure}
    \includegraphics[width = \textwidth]{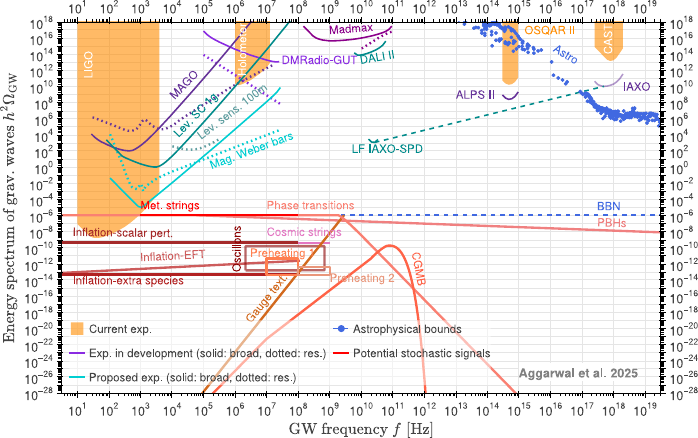}
    \caption[Sensitivity of high-frequency GW detectors to stochastic GW backgrounds.]
    { Sensitivity of high-frequency gravitational wave detectors to stochastic gravitational wave backgrounds assuming one year of integration time. The solid curves (broadband instruments) are power-law-integrated sensitivity curves, the dashed lines (resonant instruments) indicate the reach when running at fixed frequency for $t_\text{int} = \SI{1}{year}$. See text for details and caveats. In blue we indicate astrophysical constraints as discussed in \cref{sec:AstroDetectors}, where integration time varies dependent on observations \cite{Hill:2018trh}. The horizontal dashed blue line indicates the upper bound from BBN on cosmological sources, see \cref{sec:earlyU}. The remainder of the color coding is as in \cref{fig:sens_HF,fig:sens_VHF}, with orange, purple and cyan curves indicating published GW results, active R\&D efforts, and proposed concepts, respectively.}
    \label{fig:PLS}
\end{figure}

For broadband detectors, we show power-law integrated sensitivity (PLS) curves in \cref{fig:PLS}. To obtain these curves, we have fixed the integration time to 1~year and the SNR threshold to $\text{SNR}_\text{thr} = 10$, and we have then determined the power-law GW templates,
\begin{align}
    \Omega_\text{GW} = \Omega_0 (f/f_*)^\alpha \,,
\end{align}
for which \cref{eq:QuadSNR} evaluates to $\text{SNR} = \text{SNR}_\text{thr}$, using \cref{eq:OmegaShRelation} to relate $S_h$ and $\Omega_\text{GW}$.
An important exception are single photon detectors (OSQAR, ALPS, CAST and IAXO) for which the achievable senstivity to SGWBs is limited by the requirement of producing at least one photon (see \cref{eq:NsigSPD}) during the assumed detector run time of one year.

For resonant detectors, we first note that a simple scanning strategy spending an equal amount of time in each frequency bin ($t_{\text{int}, \Delta f} \sim t_\text{int,tot}/Q$ and $\Delta f_\text{det} \sim f/Q$) does not lead to any gain in SNR for a large quality factor $Q$, since the increase of the integrand of \cref{eq:QuadSNR} by a factor of $Q$ is compensated by reduced time and frequency interval per bin. However, since SGWBs typically have a broad frequency spectrum, one could consider running a resonant detector at a fixed frequency (no scanning), with $t_\text{int} = \SI{1}{year}$ as above. In this case, the sensitivity to $\Omega_\text{GW}$ scales as $Q^{-1/2}$ (with the exception of detectors limited by the single-photon threshold, which do not profit from this scaling). We show this sensitivity as dashed lines in \cref{fig:PLS}, emphasizing that this indicates the possible reach at a given sensitivity, while fully covering the entire frequency range shown would require an unrealistic amount of time, or an unrealistic number of detectors running in parallel at different resonance frequencies.

\reply{Figure~\ref{fig:PLS} shows the resulting exclusion regions, prospective sensitivities and possible signals in units of $\Omega_\text{GW} h^2$, with $h = H_0/(100 \text{ km/s/Mpc})$ denoting the dimensionless Hubble parameter.} Several comments are in order.
Firstly, we note that no proposal above the LIGO--VIRGO--KAGRA band currently reaches below the cosmological bound of $\Omega_\text{GW} h^2 \lesssim 10^{-6}$ arising from the limits on excess energy density in relativistic degrees of freedom ($N_\text{eff}$) at BBN (see \cref{sec:earlyU}). Therefore, cosmological GW sources seem currently out of reach. Scenarios detectable with current sensitivities would for the most part imply values of $\Omega_\text{GW} \gg 1$, which taken at face value would correspond to a GW dominated universe. In this context, \cref{fig:PLS} can be interpreted as (i) showing the sensitivity to local overdensities of GW energy and (ii) indicating the improvement in sensitivity needed to probe cosmological sources. Secondly, we caution that the sensitivity curves shown for laboratory detectors do not take into account the angular response function of the detectors but assume that a local overdensity of GW energy is located in the optimal position with respect to the field of view. For detectors with a broad field of view, such as interferometer or electromagnetic oscillators, the sky-averaged sensitivity is about a factor 10 smaller than this ideal sensitivity. For detectors with a very narrow field of view, such as some photon regeneration experiments, the degradation can be much more significant. Third, we note that the bandwidth of broadband detectors is limited by the data acquisition system. Here we have assumed a readout covering the entire frequency range of these detectors as shown in \cref{fig:PLS}, which in some case would require multiple layers of readout systems.

\Cref{fig:PLS} also shows cosmological and astrophysical bounds on UHF GWs. Besides the aforementioned BBN bound and a very similar bound from the CMB, not shown here, this includes limits based on GW-to-photon conversion in astrophysical environments with strong magnetic fields, see \cref{sec:AstroDetectors}. The blue points in the upper right corner of \cref{fig:PLS} correspond in particular to limits from GW-to-photon conversion in galactic magnetic fields. Additional astrophysical bounds are summarized in \cref{fig:astro} in \cref{sec:AstroDetectors}, but the galactic ones are the only limits which translate to constraints $\Omega_\text{GW} h^2 < 10^{15}$.

Finally, \cref{fig:PLS} also shows a representative selection of SGWB sources, discussed in more detail in \cref{sec:th}. The regions bounded by the colored curves illustrate the region of parameter space which may be covered by the corresponding source for appropriate parameter choices as specified below. Except for the cases of inflation with broken spatial reparametrization symmetry and the cosmic gravitational microwave background they should not be mistaken for GW spectra obtained for a fixed model parameter choice. Rather, they show the estimated envelopes of the signals obtained in different classes of models, and should thus be seen as the most optimistic estimate for possible signals.
\begin{itemize}
    \item In certain models, inflation (\cref{sec:Inflation}) can yield a signal stretching over a broad frequency range (see \cref{eq:fN}), with an amplitude determined by \cref{PGW-frequency,eq:nteft}, respectively. Here in the case of inflation with extra-species we have taken the parameter $\xi$ (defined in \cref{PGW-frequency}) to be bounded by the perturbative limit, and in the case of inflation described by an effective field theory with broken spatial reparametrization symmetry we have chosen the speed of sound and the spectral tilt to be $c_T = 1$ and $n_T = 0.2$, respectively. Moreover, inflation models with strongly enhanced scalar fluctuations ($P_\zeta \lesssim 10^{-2.5})$ can source GWs with $\Omega_{\rm GW, 0} \lesssim 10^{-9}$ at second order in cosmological perturbation theory.
    
    \item For preheating (\cref{sec:Preheating}), we show typical values for models with parametric resonance in quadratic ("preheating 1") and quartic ("preheating 2") potentials as well as oscillons. In the latter case the frequency is set by the mass of the scalar field through \cref{eq:OscillonFrequency}, where here we have chosen the mass of the scalar field to be $\SI{e10}{GeV} < m < \SI{e13}{GeV}$ with $X = 100$, while the amplitude is the typical value inferred from numerical simulations.
    
    \item For the cosmic gravitational microwave background \reply{(\cref{sec:CGMB})}, we show the spectrum given by \cref{eq:CGMB} with $T_{\max} = \SI{e16}{GeV}$, which is the upper bound on the reheating temperature set by the constraints on the tensor-to-scalar-ratio \citep{Akrami:2018odb}.
    
    \item For phase transitions (\cref{sec:PhaseTransitions}), we have obtained an envelope of curves with strength parameter $\alpha=10^2$, duration parameter $\beta/H_*=1$, and $v_w=1$. 
    
    \item As an example for topological defects (\cref{sec:TopologicalDefects}), cosmic strings lead to a broad spectrum with an amplitude given in \cref{eq:plateauStringsNG}, where the string tension for stable cosmic strings is bounded by $G \mu < 10^{-10}$  from PTA measurements, whereas for metastable cosmic strings it can be as large as $G \mu \simeq 10^{-3}$ above the LIGO frequency range. The spectrum of gauge textures is described by \cref{eq:gaugetextures}, where here we have chosen the symmetry breaking scale to be $\SI{e12}{GeV} < v < \SI{e19}{GeV}$.

    \item PBH mergers also produce a SGWB in the late universe, as discussed in \cref{sec:PBHmergers}. With the line shown in \cref{fig:PLS}, we indicate the envelope of the maximal amplitudes reached by such a SGWB, varying the assumed typical population mass $m_{\rm PBH}$, which is related to the peak frequency through \eqref{eq:fISCO-PBH}.
\end{itemize}

\section{Sources of Gravitational Waves at High Frequencies}
\label{sec:th}

This section reviews various production mechanisms for GW signals in the high-frequency regime, typically in the kHz--GHz range, that fall into two broad classes: late Universe sources and early Universe sources. The former category, which we discuss in \cref{sec:lateU}, corresponds to sources in our cosmological neighborhood, emitting coherent transient and/or monochromatic GW signals. Early Universe sources, which will be the topic of \cref{sec:earlyU}, in contrast are sources at cosmological distances which typically lead to a stochastic background of GWs. We emphasize that all proposed sources, with the notable exceptions of the neutron star mergers discussed in \cref{sec:NS-mergers} (kHz range) and the cosmic gravitational microwave background discussed in \cref{sec:CGMB}, require new physics beyond the Standard Model of particle physics to produce an observable GW signal. Thus, while being admittedly somewhat speculative, these proposals provide unique opportunities to shed light on the fundamental laws of nature, even by `only' setting upper bounds on the existence of GWs in the corresponding frequency range.


\subsection{Late Universe}
\label{sec:lateU}

In the following, we give an overview of high-frequency GW sources that are active in the late Universe. A concise summary of these sources is given in \cref{tab:summary-coherent} in \cref{sec:SummaryTable}.

\subsubsection{Known Astrophysical Systems} 

\paragraph{Core-Collapse Supernovae.}
\label{sec:CCSNe}

\begin{figure}
    \centering
    \includegraphics[width=0.45\textwidth]{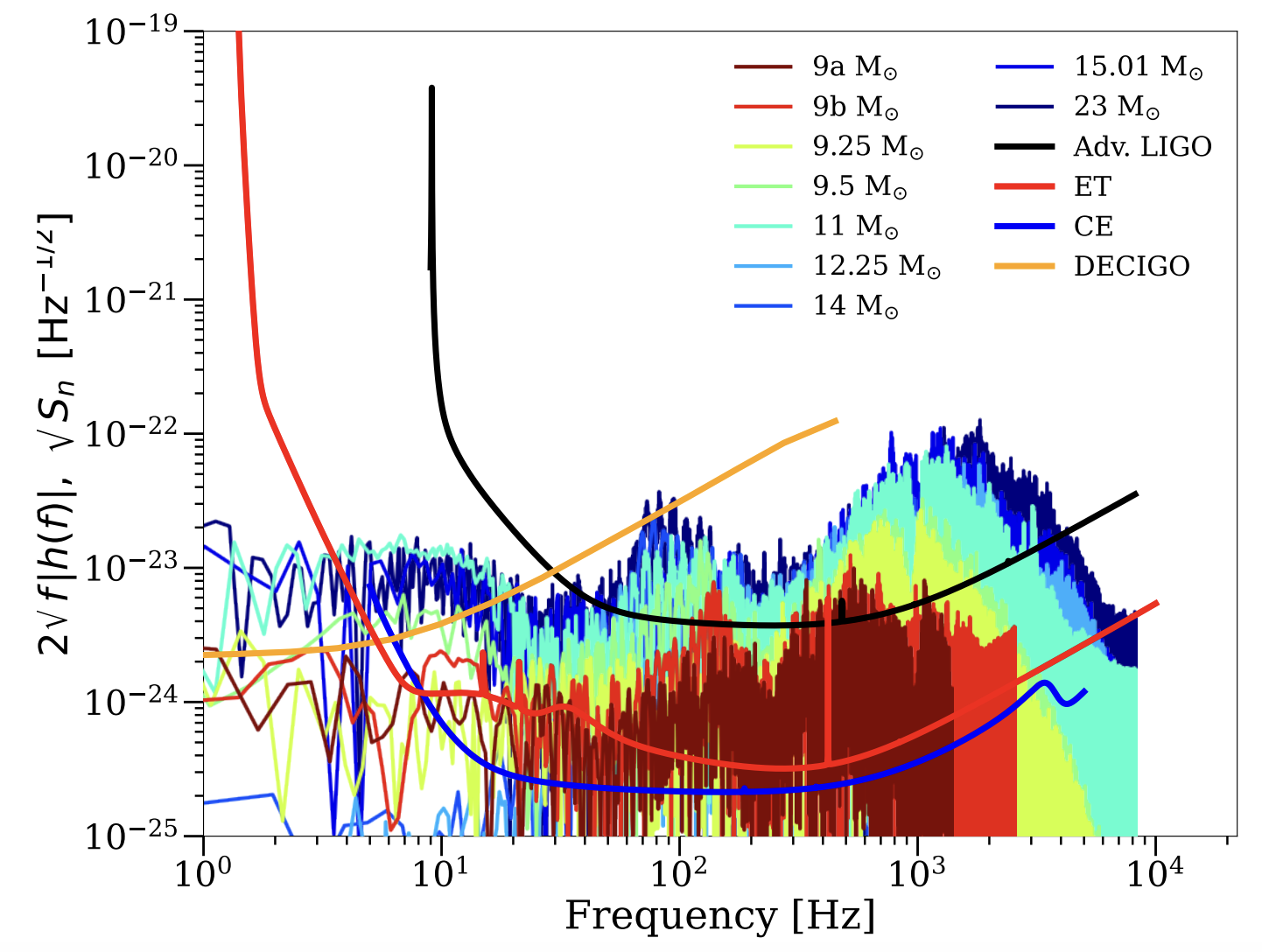}
    \raisebox{-0.05cm}{\includegraphics[width=0.51\textwidth]{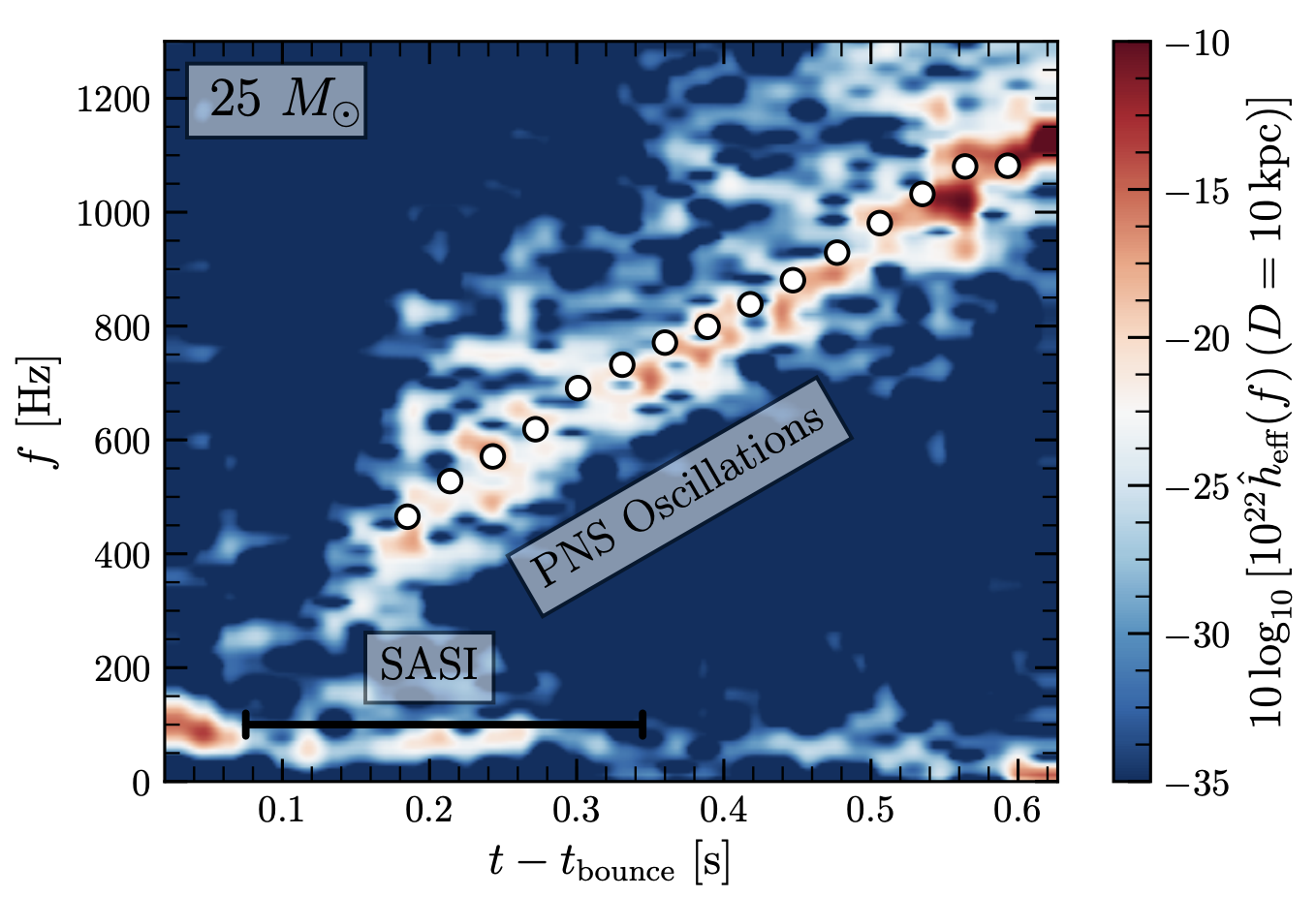}}
    \caption[Gravitational wave signals from core-collapse supernovae.]
    {{\bf Left:} GW spectrum of slowly rotating core-collapse supernovae from several different simulations compared to the sensitivities of interferometric detectors. {\bf Right:} Frequency of the signal from a core-collapse supernova of a $25 M_\odot$ progenitor star as a function of time and of the proto-neutron star's oscillatory modes. The white dots denote the eigenfrequencies associated with the quadrupolar f- and g-modes of the PNS. Asymmetric accretion produces an early subdominant peak around \SI{100}{Hz} and excites the proto-neutron star oscillations which emit the dominant peak around \SI{1}{kHz}. Figures taken from \cite{Vartanyan:2023sxm, Radice:2018usf}.}
    \label{fig:supernovae-spectrum}
\end{figure}

Massive stars reach the end of their lives by exploding in a core-collapse supernova (CCSN), giving birth to neutron stars and black holes (BHs). GW detection from CCSNe is a promising candidate to learn about the inner core dynamics and explosion mechanism, as well as the properties of nuclear matter at high densities (see \cite{Abdikamalov:2020jzn} for a review). As the core collapses, forming a proto-neutron star (PNS), it reaches supranuclear densities, and the stiffness of the PNS stops the in-fall and bounces back a shock wave that triggers the explosion.

For slowly rotating CCSNe, neutrino-driven convection, turbulent flow, and the standing-accretion shock instability (SASI) produce asymmetric flows that generate GWs at $\sim \SI{100}{Hz}$. More importantly, though, these mechanisms exciting the oscillatory modes of the PNS, which lead to much stronger GW emission at $\sim \SI{1}{kHz}$ (see \cref{fig:supernovae-spectrum}). These oscillatory modes depend solely on and the mass and equation of state of the PNS, implying that valuable information about the nuclear matter could be obtained by observing them (see e.g.\ \cite{Jakobus:2023fru, Kunjipurayil:2022zah}). For example, the frequency and the amplitude of the dominant peak both increase with the effective in-medium mass of the nucleons forming the PNS \citep{Andersen:2021vzo}. The overall signal contains additional information about the explosion, in particular, the total energy radiated is strongly correlated with the energy in turbulent flow as well as with the compactness of the original star \citep{Vartanyan:2023sxm, Radice:2018usf}.

For fast-rotating stars, the PNS is born with an asymmetry, determining the dominant pulsations. Rotation enhances the GW signal strength until centrifugal forces become too strong and prevent the PNS from acquiring larger densities \citep{Abdikamalov:2013sta}. Furthermore, instabilities associated with rotation produce new signatures in the \SI{100}{Hz}--\SI{1}{kHz} band \citep{Shibagaki:2019mlq, Hsieh:2023djs}.

If the mass of the PNS is too large, it will eventually collapse into a black hole. In this case, a sudden drop in frequency after the signal peaks at $\sim \SI{1}{kHz}$ is observed as a signature of the collapse \citep{Cerda-Duran:2013swa}.

We see that CCSNe are expected to emit GWs at the upper high end of the frequency range covered by ground-based interferometers. However, even higher-frequency GWs could be radiated. For example, if the nuclear matter in the PNS undergoes a first-order phase transition into quark matter, a rapid contraction and second bounce of the core is expected. In this case, the peak of the GW signal at $\sim \SI{1}{kHz}$ is shifted to higher frequencies, $\sim 2$--\SI{4}{kHz}, associated with the quadrupolar pulsation modes of a more compact body \citep{Abdikamalov:2008df, Zha:2020gjw}. In addition, the dynamics of the phase transition may give rise to a signal in the MHz band \citep{Cao:2018tzm, Casalderrey-Solana:2022rrn}. Both signals would offer invaluable information about the properties of dense QCD matter.

\paragraph{Neutron Star Mergers.}
\label{sec:NS-mergers}

The collisions of neutron stars provide perfect environments for probing the warm and dense region of the QCD phase diagram. The process starts with a long inspiral phase, followed by the post-merger dynamics (see \cite{Baiotti:2016qnr, Sarin:2020gxb, Lovato:2022vgq} for reviews).

\begin{figure}
    \centering
    \includegraphics[width=0.6\textwidth]{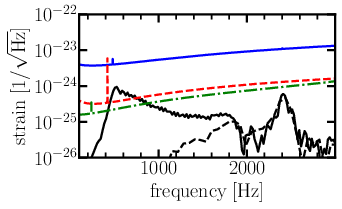}
    \caption[Gravitational wave spectrum of a binary neutron star merger.]
    {GW spectrum of a binary neutron star merger, including the inspiral and the post-merger dynamics. The post-merger emission is in the kHz band and exhibits three characteristic peaks independently of the choice of EoS. This contribution is singled out by the dashed line. Figure taken from \cite{Sarin:2020gxb}.     Colored diagonal lines indicates the forecasted sensitivity of future ground based interferometers Advanced LIGO (blue), Einstein Telescope (red) and Cosmic explorer (red).}
    \label{fig:NSM-spectrum}
\end{figure}

The full GW spectrum from a binary neutron star merger is shown in \cref{fig:NSM-spectrum}. The inspiral phase leads to the emission of a relatively low frequency (hundreds of Hz) GW signal, which carries information about the quadrupole tidal deformability of the stars (and therefore the matter equation of state), the compactness of the stars, and the binary mass ratio \cite{Hinderer:2007mb, Read:2013zra, Bernuzzi:2014kca}. 

The post-merger dynamics is the process during which most of the GW energy is radiated. The signal features depend more strongly on the underlying equation of state (EoS), including finite-temperature effects. The post-merger signal is present as long as prompt collapse into a black hole is avoided, and it depends on the dynamics of the metastable (or stable) rotating remnant. Simulations for a wide range of EoS show that three peaks are characteristic in this phase \citep{Takami:2014zpa, Bauswein:2015vxa}. The dominant peak, at frequency $f_{\mathrm{peak}}$, is associated with the fundamental quadrupolar fluid mode, which has been shown to be correlated with the maximum radii of a non-rotating star a given EoS could support \citep{Bauswein:2015vxa}. The subdominant part of the spectrum also encodes non-trivial information about the EoS. In particular, a second subdominant peak is produced by the orbital motion of antipodal bulges at the surface of the remnant right after the merger \citep{Bauswein:2015vxa, Bauswein:2018bma}, while additional features were identified with the coupling to the quasi-radial mode, see \cite{Bauswein:2015vxa}.

Whether the remnant eventually collapses to a BH or not is difficult to conclude from the post-merger GW signal. The eventual collapse induces an excess of power at higher frequencies, around the ringdown frequency of the produced black hole, which is absent if the remnant is stable \citep{Dhani:2023ijt}. If the merger leads to a prompt collapse, the post-merger emission is shut down and taken over by the ringdown signal of the corresponding rotating black hole. The peak frequency is then shifted towards higher frequencies, up to \SI{10}{kHz} \citep{PhysRevD.40.3194, Dhani:2023ijt}, making it possible to distinguish mergers that lead to a prompt collapse from those that only lead to a delayed collapse, or no collapse at all. Discerning among all these cases would have strong implications on our understanding of the EoS of dense nuclear matter, including the possibility of first-order phase transitions to quark matter in the core \citep{Most:2018eaw, Tootle:2022pvd, Demircik:2022uol}, which we discuss next.

\paragraph{First-order Phase Transitions in Neutron Stars.}

An additional potential high-frequency GW signal associated with binary neutron star mergers could arise from the dynamics of a first-order QCD phase transition (FOPT) occurring during the merger \citep{Casalderrey-Solana:2022rrn}. In such a phase transition, the core of the star would transition from the hadronic matter phase into a quark matter phase or into a color superconductor.\footnote{A similar phenomenon could take place in a neutron star that undergoes quick gravitational collapse during supernova explosion, see \cite{Cao:2018tzm}.} Whether or not this phase transition is accessible at the densities and temperatures realized in a neutron star merger, and whether it is first order, is currently unknown, though indications for a first-order nature exist \citep{Berges:1998rc, Buballa:2003qv}.  We will assume here that both conditions are met, so that GW emission can occur. Given the adiabaticity of the merger timescale compared to the timescales of the underlying microscopic nuclear processes, ($\SI{1}{ms} \gg \SI{e-20}{ms} \simeq \SI{1}{fm}$), a realization of the phase transition through bubble nucleation, expansion, and collision is expected, similar to the dynamics of cosmological first-order phase transitions (see e.g.\ \cite{Hindmarsh:2020hop} for a review).

The peak frequency of the GW signal from a FOPT inside a neutron star is determined by the average size, $R$, of the quark matter bubbles at the time they collide. $R$ is set by the speed of the bubble walls, $v_w$, together with the duration of the transition, $\beta^{-1}$,
\begin{align}
    f_\text{peak} = R^{-1} = (8\pi)^{-1/3} v_w^{-1} \beta \,.
\end{align}
Most of the uncertainty in the GW spectrum originates from the wall speed as it is a challenging property to compute from first principles for a given theory (see \cite{Moore:1995si, Dorsch:2018pat, Lewicki:2021pgr, Laurent:2022jrs, Jiang:2022btc, Bigazzi:2021ucw, Bea:2021zsu, Bea:2022mfb, Janik:2022wsx, Sanchez-Garitaonandia:2023zqz} for some computations at weak and strong coupling). The duration of the transition $\approx \beta^{-1}$ can be estimated from the ratio between the microscopic scale $\Lambda$ and the macroscopic one, $\tau$ \citep{Casalderrey-Solana:2022rrn, Hindmarsh:2020hop}, leading to
\begin{equation}
    f_{peak} \simeq
        \bigg(\! \frac{0.1}{v_w} \bigg) \!
        \bigg(\! \frac{\SI{1}{ms}}{\tau} \bigg) \!
        \bigg(\! 0.62 + \frac{\num{2e-3}}{\pi^{1/3}} \log\bigg[
            \bigg(\!\frac{v_w}{0.1}\bigg)^3 \!
            \bigg(\!\frac{\tau}{\SI{1}{ms}}\bigg)^4 \!\!
            \frac{\Lambda^4}{\SI{1}{GeV/fm^3}}\bigg] \bigg) \si{MHz}
\end{equation}
Taking $\Lambda \simeq \SI{1}{GeV/fm^3}$ based on dimensional arguments \citep{Annala:2019puf}, and $v_w \sim 0.1$, the peak frequency falls into the Mega-Hertz band, $f_{peak} \approx 0.6$ MHz, two orders of magnitude above the signal from macroscopic oscillations of the neutron star, discussed above.

The estimation of the strain is subject to several uncertainties, but a rough approximation can be obtained using results from the cosmological phase transitions literature for the total energy radiated (see e.g.\ \cite{Hindmarsh:2015qta}). This leads to the following expression for the observed strain \citep{Casalderrey-Solana:2022rrn} (written following the notation from \cref{eq:hcStoch_Omegag}),
\begin{align}
    h_{c,{\rm sto}} \simeq
        1.8\times 10^{-25} v_f^2 \times
        \bigg(\! \frac{\Lambda^4}{\SI{1}{GeV/fm^3}} \bigg)
        \bigg(\! \frac{L}{\SI{1}{km}} \!\bigg)^{3/2}
        \bigg( \frac{\SI{1}{MHz}}{f_{\text{peak}}} \bigg)^{3/2}
        \bigg( \frac{\SI{100}{Mpc}}{D} \bigg) ,
    \label{eq:hc-ns-pt-1}
\end{align}
with $D$ the luminosity distance to the NS merger, $L$ the size of the region in the NS that undergoes the transition, and $v_f$ the typical velocity of the fluid after the collision of all bubbles. Using the same numerical parameter values as before, and $L \simeq \SI{5}{km}$ \citep{Tootle:2022pvd, Demircik:2022uol}, \cref{eq:hc-ns-pt-1} reduces to
\begin{align}
    h_{c,{\rm sto}} \simeq 1.5 \times 10^{-24} v_f^2 \times
        \bigg( \frac{\SI{100}{Mpc}}{D} \bigg) \,.
\end{align}
These estimates are based on the assumption that GWs are acoustically generated after the bubble collisions. Sound waves are expected to have a lifetime of order a millisecond, setting the duration of the emission. Simulations show that during the merger several regions that undergo the transition cross back to the initial phase later \citep{Tootle:2022pvd, Demircik:2022uol}, thereby undergoing an additional transition. This implies that several signals are expected to come from a single merger, all with a peak frequency around the MHz band.

The detection of such a signal would imply that a FOPT is present in nuclear matter at high densities, it would constrain the location of this phase transition in the QCD phase diagram, and it would elucidate its dynamics. It would therefore provide major insights into the physics of strong interactions that are very difficult to obtain in any other way.

\paragraph{Disks around Supermassive BHs.}

In \cite{Saito:2021sgq}, it was shown that photons emitted from accretion disks around supermassive black holes can be converted into gravitational waves in the black hole's magnetosphere through the Gertsenshtein effect \cite{Gertsenshtein}, inducing a high-frequency GW signal, which experimentally would manifest \reply{itself} as a stochastic background.  In practice, photons from the accretion disk steadily accumulate around the photon sphere. If their frequency matches the resonance frequency at which the effective photon mass (that receives opposite-sign contributions from plasma effects and magnetic field effects) vanishes, they are efficiently converted into gravitons of the same frequency by the magnetic field. The characteristic frequency of the resulting GWs is therefore \citep{Saito:2021sgq}
\begin{align}
    f \sim \frac{1}{2 \pi}
           \bigg( \frac{45}{12 k \alpha} \frac{m_e^3}{m_p c^2} \bigg)^{1/2}
      \sim  \SI{3.3e19}{Hz} \,,
\end{align} 
where $m_e$ and $m_p$ are the electron and proton mass, respectively, while $\alpha$ is the electromagnetic fine structure constant, and $k$ ($=2$ or $7/2$ for the $+$ and $\times$ polarizations, respectively) controls the magnetic field contribution to the effective photon mass. Interestingly, $f$ does not depend on the supermassive BH mass and the magnetic field and inevitably falls in the UHF-GW window. 

\reply{
The GW luminosity can be estimated based on the conversion probability. By integrating the emission from all supermassive black holes in the Universe, one predicts a stochastic gravitational wave background with energy density
\begin{align}
    \Omega_{\rm GW} \simeq 2 \times 10^{-12} \, \xi \,,
    \label{eq:Omega-bh-disk}
\end{align}
where $\xi \lesssim 1$ is the dimensionless ratio between the black hole horizon area and the accretion disk area. 
This estimate assumes a small tilt of the SMBH mass function, meaning that the mass is taken to follow a mild power-law dependence 
\begin{align}
n(M) \propto M^{-\beta},
\end{align}
with $\beta \ll 1$, consistent with current SMBH mass function measurements. Here, $n(M)$ denotes the comoving number density of SMBHs of mass $M$, which we take to be spatially homogeneous across the observable Universe.
The observed SMBH mass range spans $M \sim 10^{6} M_{\odot} - 10^{11} M_{\odot}$, and using this mass function with a small tilt leads a parametric dependence omitted in Eq.~\eqref{eq:Omega-bh-disk} for simplicity, see \cite{Saito:2021sgq} for more detail. For $\beta \ll 1$, one maximises the amplitude of the signal, which yield $\Omega_{\rm GW}$ at the level of $10^{-12}$.
Important uncertainties remain due to the unknowns in the SMBH population, particularly the precise number density and mass distribution of these objects across cosmic time.}

\paragraph{GW Spectrum of the Sun.}
The high-temperature plasma within stellar interiors generates stochastic GWs~\citep{Weinberg:1972kfs,Gould:1985, Garcia-Cely:2024ujr}, with frequencies roughly determined by the temperature at the core. For the Sun, this results in a spectrum spanning the range $10^{12}$–\SI{e19}{Hz}, peaking at \SI{e18}{Hz}. These GWs are produced through two primary mechanisms:
\begin{itemize}
    \item Hydrodynamic fluctuations. These are sourced by tensor fluctuations of the energy--momentum tensor of the solar plasma and are proportional to the shear viscosity $\eta$~\citep{Ghiglieri:2015nfa}. The resulting GW emission power is given by \cite{Garcia-Cely:2024ujr}
    \begin{align}
        \frac{dP}{d\omega} \bigg|_{\text{Hydrodynamics}}
            = \frac{16 G \omega^2}{\pi} \int_{\text{Sun}} d^3r \, \eta T \,,
    \end{align}
    where $T$ is the temperature of the solar plasma and $\omega = 2 \pi f$.
    
    \item Graviton emission from  particle collisions. In contrast to hydrodynamical fluctuations, these correspond to frequencies higher than those of collisions in the solar plasma, so that there is sufficient time for them not
to interfere with each other. In this case \cite{Garcia-Cely:2024ujr}
    \begin{align}
        \frac{dP}{d\omega} \bigg|_{\text{Collisions}}
            = \int_\text{Sun} \! d^3r \sum_i \omega \,
              \bigg\langle \frac{d\Gamma^{(i)}(r)}{d\omega \, dV} \bigg\rangle \,,
        \label{eq:sunPC}
    \end{align}
    where $\langle\cdot\rangle$ denotes a thermal average and $\Gamma^{(i)}$ is the graviton emission rate for each process: 
    {\it i)} photoproduction $\gamma Z \to e h$ and $\gamma e \to e h$;
    {\it ii)} bremsstrahlung $e Z \to e Z h$;
    {\it iii)} bremsstrahlung $ee \to ee h$. 
\end{itemize}
The characteristic strain amplitude $h_c(f)$ of the stochastic gravitational wave background from the Sun can be expressed as
\begin{align}
    h_{c,{\rm sto}} = \frac{1}{D_{\odot}}
        \left( \frac{2 G \, dP/d\omega}{\omega} \right)^{1/2}
    \simeq 10^{-42},
    \label{eq:strainfromsun}
\end{align}
where $D_{\odot}$ is the distance from the Earth to the Sun. While we use here the same notation as in \cref{eq:hcStoch_Omegag}, it should be kept in mind that the stochastic GW signal from the Sun is highly anisotropic and defined by an integration over the solid angle under which we see the Sun.

In analogy to the Sun, also the other main-sequence stars in the galaxy are expected to emit a similar GW signal; the characteristic strain of their integrated emission has been found to be a few orders of magnitude lower than the one in \cref{eq:strainfromsun}~\citep{Gould:1985, Garcia-Cely:2024ujr}.

\subsubsection{Light Primordial Black Holes}
\label{sec:PBHmergers}

The detection of BH mergers by LIGO and Virgo has revived the interest in primordial BHs (PBHs) in the mass range (1--100)\,$M_\odot$ \citep{Bird:2016dcv, Clesse:2016vqa, Sasaki:2016jop}, which could constitute a relevant fraction of the observed dark matter abundance. 
In this context, detecting a sub-solar mass compact object, and provided large tidal effects are excluded \citep{Crescimbeni:2024cwh}, would point to a primordial origin.\footnote{
See, however, \cite{Kouvaris:2018wnh, Takhistov:2020vxs, Dasgupta:2020mqg, Chakraborty:2024eyx} for other formation channels of sub-solar BHs, such as white dwarf or neutron star transmutation triggered by accretion of dark matter.}
PBHs can form in a much wider range of masses than what is expected from astrophysical formation mechanisms (see e.g.\ \citep{LISACosmologyWorkingGroup:2023njw, Carr:2023tpt} recent reviews), with their size typically related by ${\cal O}(1)$ factors to the mass contained within one Hubble sphere at the time of production in the early Universe.
Many constraints were set on the abundance of PBHs (usually parameterized as as fraction of the total DM abundance, $f_{\rm PBH} \equiv \Omega_{\rm PBH} / \Omega_\text{DM}$) across many orders of magnitude in mass, while the so-called asteroid mass range, $m \sim 10^{-12} M_\odot$, currently remains very challenging to probe \citep{Katz:2018zrn, Carr:2020gox}.  UHF-GWs may allow us to set unprecedented constraints on this elusive population of objects, potentially addressing the question of whether they compose a significant fraction of dark matter.

\paragraph{PBH Mergers.}

The GW emission from a binary inspiral is close to maximal at the innermost stable circular orbit (ISCO), which marks the end of the inspiral phase and the beginning of the merger phase.\footnote{Slightly larger strains are reached during the merger, but we focus on the ISCO here to allow for analytic estimates of the strain.} The ISCO frequency is given by
\begin{align}
    f_{\rm ISCO} = \SI{4400}{Hz} \, \frac{M_\odot}{M} \,,
    \label{eq:fISCO-PBH}
\end{align}
where we have introduced the total mass of the binary $M = m_1 + m_2$ and $M_\odot$ denotes the solar mass.
Frequencies in the range $10^4$--$\SI{e15}{Hz}$ correspond to a primordial BH mass range $10^{-12}$--$\SI{e-1}{M_\odot}$. In particular, the planetary-mass range, in which recent detections of star and quasar microlensing events \citep{Niikura:2019kqi, Hawkins:2020zie, 2019ApJ...885...77B, Mroz:2024mse} allow a PBH fraction of $f_{\rm PBH} \sim 0.01$, could be probed in a novel and independent way with GWs.

A good estimate of the GW strain produced by a circular PBH binary at a given frequency $f$ can be obtained at zeroth post-Newtonian (0-PN) order \citep{Maggiore:1900zz,Antelis:2018sfj}:\footnote{We assume that GW emission is the dominant effect driving binary evolution. While accretion can speed up binary evolution (also enhancing the merger rates \citep{Ali-Haimoud:2017rtz,DeLuca:2020qqa}), it is typically small in the subsolar mass range of interest here (see e.g.\ \cite{Ricotti:2007au}).}
\footnote{\reply{Throughout this section we neglect cosmological redshift effects. As illustrated in Fig.~\ref{fig:sharkfin}, the mergers considered here occur at distances of at most $\mathcal{O}(10)$~Mpc, corresponding to redshifts $z \sim 10^{-3}$. At such low redshifts, cosmological corrections (e.g., redshifting of the chirp mass or luminosity distance) are negligible.
}}
\begin{align}
    h(f) &= \bigg(\frac{5}{24}\bigg)^{1/2} \frac{1}{\pi^{2/3}} \frac{1}{D}
            (G M_c)^{5/6} f^{-7/6} e^{i \psi} Q(\theta,\phi,\varphi) 
                                \nonumber \\
    &\approx \SI{2e-37}{sec} \,
        \bigg( \frac{\rm kpc}{D} \bigg)
        \bigg( \frac{m_{\rm PBH}}{\SI{e-12}{M_\odot}} \bigg )^{5/6}
        \bigg( \frac{f}{\rm GHz} \bigg )^{-7/6}\ ,
    \label{eq:strainPBHs}
\end{align}
where $G$ is Newton's constant; $M_c \equiv (m_1 m_2)^{3/5}/(m_1+m_2)^{1/5}$ is the chirp mass of a binary with constituent masses $m_1$, $m_2$;  $D$ is the luminosity distance from the binary to the observer; $\psi$ is a phase; and $Q(\theta,\phi,\varphi)$ is a function that depends on the position of the binary with respect to the detector, and the angle $\varphi$ between the normal of the orbit and the line of sight.
In the second line of \cref{eq:strainPBHs} we have fixed $m_1 = m_2 \equiv m_{\rm PBH}$.
\reply{Throughout this section we consider quasi-circular orbits. This is justified by the fact that PBH binaries typically form at high redshift and undergo long periods of GW-driven evolution, which efficiently circularizes their orbits before they become observable.}
This modeling of the GW signal only describes the inspiral phase of the binary roughly until the ISCO frequency is reached. 
While it neglects the merger and the ringdown part of the signal, it is sufficient for the present purposes as only the GW signal produced during the inspiral phase can last for a sufficiently long time to allow for a potential detection.

A crucial quantity for determining detection prospects for GWs from PBH binaries is the time, or the number of orbital cycles $N_\text{cycles}$, the GW signal spends within a given frequency interval. As discussed in \cref{sec:sensitivity-transients}, this time may in particular be shorter than the integration time of the detector, limiting the sensitivity. For an equal mass PBH binary ($m_1 = m_2 = m_{\rm PBH}$) and assuming energy loss is dominated by GW emission,\footnote{Environmental effects such as the presence of accretion disks could speed up binary evolution, but are expected to be subdominant in the subsolar mass range.} $N_{\rm cycles}$ is given by \citep{Moore:2014sen}
\begin{align}
    N_\text{cycles} = \frac{f^2}{\dot{f}}
                      \simeq 2.2 \times 10^6 \,
                      \bigg( \frac{f}{\si{GHz}} \bigg)^{-5/3}
                      \bigg( \frac{m_{\rm PBH}}{\SI{e-9}{M_\odot}} \bigg)^{-5/3} \,,
    \label{eq:Ncycles}
\end{align}
where we have used \citep{Maggiore:1900zz}
\begin{align}
    \dot{f} = \frac{96}{5} \pi^{8/3} \bigg(\frac{G M_c}{c^3}\bigg)^{5/3} f^{11/3} 
    \simeq \SI{4.6e11}{Hz^2} \bigg( \frac{m_{\rm PBH}}{\SI{e-9}{M_\odot}} \bigg)^{5/3}
                             \bigg( \frac{f}{\si{GHz}} \bigg)^{11/3} \,.
    \label{eq:dotf}
\end{align}
Note that only close to the ISCO frequency, namely in the final phase of the inspiral, the number of cycles becomes of order unity. Note also that $N_\text{cycles}$ determines whether the signal can be approximated as nearly monochromatic, which is the case when $N_\text{cycles} \gg 1$.

A useful quantity closely related to $N_\text{cycles}$ is the time to coalescence, which is given by \citep{Maggiore:1900zz}
\begin{align}
    \tau(f) \approx \SI{83}{sec} \,
                    \bigg( \frac{m_{\rm PBH}}{\SI{e-12}{M_\odot}} \bigg)^{-5/3} \bigg( \frac{f}{\si{GHz}} \bigg)^{-8/3} \,.
    \label{eq:CoalescenceTime}
\end{align}

\paragraph{Formation Channels for PBH Binaries.}

There are two main formation channels for primordial BH binaries (see e.g.\ \cite{Raidal:2024bmm}):
\begin{enumerate}
    \item \textit{Primordial binaries.} These are pairs of PBHs that were formed sufficiently close to each other for their dynamics to decouple from the expansion of the Universe before the time of matter--radiation equality \citep{Nakamura:1997sm, Sasaki:2016jop}. The gravitational influence of one or several PBHs nearby prevents the two BHs from merging directly, leading to the formation of a binary.  Typically, the binaries are sufficiently stable, \reply{i.e. are not disrupted by interaction with the surrounding environment}, and a large fraction of them merge on a timescale on the order of the age of the Universe. If the PBHs have a mass spectrum $\rho(m)$ and are randomly distributed spatially, and assuming that early formation of PBH clusters does not impact the lifetime of these primordial binaries (a criterion satisfied for $f_{\rm PBH} \lesssim 0.1$) \citep{Raidal:2018bbj}, then the present day merger rate is approximately given by \citep{Kocsis:2017yty, Raidal:2018bbj, Gow:2019pok}
    \begin{multline}
        \frac{dR_{\rm PBH}}{d(\ln m_1) \, d(\ln m_2)}
            = \SI{0.0038}{kpc^{-3} yr^{-1}} \times f_{\rm PBH}^{\frac{53}{37}}
              \bigg( \frac{t}{t_0} \bigg)^{-\frac{34}{37}}
              \bigg( \frac{M}{\SI{e-12}{M_\odot}} \bigg)^{-\frac{32}{37}}
                \\ \times
              \bigg[ \frac{m_1 m_2}{(m_1+m_2)^2} \bigg]^{-\frac{34}{37}} 
              S(M, f_{\rm PBH}, \psi) \rho(m_1) \rho(m_2) \,,
        \label{eq:PBHrate}
    \end{multline}
    where $f_{\rm PBH}$ is the integrated dark matter fraction made of primordial BHs, $m_1$ and $m_2$ are the masses of the two constituent BHs of the binary, and $\rho(m)$ is the PBH mass function normalized to one ($\int \rho(m) \, d\ln m = 1$).
    \reply{Here $t$ stands for the universe age at the time of merger, and can be approximated with $t_0$ at sufficiently low redshift. }
    The suppression factor $S(M, f_{\rm PBH}, \psi)$ in \cref{eq:PBHrate} corrects the merger rate by introducing the effect of binary interactions with the surrounding environment in both the early- and late-time Universe (see \cite{Hutsi:2020sol} for its analytical parametrization), informed by the numerical simulations performed in \cite{Raidal:2018bbj}).  

    \item \textit{Capture in primordial BH halos} The second PBH binary formation channel is through dynamical capture in dense primordial halos. As with any other dark matter candidate, PBHs are expected to form halos during cosmic history. Even more so if they compose a large fraction of dark matter, as structure formation at small scales is boosted by the initial Poisson perturbations in the PBH distribution \citep{Inman:2019wvr, DeLuca:2020jug}. For a generic PBH mass function $\rho(m)$, an effective formula for the merger rate of binaries formed in the late-time universe is~\citep{Clesse:2020ghq,Carr:2019kxo}
    \begin{align}
         \frac{d R_{\rm PBH}}{d(\ln m_1) \, d(\ln m_2)}
             &\approx R_{\rm clust} f_{\rm PBH}^2 \, \rho(m_1) \, \rho(m_2)
                      \frac{(m_1 + m_2)^{10/7}}{(m_1 m_2)^{5/7}} \si{yr^{-1} Gpc^{-3}} \,,
         \label{eq:ratescatpure2}
    \end{align} 
    where $R_{\rm clust}$ is a scale factor that depends on the PBH clustering properties, the small-scale halo mass function, and the velocity distribution.  This formula assumes that the time it takes for the binary to merge is much shorter than the age of the Universe, as is the case for hard binaries formed through this mechanism \citep{Raidal:2024bmm}.  For stellar mass PBHs, one finds $R_\text{clust} \sim 10^{2 \div 3}$ \citep{Clesse:2020ghq,DeLuca:2020jug}, with weak scaling with the typical PBH mass, $R_{\rm PBH}^\text{cap} \sim m_{\rm PBH}^{-11/21}$ \citep{Franciolini:2022ewd}. 
\end{enumerate}
As a formation channel for PBH binaries, capture in dense halos is typically subdominant compared to primordial binary formation, at least if one assumes a relatively narrow PBH mass distribution.  
Let us mention, however, that whether this conclusion remains valid in the case of a very wide PBH mass function (spanning multiple decades in mass) is still subject to uncertainties, related especially to the amount of binaries disrupted by interactions with light PBHs. 
For definiteness, in the following, we will restrict our analysis to narrow PBH mass functions and therefor retain only the contribution from early binaries. 

When considering the merger rate of PBH within $\mathcal{O}(\SI{100}{kpc})$ from Earth, the effect of the local dark matter overdensity needs to be taken into account (see e.g.\ \cite{Pujolas:2021yaw}).
We model the Milky Way's dark matter halo as a Navarro--Frenk--White density profile~\citep{Navarro:1995iw, Navarro:1996gj}, $\rho_\text{DM}(r) = {\rho_0} r_0 / [r \left(1+{r}/{r_0}\right)^{\!2} ] \,$, with $\rho_\text{DM}(r=r_\odot) = \SI{7.9e-3}{M_\odot/pc^3}$~\citep{Cautun:2019eaf} at the location of the solar system, $r_\odot \simeq  \SI{8.0}{kpc}$, and with $r_0 = \SI{15.6}{kpc}$. 
The average overdensity within a shell at a distance $r$ from the observer location $r_\odot$ can then be approximated by
\begin{align}
	\rho(r) =
        \begin{cases}
		      \rho_\text{DM}(r_\odot)    &\quad r < r_\odot \,,	\\
		      \rho_\text{DM}(r) &\quad r \gtrsim r_\odot \,.
        \end{cases}
\end{align}
As we expect the distribution of PBH binaries to roughly follow the dark matter overdensities, the local merger rate is enhanced by an overall factor
\begin{align}
	R_{\rm PBH}^\text{local} (r) = \delta(r) R_{\rm PBH} \,,
    \label{eq:Rlocal}
\end{align}
where we defined the overdensity factor $\delta(r) \equiv \rho_\text{DM}(r)/\bar\rho_\text{DM}$, with $\bar\rho_\text{DM}$ the average cosmological dark matter density.
The correction is of order $\delta(r) \subset(1 \div 2 \times 10^5) \,	$.

Accounting for this local enhancement, one can define the volume $V_\text{yr}$,  corresponding to a distance $d_\text{yr} \equiv (3 V_\text{yr} / 4 \pi)  ^{1/3}$ enclosing the region where on average at least one merger per year takes place \citep{Domcke:2022rgu,Franciolini:2022htd}. 
The number of events per year $N_\text{yr}$ within the volume $V_\text{yr}$ is defined as
\begin{align}
    N_\text{yr} \equiv \Delta t \int_0^{d_\text{yr}} \! dr \, 4 \pi r^2
                       R^\text{local}_{\rm PBH}(r)
\,,
    \label{eq: dyr PBH mergers}
\end{align}
where we set $\Delta t = 1\,{\rm yr}$.

\begin{figure}[t!]
    \centering
    \includegraphics[width=0.7\textwidth]{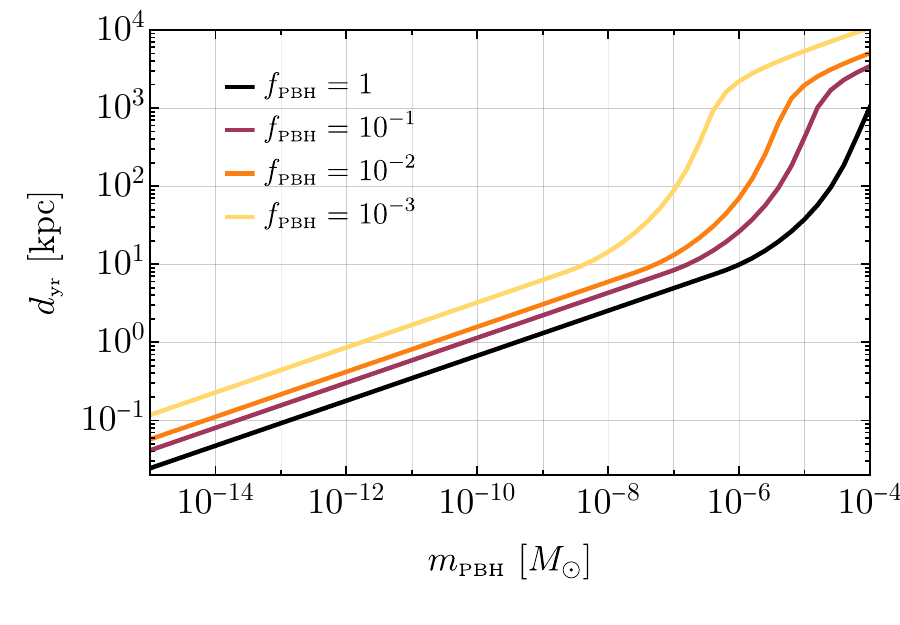}
    \caption[Distance from Earth within which one PBH merger event is expected per year.]
    {Distance from Earth within which on average one PBH merger event is expected per year. The change in slope around $\SI{e-5}{M_\odot}$ is due to the local dark matter overdensity in the Milky Way, which is relevant at distances $r \lesssim r_\odot$, relevant for light PBHs, but less important at larger distances. Figure adapted from \cite{Franciolini:2022htd}.}
    \label{fig: dist}
\end{figure}

In \cref{fig: dist}, we show the distance $d_\text{yr}$ as a function of the PBH mass and abundance for $N_\text{yr} = 1$, assuming equal mass binaries.
Due to the galactic DM overdensity, $d_\text{yr}$ is smaller than it would be based on the average cosmological density at small $m_{\rm PBH}$.

A further convenient way of parameterizing the sensitivity of detectors to inspiralling PBHs is the distance reach, $d$, for a fixed SNR, defined for generic transient sources in \cref{eq:reach}. For PBH mergers it is given by~\citep{Maggiore:1900zz}
\begin{align}
    d = \bigg(\frac{5}{24} \bigg)^{1/2} \frac{1}{\pi^{2/3}}
        (G M_c)^{5/6} 
        \Delta t^{-(k-1)/4}
        \bigg[
            \frac{2}{\text{SNR}} \int_{f_{\rm min}}^{f_{\rm max}} \! df \,
                \bigg( \frac{f^{-7/3}}{S_h^{\rm noise}(f)} \bigg)^k
        \bigg]^{1/2k} \,,
    \label{eq:reachPBH}
\end{align}
with $k=1,2$ for linear or quadratic detectors, respectively, with $S_h^{\rm noise}(f)$ the detector's noise equivalent strain PSD and $ \Delta t$ the integration time interval.\footnote{For photon (re-)generation experiments, \cref{eq:reachPBH} should contain an extra Heaviside $\theta$-function ensuring that the number of signal photons to be larger then one. See \cref{sec:PhotonRegen} for more details. } 
\Cref{eq:reachPBH} is valid under the assumption of an optimally oriented source.
The integration limits $f_{\rm min}$ and $f_{\rm max}$ depend on the detector's broadband sensitivity as well as the source properties. In practice, as we only integrate over the inspiral phase of the signal, we fix $f_{\rm max}$ to be the smallest frequency between $f_\text{ISCO}$ and the detector's maximum observable frequency. 
If $f_\text{\tiny ISCO}$ is smaller then the minimum observable frequency, the binary never enters the detector's frequency band and the sensitivity is zero.
We have already seen the distance reach for different detector designs in \cref{fig:sharkfin} above.
To gain an understanding of the detection prospects for a specific detectors, $d$ should be compared to $d_\text{yr} $ defined above.

\paragraph{Stochastic Gravitational Wave Background from PBH Binaries.}

The superposition of the GW signals from many PBH binaries generates a stochastic GW background (SGWB). Its frequency spectrum is
\begin{align}
    \Omega_{\rm GW}(f) = \frac{f}{\rho_c} \iint \! dm_1 \, dm_2 \int_0^{{f_\text{cut}}/f-1} \!
                  \frac{dz}{(1+z) H(z)} \frac{d^2R_{\rm PBH}(z)}{dm_1 \, dm_2} 
                  \frac{dE_\text{GW}(f_s)}{df_s} \,,
    \label{eq:stocOmegaPBH}
\end{align}
with the redshifted source frequency $f_s = f (1+z)$, the critical density of the Universe today, $\rho_c = 3 H_0^2/8\pi G$ (where $H_0$ is the Hubble constant), and the GW energy spectrum of a single binary, ${dE_\text{GW}(f_s)}/{df_s}$. As before, $R_{\rm PBH}$ accounts for the local overdensity.
The upper boundary of the redshift integral is given by the maximum $z$ from which GWs with redshifted frequency $f$ can come if the maximum frequency of the source spectrum is $f_\text{cut}$.

The GW energy spectrum emitted by the binary is composed of inspiral, merger, and ringdown contributions. 
Assuming circular orbits, we adopt for the individual contributions the parameterization from~\cite{Bavera:2021wmw},
\begin{align}
    \frac{dE_\text{GW}(f)}{df} = \frac{(G \pi)^{2/3} \mathcal{M}^{5/3}}{3} 
        \begin{cases}
            f^{-1/3} \nu_1^2         & f < f_\text{merger}, \\
            \omega_1 f^{2/3} \nu_2^2 & f_\text{merger} \leq  f < f_\text{ringdown}, \\
            \omega_2 \nu_3^2         & f_\text{ringdown} \leq f < f_\text{cut}.
        \end{cases}
    \label{eq:dEdnu}
\end{align}
The explicit expressions for the dimensionless coefficients $\nu_{1,2,3}$ as well as for $f_\text{merger}$, $f_\text{ringdown}$, and $f_\text{cut}$ can be found in Ref.~\cite{Bavera:2021wmw} (see also~\cite{Ajith:2009bn, Zhu:2011bd}). Parametrically, one expects $\nu_i \sim {\cal O}(1)$, while the other characteristic frequencies scale as $\approx 1/(\pi M G)$, where $M = m_1 + m_2$, with prefactors that depend on the binary mass ratio and individual spins. 
One can also translate the energy spectrum $\Omega_{\rm GW}$ to a characteristic strain using \cref{eq:hcStoch_Omegag}.
The contribution from inspiralling circular binaries, whose evolution is dominated by GW emission, leads to a low-frequency tail that scales as $\Omega_{\rm GW}(f) \sim f^{2/3}$ \citep{Moore:2014sen}, or equivalently a characteristic strain scaling as $h_c(f) \sim f^{-2/3}$. 

Unlike for individual transients, the stochastic signal from binary mergers is stationary, and the available observation time within a frequency band is only dictated by the detector properties, see related discussion in \cref{sec:characterizing-sources}.
As the SGWB is mostly emitted in the late-time universe, with most of the contribution to the integral in \cref{eq:stocOmegaPBH} coming from redshifts $z \sim {\cal O}(10)$, it is not subject to bounds coming from the number of effective relativistic degrees of freedom in the early universe. However, for realistic PBH populations with $f_{\rm PBH} \approx 1$, the amplitude $\Omega_{\rm GW}$ still falls below that range \citep{Franciolini:2022htd}. 
The SNR for these signal can be computed in analogy to the SNR for relic GW backgrounds from the early Universe, adopting \cref{eq:SNRGWBdensity}.
Notice, however, that the stochastic signal from PBH mergers would be characterized by potentially much larger anisotropies than a primordial background due to the inevitable Poisson noise in the distribution of the PBH binaries dominating the GW emission.

\paragraph{PBH Encounters.}

A fraction of PBH encounters will not lead to the formation of bound systems, which would then inspiral, but will rather produce single scattering events via a hyperbolic encounter.
This could happen for instance if the relative velocity or relative distance of the two PBHs is large enough that capture is not possible.
\reply{We will come back to the rate of such events in the following. }
The emission of GWs in close encounters of compact bodies has been 
extensively studied in the literature since the seminal works~\citep{1974SvA....18...17Z,1977ApJ...216..610T}. It is worth noting that the memory effect, to be discussed below, was first discussed in this context~\citep{Braginsky:1987kwo}. With the advent of interferometric GW detectors,  the GW emission from such encounters has been revisited in \citep{Kocsis:2006hq, OLeary:2008myb, Capozziello:2008ra, DeVittori:2012da, Garcia-Bellido:2017knh, Garcia-Bellido:2017qal, Grobner:2020fnb, Mukherjee:2020hnm, Morras:2021atg, Bini:2023gaj, Kerachian:2023gsa, Codazzo:2023kcx, Dandapat:2023zzn, Teuscher:2024xft}. The waveform and characteristic parameters of the GW emission in such encounters are different from those of the inspiralling binaries, and both provide complementary information that can be used to discover, as well as determine, the mass distribution of PBHs as a function of redshift and their spatial distribution in the clustered scenarios.
Hyperbolic encounters generate bursts of GWs, where the majority of the energy is released near the point of closest approach. This leads to a characteristic ``tear-drop" shape of the emission in the time-frequency domain. In the Newtonian limit, the frequency of the emitted GWs peaks at periapsis, and the peak frequency is a function of only three variables: the impact parameter $b$, the eccentricity $e$ and the total mass of the system. The duration of such events is on the order of a few milliseconds to several hours, depending on those parameters.

\begin{figure}
    \centering
    \includegraphics[width=0.52\textwidth]{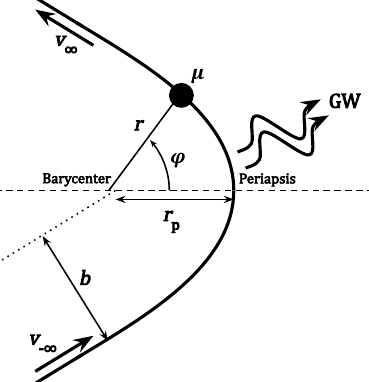}
    \caption[Geometry of black hole--black hole scattering.]
    {The scattering of two black holes induces the emission of gravitational waves whose emitted power is maximal at the point of closest approach.}
    \label{fig:hyperbolic}
\end{figure}

More precisely, the peak frequency at periapsis given by \cite{Teuscher:2024xft}
\begin{align}
    f_{\rm p} = \frac{1}{2\pi} \sqrt{\frac{G M (e+1)}{r_p^3}}
              \simeq \SI{1.6}{GHz} \times \bigg( \frac{\SI{e-5}{M_\odot}}{M} \bigg)
                                          \bigg( \frac{R_S}{r_p} \bigg)^{\frac{3}{2}}
                                          \sqrt{\frac{e+1}{2}} \,, 
\end{align}
where $r_p$ is the periapsis radius (or the distance to the hyperbola's focus point at closest approach) and $R_S = 2 G M / c^2$ is the Schwarzschild radius of the system with total mass $M = m_1 + m_2$. Note that $f_p$ depends only on $M$, on the ratio $R_S/r_p$, and on the eccentricity of the hyperbolic orbit $e = \sqrt{1+b^2 v_0^4 / G^2 M^2}$, where $v_0$ is the asymptotic relative velocity of the encounter, and $\beta \equiv v_0 / c$.
Introducing $G(e) \equiv e+2 / (e+1)^{1/3}$ and $q = m_1 / m_2$, the maximum strain and power of the GW burst at periapsis are respectively given by
\begin{align}
    h_{p} &= 3.6 \times 10^{-25} \times \frac{4 q}{(1+q)^2} \frac{G(e)}{G(1)}
             \bigg( \frac{M}{\SI{e-5}{M_\odot}} \bigg)^{\frac{5}{3}}
             \bigg( \frac{f_p}{\SI{1.6}{GHz}} \bigg)^{\frac{2}{3}}
             \bigg( \frac{\SI{1}{Mpc}}{D} \bigg) \,, \nonumber\\
    P_{p} &= \SI{3.7e24}{L_\odot} \times \frac{1}{(e+1)^{\frac{2}{3}}}
             \bigg( \frac{4 q}{(1+q)^2} \bigg)^2
             \bigg( \frac{M}{\SI{e-5}{M_\odot}} \bigg)^{\frac{10}{3}}
             \bigg( \frac{f_p}{\SI{1.6}{GHz}} \bigg)^{\frac{10}{3}} \,,
\end{align}
where  $L_\odot$ is the solar luminosity and $D$ is the distance of the event from Earth.
The signal duration in a detector operating at frequency $f$ and having a frequency bandwidth $\Delta f$ can be computed from the conservation of angular momentum and reads
\begin{align}
    t_{\Delta f}(f, \Delta f, e) = \frac{1}{\pi f} \sqrt{1 + \frac{1}{e}}
                                   \sqrt{\frac{\Delta f}{f}} \,.
\end{align}
This estimate shows that the duration of the GW signal from PBH encounters is close to the inverse of the peak GW frequency. This is similar to what happens for PBH mergers close to the ISCO frequency. However, for hyperbolic encounters, the inspiral phase associated with GWs with smaller frequencies and slower frequency evolution is absent.

The rate of close encounters remains rather uncertain. We report an estimate based on the cross-section of a close hyperbolic encounter event, which is given by 
$\sigma = \pi b^2 = \pi (GM/v_0^2)^2(e^2-1)$. This leads to~\citep{Garcia-Bellido:2017knh,Garcia-Bellido:2017qal,Garcia-Bellido:2021jlq}
\begin{align}
    \frac{dR_{\rm PBH}^{\rm enc}}{d(\ln m_1) \, d(\ln m_2)} 
        \approx \frac{\num{1.4e-8}}{\si{yr\,Gpc^{3}}} \,
                \rho(m_1) \, \rho(m_2)
                \bigg( \frac{\delta_{\rm loc}}{10^8} \bigg)
                \frac{M^2}{m_1\,m_2}
                \frac{e^2-1}{(v_0/c)^3}\,.
    \label{Eq:CHErate}
\end{align}
In this estimate, $\delta_{\rm loc}$ characterizes the PBH overdensity compared to the mean DM density today.
Notice $\delta_{\rm loc}$ is at least as large as the one introduced in Eq.~\eqref{eq:Rlocal}, accounting for the local DM concentration at around the solar system in the galaxy, but can also reach larger values due to the small-scale structure induced in light PBH DM scenarios (see e.g.~\citep{Inman:2019wvr}), potentially boosting the encounter rates. This latter effect does not impact the rate of mergers \eqref{eq:PBHrate} as it is dominated by binaries formed at high redshift. 
Also, in Eq.~\eqref{Eq:CHErate}, we introduced $v_0$ as the virial velocity of PBHs in a cluster. Given the scaling $\sim v^{-3}$, the rate of mergers is dominated by light clusters for which $v_0$ is small. 
While a complete determination of the rate would require averaging \eqref{Eq:CHErate} over encounter parameters such as eccentricity and PBH cluster properties, one expects this rate to be subdominant compared to the one of mergers \eqref{eq:PBHrate}.

\paragraph{Stochastic Gravitational Wave Background from PBH Encounters.}

Overlapping GW signals from close PBH encounters can also form a stochastic GW background, in analogy to the SGWB from PBH mergers discussed above. The energy density of this background can be estimated in analogy to the PBH merger case, \cref{eq:stocOmegaPBH}, accounting for the different event rate and GW energy spectrum for hyperbolic encounters compared to mergers. 
For hyperbolic encounters, the energy emitted per logarithmic frequency bin is given by (see e.g. \citep{Garcia-Bellido:2021jlq})
\begin{align}
    \frac{dE_{\rm GW}}{d\ln f} =
        \frac{4\pi}{45} 
        \frac{G^{7/2} m_1^2 m_2^2 (m_1+m_2)^{1/2}}{a^{7/2}} \, \nu^5 F_{\rm e}(\nu) \,,
\end{align}
where $a = G M/v_0^2$ is the semi-major axis and $v_0$ is the initial relative velocity. We have moreover defined $\nu^2 \equiv 4\pi^2 f^2 a^3/GM$. The function $F_{\rm e}(\nu)$ describes the dependence on eccentricity $e$ and is given by
\begin{align}
    F_{\rm e}(\nu) \simeq \frac{12 \big(1 - y^2 - 3\nu y^3 + 4y^4 + 9\nu y^5 + 6\nu^2 y^6 \big)}
                               {\pi \nu^3 y (1+y^2)^2} \, e^{-2 \nu \xi(y)} \,,  
    \label{Eq:CHEpower}
\end{align}
with $\xi(y) = y - \arctan y$ and $y = \sqrt{e^2 - 1}$. The amplitude of the SGWB background induced by hyperbolic PBH encounters is typically smaller than the one from PBH binaries from the same population, but can lead to additional features due to the different frequency dependence \citep{Garcia-Bellido:2021jlq}. In particular, the low frequency tail of the SGWB from close encounters can fall as $f^2$ instead of the $f^{2/3}$ scaling of the background due to binary mergers.
Note that parabolic encounters, $e \to 1$, yield the highest emission rate within the Newtonian approximation, while the cross section in the same limit approaches zero (see Eq.~\eqref{Eq:CHErate}).  This suggests that relativistic effects may play an significant role, particularly dynamical capture must be accounted for.\footnote{Dynamical capture systems are those that, under Newtonian gravity, would scatter along hyperbolic trajectories but instead merge due to radiation reaction effects in general relativity. The focus on these systems is motivated by their substantial observational interest, at least at low frequencies~\citep{Gamba:2021gap}. For classical work on this topic, see \cite{East:2012xq, Gold:2012tk}.}

A final assessment of the magnitude of the SGBW signal from PBH encounters requires a careful population study accounting not only for the PBH number density (and hence $\delta_\text{loc}$) but also the distribution of eccentricity $e$ across the binaries, in addition to the inclusion of relativistic effects. However, given the result for the merger rate~\eqref{Eq:CHErate}, the result will likely be subdominant compared to the SGWB signal from PBH mergers of the same population.

\subsubsection{Memory Effects}

The gravitational memory effect occurs when the metric perturbation long after the passage of a GW is different from the metric long before the passage for at least one of the GW polarizations \citep{Braginsky:1985vlg, 1974SvA....18...17Z, Braginsky:1987kwo}. In other words, the effect is characterized by the quantity
\begin{align}
    \delta h_{+,\times}^{\rm mem} = \lim_{t\to +\infty} h_{+,\times}(t)
                                  - \lim_{t\to-\infty} h_{+,\times}(t),
    \label{eq:memdef}
\end{align}
being non-zero. Here, $t$ is the observer's coordinate time. The gravitational memory effect thus induces a permanent displacement of free-falling test masses.

While two types of memory exists, related to linear and non-linear effects, we will focus here on the non-linear memory induced by a BH merger \citep{Christodoulou:1991cr, PhysRevD.44.R2945, Blanchet:1992br, Favata:2008ti, Favata:2009ii, Pollney:2010hs, Lasky:2016knh, Hubner:2019sly, Ebersold:2020zah, Zhao:2021hmx, Gasparotto:2023fcg}. A linear signal can originate or instance from close hyperbolic PBH encounters \citep{Favata:2010zu, Caldarola:2023ipo}, with similar phenomenology.

If we define $h_0$ as the primary GW strain, the memory strain~$\delta h$ is computed as (see e.g.~\cite{Ebersold:2020zah})
\begin{multline}
    \delta h^{\ell m} = -D \sum_{\ell',\ell'' \geq 2} \, \sum_{m',m''}
        \sqrt{\frac{(\ell-2)!}{(\ell+2)!}}  
            \int \! d\Omega \, Y^{\ell m *}(\Omega) \,
                               Y^{\ell' m'}_{-2}(\Omega) \, 
                               Y^{\ell''m'' *}_{-2}(\Omega)   \\
            \times \int_{-\infty}^t \! dt' \, \dot{h}_0^{\ell' m'} (t') \, 
                                              \dot{h}_0^{\ell'' m'' *}(t') \,,
    \label{eq:memorymodes}
\end{multline}
with $D$ being the distance to the source, a dot indicates a derivative with respect to time, and we have introduced the spin-weighted spherical harmonics decomposition 
\begin{align}
    h_+ - i h_\times \equiv \sum_{\ell \geq 2} \, \sum_{|m| \leq \ell}
                            h^{\ell m} Y^{\ell m}_{-2}.
    \label{eq:mode_decomp}
\end{align}
The functions $Y^{\ell m}_{-2}$ are the spin-weighted spherical harmonics, defined for instance in \cite{Boetzel:2019nfw}.

Phenomenologically, the strain of non-linear memory behaves approximately as
\begin{align}
    h (f) \simeq \frac{\delta h^\text{mem}}{2 \pi f}  \Theta(f_\text{cut} - f).
\end{align}
where the UV cut-off is approximately placed close to the ISCO frequency at $f_{\rm cut} \sim 1/(60 M)$ (see e.g. \citep{Gasparotto:2023fcg}). This description neglects features induced by the non-linear dynamics close to the merger, but captures the main properties of the signal at low frequencies.  The typical value of the memory strain amplitude at $f_\text{ISCO}$, averaged over source orientations and sky positions, can be related to the amplitude of the GW signal at its peak frequency by the factor $\kappa = \sqrt{{\langle h_\text{mem}^2\rangle}/{\langle h_\text{osc}^2 \rangle}} \simeq 1/20$ \citep{McNeill:2017uvq}, where $h_{\rm osc}$ is the amplitude of the oscillating primary GW signal. 	
The GW memory strain then turns out to be
\begin{align}
    h(f) \sim \SI{1.2e-24}{sec} \bigg( \frac{f_{\rm ISCO}}{f} \bigg)
                                   \bigg( \frac{M}{\SI{e-5}{M_\odot}} \bigg)
                                   \bigg( \frac{D}{\si{kpc}} \bigg)^{-1} \,,
\end{align}
where we assumed, for simplicity,  an equal mass PBH binary with total mass $M$. 

The peculiar feature of the GW memory is that it extends to frequencies that are much smaller than $f_\text{\tiny ISCO}$. 
This implies that low-frequency interferometers could detect memory signals from UHF-GW sources \citep{McNeill:2017uvq,Lasky:2021naa}. However, the memory effect of PBH of mergers with masses $m \lesssim 10^{-4} M_\odot$ at a distance $d_\text{yr}$ (as defined in \cref{eq: dyr PBH mergers})  would fall much below the forecasted sensitivity curves of both LISA and third generation ground-based detectors, motivating searches based on UHF-GW observatories \citep{Franciolini:2022htd}. 
It is also worth noticing that for binary PBHs, the early inspiral phase is associated with larger strain signals in the sub-kHz range, scaling as $h_c \sim f^{-7/6}$ down to a very small minimum frequency. Therefore, if the available observation time at the GW observatory is sufficient to map out this signal over long time periods, the low-frequency strain from the inspiral phase may be easier to detect than the one induced by non-linear memory. This conclusion, however, depends on the distribution of binary parameters in a population of PBH inspiralling sources, and deserves further investigation. 

We conclude this section by mentioning that, although the memory strain could eventually cross some of the sensitivity bands of UHF-GW detectors, the memory signal is very different from other HFGW signals (such as plane monochromatic GWs or a stationary stoachastic GW background). Therefore, dedicated studies on the sensitivity to GW memory are required.

\subsubsection{Exotic Compact Objects }
\label{sec:ECOs}

Beyond the well-known compact astrophysical objects, namely black holes and neutron stars, there are several candidates for stable (or long-lived) exotic compact objects, composed of particles beyond the Standard Model \citep{Giudice_2016, Cardoso:2019rvt}. For instance, they can be composed of exotic fermions such as gravitinos in supergravity theories, giving rise to gravitino stars \citep{Narain_2006}, or of dark quarks \citep{Witten:1984rs, Hong:2020est, Gross:2021qgx}. They can also be composed of bosons, such as moduli in string compactifications and supersymmetric theories \citep{Krippendorf_2018}. Depending on the mechanism that stabilizes bosonic compact objects, they have specific names such as Q-balls, boson stars, oscillatons, and oscillons. Additional proposals include fermion--boson stars~\citep{Lee:1986tr, DelGrosso:2023trq, Diedrichs:2023trk} and anisotropic stars~\citep{Raposo:2018rjn}, as well as gravastars \citep{mazur2001gravitational}. 

The masses and radii of the compact objects depend on their constituents, and in particular on their internal pressure that is needed to counterbalance gravity. In regular astrophysical stars, this pressure is thermal, while in stars composed of fermions it is the Fermi degeneracy pressure. For bosons, in contrast, the star is stabilized by the quantum property dictating that particles cannot be localized to scales below their Compton wavelength. Stable configurations that do not collapse can be found below a maximum mass $M_{\rm max}$. For example, for stars composed of bosons with negligible self-interactions one finds \citep{Kaup:1968zz}
\begin{align}
    M_{\rm max} = 0.633 \frac{M_{\rm Pl}^2}{m_B}
                \approx {\rm M}_\odot \bigg( \frac{\SI{e-10}{eV}}{m_B} \bigg) ,
\end{align}
where $m_B$ is the boson mass. This means that UHF-GW experiments searching for mergers of subsolar mass objects could detect signals from compact objects composed of ultralight boson particles heavier than around \SI{e-10}{eV}. 
In the presence of a quartic interactions $\lambda \phi^4$, this scaling relation is modified to \citep{Colpi:1986ye}
\begin{align}
    M_{\rm max} = 0.06 \sqrt{\lambda} \frac{M_{\rm Pl}^3}{m_B^2}
                \approx \SI{10}{M_\odot} \sqrt{\lambda} \bigg( \frac{\SI{100}{MeV}}{m_B} \bigg)^2,
\end{align}
where $M_{\rm Pl}$ is the Planck mass. A similar relation can be found for other models (see \cite{Cardoso:2019rvt} for a review), and high-frequency GW detection would generically allow access to new regions of parameter space.

Gravitational wave emission from exotic compact objects will be indistinguishable from the signal of conventional black hole or neutron star mergers during the early inspiral phase, where frequency evolution is slow and allows for long observation time within the experimental frequency band. Close to the merger, on the other hand, the exotic objects' potentially much larger size and tidal deformability comes into play and may lead to significant differences. Therefore, to distinguish different types of compact objects, it will be crucial to observe the final stages of the binary evolution.

The ISCO frequency for a binary system of two exotic compact objects with mass $M$ and radius $R$ is given by \citep{Giudice_2016}
\begin{align}
    f_{\rm ISCO} = \frac{1}{6 \sqrt{3} \pi} \frac{C^{3/2}}{G M}
                 \simeq \SI{1}{MHz} \times C^{3/2} \bigg( \frac{\SI{6e-3}{M_\odot}}{M} \bigg) \,,
    \label{eq:fISCO-ECO}
\end{align}
where $C = G M/R$ is the compactness of the exotic compact object. This expression is only slightly modified for a boson star binary with two different values of the masses. Note that for a BH the radius is given by the Schwarzschild radius $R_{\text{S}} = 2 G M$, therefore $C = 1/2$ is the maximum attainable value for the compactness.
The GW strain for an equal-mass binary of exotic objects during the inspiral phase can be calculated as in \cref{sec:PBHmergers}, see in particular \cref{eq:strainPBHs}.

The exact waveform produced by the merger of two exotic compact objects is in general different from the one of black holes or neutron stars. It depends on microphysics details, in particular through tidal deformability effects \citep{Giudice_2016, Palenzuela_2017}.\footnote{See however \cite{Helfer:2018vtq} for more details on the initial conditions.} Hence, the detection of GWs close to the ISCO frequency from an exotic compact object merger can give valuable information on physics beyond the Standard Model. Additional information on the nature of the exotic objects could be obtained by using mergers to infer their mass function. It is important to keep in mind, though, that cosmological formation scenarios and the expected merger rate of exotic compact objects remain uncertain (see e.g.\ \cite{Frieman:1989bx, Bai:2022kxq, Croon:2022tmr, Gorghetto:2022sue, Banks:2023eym} for some estimates in this direction).

\subsubsection{Black Hole Superradiance}
\label{sec:Superradiance}

Boson clouds created by gravitational superradiance of BHs are a powerful GW source \citep{Ternov:1978gq, Zouros:1979iw, Arvanitaki:2009fg, Arvanitaki:2010sy, arvanitaki:2016gw, aggarwal2020searching, Detweiler:1980uk, Yoshino:2013ofa, Arvanitaki:2014wva, Brito:2014wla, Brito:2015oca, Sprague:2024lgq}. Superradiance is an enhanced radiation process that is associated with bosonic fields around rotating objects with dissipation. The event horizon of a spinning BH is one such example that provides conditions particularly suitable for this phenomenon to occur \citep{Arvanitaki:2014wva}. 

For a bosonic field of mass $\mu$ in the vicinity of a rotating BH, there exists a set of quasibound states whose oscillation frequency  $\omega_R \sim \mu$ satisfies the superradiance condition $\omega_R < m \Omega_H$, where $m$ is the azimuthal quantum number (with respect to the BHs rotation axis) and $\Omega_H$ the angular velocity of an observer at the horizon, as measured by a static observer at infinity. If the superradiance condition is saturated, the boson occupation number grows rapidly over a timescale $1/\Gamma = 1/(2 \omega_I)$, where $\omega_I$ is the imaginary part of the boson's oscillation frequency. (Due to the special boundary conditions in the space-time around a black hole, the solution to the Klein--Gordon equation acquires an imaginary part.) Superradiance is strongest when the Compton wavelength of the bosonic field is on the order of the BH radius, $G m_\text{PBH} \mu \equiv  \alpha  \sim  \mathcal{O}(1)$, with $m_\text{PBH}$ being the BH mass. 

The BH and its superradiant boson cloud form a gravitationally bound `atom', with different atomic `levels' occupied by exponentially large numbers of particles. As the bosonic cloud is non-spherical, it emits nearly monochromatic gravitational waves at a frequency \citep{Brito:2015oca}
\begin{align}
    f_{\text{GW}} \sim \omega_R / \pi 
                  \sim \SI{5}{MHz} \bigg( \frac{\mu}{\SI{e-8}{eV} } \bigg) 
                  \sim \SI{5}{GHz} \bigg( \frac{\SI{e-5}{M_\odot}}
                                               {m_\text{PBH}} \bigg) 
                                   \bigg(\frac{G m_\text{PBH} \mu}{0.1} \bigg).
    \label{eq:freqSuperradiance}
\end{align}
In the last step we have fixed $G m_\text{PBH} \mu$ to the typical value required by the resonant condition. In a particle physics context, GW emission from superradiant boson clouds can be interpreted as originating from annihilations (or decays) of the boson field into gravitons.

The peak gravitational strain from a source at luminosity distance $D$ is approximately \citep{Brito:2015oca}\footnote{These expressions are obtained in the $G \, m_\text{PBH} \, \mu \ll 1$ limit, but they still provide good estimations when $G \, m_\text{PBH} \, \mu \gtrsim 0.1$. A detailed analysis has been performed in \cite{Isi:2018pzk} for scalar bosons.}
\begin{align}
    h_S \approx \num{5e-30}
                \bigg( \frac{m_\text{PBH}}{\SI{e-5}{M_\odot}} \bigg) 
                \bigg( \frac{G m_\text{PBH} \mu_S}{0.1} \bigg)^7
                \bigg( \frac{\chi_i - \chi_f}{0.5} \bigg)
                \bigg(\frac{\si{kpc}}{D} \bigg) \,, \\
    h_{V,T} \approx \num{e-26}
                    \bigg( \frac{m_\text{PBH}}{\SI{e-5}{M_\odot}} \bigg)
                    \bigg( \frac{G m_\text{PBH} \mu_{V,T}}{0.1} \bigg)^5
                    \bigg( \frac{\chi_i - \chi_f}{0.5} \bigg)
                    \bigg(\frac{ \si{kpc}}{D} \bigg) \,,
\end{align}
where the subscripts $S$, $V$ $T$ refer to scalar, vector and tensor depending on the spin of the boson field. The parameters $\chi_i$ and $\chi_f$ stand for the dimensionless BH spin evaluated at the beginning and end of the superradiant growth.

The duration of the gravitational wave signal can be estimated by the time it takes to radiate away half of the cloud's rest energy. The approximate result is \citep{Brito:2015oca}
\begin{align}
    \tau_S \approx \SI{0.13}{yrs}
                   \bigg( \frac{m_\text{PBH}}{\SI{e-5}{M_\odot}} \bigg)
                   \bigg( \frac{0.1}{G m_\text{PBH} \mu_S} \bigg)^{15}
                   \bigg( \frac{0.5}{\chi_i - \chi_f} \bigg) \,,
    \label{eq:tau-superrad-S} \\
    \tau_{V,T} \approx \SI{0.17}{sec}
                       \bigg( \frac{m_\text{PBH}}{\SI{e-5}{M_\odot}} \bigg)
                       \bigg( \frac{0.1}{G m_\text{PBH} \mu_{V,T}} \bigg)^{11}
                       \bigg( \frac{0.5}{\chi_i - \chi_f} \bigg) \,.
    \label{eq:tau-superrad-VT} 
\end{align}
These estimates, in particular \cref{eq:tau-superrad-S}, show that, unlike other astrophysical or cosmological sources discussed in this document, scalar boson clouds around rotating black holes can be considered continuous sources, similar to verification binaries for LISA or pulsars for interferometers in the LIGO / VIRGO / KAGRA range.

As mentioned above, GW emission due to superradiance is expected to be almost-monochromatic and coherent. However, given the potentially long signal durations, one may expect very small frequency drifts which should be taken into account in the search strategy. For the case of a scalar cloud with a small self-interaction, \cite{Baryakhtar:2020gao} finds a frequency drift
\begin{align}
    \frac{\dot{f}}{f^2} \simeq 3 \times 10^{-20} \bigg ( \frac{\alpha}{0.1} \bigg)^{17}.
    \label{eqn:gfdrift}
\end{align}

If nonlinear effects, for instance due to boson self-interactions, become important, the GW signature changes. In this case, periodic collapses of the boson cloud are expected, similarly to Bose-Einstein condensate bosenovae. In these explosive events, part of the boson cloud escapes to infinity, accompanied by a gravitational wave burst. Focusing on the QCD axion, the primary frequency component of a bosenova GW burst is \citep{Arvanitaki:2014wva}
\begin{align}
    f_{\text{bn}} \approx \SI{30}{MHz}
                  \bigg( \frac{16}{c_{\text{bn}}} \bigg)
                  \bigg( \frac{m_\text{\PBH}}{\SI{e-5}{M_\odot}} \bigg)^{-1}
                  \bigg( \frac{G m_\text{\PBH} \mu_a / \ell}{0.4} \bigg)^2 ,
\end{align}
where $\ell$ is the orbital quantum number and $c_{\text{bn}}$ parametrizes the collapse timescale. (The infall time is $t_{\text{bn}} = c_{\text{bn}} r_\text{cloud}$, where $r_\text{cloud}$ the typical distance between the boson could and the black hole \citep{Arvanitaki:2014wva}.) For quadrupole radiation, the strain can be estimated as \citep{Arvanitaki:2014wva}
\begin{align}
    h (f) \approx \SI{e-27}{sec}
                  \bigg( \frac{\sqrt{\epsilon}/c_{\text{bn}}}{10^{-2}} \bigg)^2
                  \bigg( \frac{G {m_\text{\PBH}} \mu / \ell}{0.4} \bigg)
                  \bigg( \frac{{m_\text{\PBH}}}{\SI{e-5}{M_\odot}} \bigg)
                  \bigg( \frac{f}{f_a^{\text{max}}} \bigg)^2
                  \bigg( \frac{\si{kpc}}{D} \bigg) ,
\end{align}
with $\epsilon \sim 5\%$ being the fraction of the cloud that plunges into the black hole, and $f_a^{\text{max}}$ the largest value of the QCD axion decay constant for which bosenovae take place.


\subsection{Early Universe}
\label{sec:earlyU}

We now turn to sources emitting GWs at cosmological distances, i.e., in the early Universe. For a summary of these sources see \cref{fig:PLS} and \cref{tab:summary-stochastic} in \cref{sec:SummaryTable}. They are associated to events in our cosmological history which are triggered, for instance, by the decreasing temperature $T$ of the thermal bath and typically occur everywhere in the Universe at (approximately) the same time. This results in a stochastic background of GWs which is a superposition of GWs with different wave vectors.

The total energy density of such a GW background,
\begin{align}
 \rho_{\rm GW} = \int \! d\log k \, \frac{d\rho_{\rm GW} }{ d\log k},
\end{align}
with characteristic wavelengths well inside the horizon, decays with the expansion of the Universe as $\rho_{\rm GW} \propto a^{-4}$, as expected for relativistic degrees of freedom. This implies that a GW background acts as an additional radiation field contributing to the background expansion rate of the Universe. Observables that can probe the background evolution of the Universe can therefore be used to constrain $\rho_{\rm GW}$. In particular, two events in cosmic history yield precise measurements of the expansion rate of the Universe: Big Bang Nucleosynthesis (BBN) at temperatures $T_{\rm BBN} \sim \SI{0.1}{MeV}$ and the decoupling of the cosmic microwave background (CMB) at recombination ($T_{\rm CMB} \sim \SI{0.3}{eV}$). An upper bound on the total energy density of a GW background present at the time of BBN or recombination can therefore be derived from the constraint on the amount of radiation tolerable at these cosmic epochs. Obviously, such bounds apply only to GW backgrounds that are present before the epoch considered (BBN or recombination).

Constraints on the presence of `extra' radiation are usually expressed in terms of an effective number of neutrino species, $N_{\rm eff}$, after electron--positron annihilation and neutrino decoupling. The total number of Standard Model relativistic degrees of freedom after $e^+e^-$ annihilation is $g_*(T < T_{e^+e^-}) = 2 + \frac{7}{4}\,N_{\rm eff} \left(\frac{4}{11}\right)^{4/3}$, with $N_{\rm eff} = 3.043$ \citep{Cielo:2023bqp}. As the energy density for thermalized relativistic degrees of freedom in the Universe is given by $\rho_{\rm rad} = \frac{\pi^2}{30} \, g_{*}(T) \, T^4$, an extra amount of radiation, $\Delta \rho_\mathrm{rad}$, can be parametrized by $\Delta N_{\rm eff}$ extra neutrino species using
\begin{align}
    \Delta \rho_\mathrm{rad} = \frac{\pi^2}{30}\, \frac{7}{4} \, \bigg(\frac{4}{11}\bigg)^{4/3}
                               \Delta N_{\rm eff} \, T^4 \,.
\end{align}
This is independent of whether the extra radiation is in a thermal state or not, as $N_{\rm eff}$ is only a parametrization of the total energy density of the extra component, independent of its spectrum. Since the energy density in GWs must satisfy $\rho_{\rm GW}(T) \leq \Delta \rho_\mathrm{rad}(T)$, we obtain the limit
\begin{align}
    \frac{\rho_\mathrm{GW}}{\rho_\gamma} \bigg|_{T\leq\si{MeV}}
        \leq \frac{7}{8}\, \bigg(\frac{4}{11} \bigg)^{4/3} \Delta N_{\rm eff}^\text{max} \,,
\end{align}
with $\rho_\gamma$ denoting the energy density in photons and $\Delta N_{\rm eff}^\text{max}$ the constraint on $\Delta N_{\rm eff}$ from either BBN or the CMB. Writing the fraction of GW energy density today as\footnote{We write the current value of the Hubble parameter as $H_0 = h \times \SI{100}{km\,sec^{-1}\,Mpc^{-1}}$, following standard conventions in cosmology. We will avoid using in contexts where there could be any confusion with the GW strain, also denoted by $h$. Early Universe and late time observations report slightly different values for the Hubble parameter, see \cite{Bernal:2016gxb} for a discussion. For our purposes, we will assume $h = 0.7$ when needed.}
\begin{align}
    \frac{\rho_{\rm GW} \, h^2}{\rho_c} \bigg|_0
        = \Omega_{\text{rad},0} \, h^2 \,
          \bigg( \frac{g_S(T_0)}{g_S(T)} \bigg)^{4/3} \frac{\rho_{\rm GW}(T)}{\rho_\gamma(T)} \,,
\end{align}
we obtain a constraint on the redshifted GW energy density today, in terms of the number of extra neutrino species at BBN or at recombination \citep{Caprini:2018mtu}
\begin{align}
   \frac{\rho_{\rm GW} \, h^2}{\rho_c} \bigg|_0 \leq \Omega_{\text{rad},0} \, h^2 \times
                            \frac{7}{8}\, \bigg(\frac{4}{11} \bigg)^{4/3}
                            \Delta N_{\rm eff}^\text{max}
                 = 5.6 \times 10^{-6} \, \Delta N_{\rm eff}^\text{max} \,,
    \label{eq:ConsRhoBBN}
\end{align}
where we have inserted $\Omega_{\text{rad},0} \, h^2 = (\rho_{\gamma}/\rho_c)_0 \, h^2 = 2.47 \times 10^{-5}$.  We recall that this bound applies only to the total GW energy density, integrated over wavelengths well inside the Hubble radius (for super-horizon wavelengths, tensor modes do not propagate as a wave, and hence they do not affect the expansion rate of the Universe). Except for GW spectra with a very narrow peak of width $\Delta f \ll f$, the bound can be interpreted as a bound on the amplitude of a GW spectrum as defined in \cref{eq:Omega_g}, $\Omega_{\text{GW},0}(f) \, h^2 \lesssim 5.6 \times 10^{-6} \Delta N_{\rm eff}$, over a wide frequency range.

Current limits on $\Delta N_{\rm eff}$ from BBN and from the CMB are similar. In particular, \citep{Cyburt:2015mya} find $\Delta N_{\rm eff} < 0.2$ at 95\% confidence level from BBN, while \citep{Smith:2006nka,Sendra:2012wh,Pagano:2015hma, Clarke:2020bil} find similar bounds based on the Hubble rate at CMB decoupling. A recent combined analysis of CMB and BBN constraints \citep{Yeh:2022heq} gives $\Delta N_{\rm eff} < 0.18$ (95\% confidence level), which, when plugged into \cref{eq:ConsRhoBBN} lead to
\begin{align}
    \Omega_{\text{GW},0} h^2 < 1.1 \times 10^{-6}\,.
    \label{eq:BBNbound}
\end{align}
This constraint applies to stochastic GW backgrounds produced before BBN, with wavelengths inside the Hubble radius at the onset of BBN, corresponding to present-day frequencies $f \geq \SI{1.5e-12}{Hz}$.
Even lower-frequency backgrounds, down to $f \gtrsim \SI{e-15}{Hz}$, can be constrained using CMB-only limits on $\Delta N_{\rm eff}$, which translate into
\begin{align}
    \Omega_{\text{GW},0} h^2 < 2.9 \times 10^{-7} \,,
    \label{eq:CMBbound}
\end{align}
for GWs with homogeneous initial conditions (i.e., GW backgrounds with no initial density perturbations)~\citep{Clarke:2020bil}. 
The current theoretical uncertainty on the SM prediction for $N_{\rm eff}$ is of order $10^{-3}$. If CMB experiments were to reach this level of precision, one would obtain an upper bound of $\Omega_{\text{GW},0} h^2 < 5.6 \times 10^{-9}$.

Since high-frequency GWs carry a lot of energy, $\Omega_\text{GW} \propto f^3 \, S_h$, the above bounds impose severe constraints on possible cosmological sources of high-frequency GWs.

\subsubsection{Inflation}
\label{sec:Inflation}

Under the standard assumption of scale invariance, the amplitude of GWs produced during inflation is too small ($\Omega_{\rm GW, 0} \lesssim 10^{-16}$) to be observable with current technology.\footnote{However, note that the proposed space-borne detectors Big Bang Observer (BBO, \cite{BBO}) and the deci-Hertz Interferometer Gravitational wave Observatory (DECIGO, \cite{Seto:2001qf}) may reach the necessary sensitivity, assuming that astrophysical GW foregrounds can be subtracted to this accuracy}.

Various inflationary mechanisms have been studied in the literature that can produce a significantly blue-tilted GW signal (that is, a signal with a spectrum that increases towards higher frequency), or a localized bump at some given (momentum) scale, with a potentially visible amplitude. A number of these mechanisms have been explored in \cite{Bartolo:2016ami} with a focus on the LISA experiment and therefore on GW signals in the mHz range. However, these mechanisms can be easily extended to higher frequencies. Assuming an approximately constant Hubble parameter $H$ during inflation, a GW signal generated $N$ Hubble times (e-folds) before the end of inflation with frequency $H$ is redshifted to a frequency $f$ today according to
\begin{align}
    \ln \bigg[ \frac{f}{\SI{e-18}{Hz}} \bigg] \simeq N_{\rm CMB} - N \,,
    \label{eq:fN}
\end{align}
where $N_{\rm CMB}$ is the number of e-folds at which the CMB modes (in particular the conventionally chosen pivot scale of \SI{0.5}{Mpc^{-1}}) exited the horizon. The numerical value of $N_{\rm CMB}$ depends logarithmically on the energy scale of inflation, which is bounded from above by the upper bound on the tensor-to-scalar ratio \citep{Akrami:2018odb}, $H \lesssim \SI{6e13}{GeV}$. Saturating this bound implies $N_{\rm CMB} \simeq 60$, and a peak at $f = 1 \; {\rm MHz}$ then corresponds to the $N = 4.7$, while LIGO frequencies $f_{\rm LIGO} \simeq \mathcal{O}(\SI{e2}{Hz})$ correspond to $N \simeq 14$. These late stages of inflation are not accessible to electromagnetic probes, making high-frequency GW observations unique.

\cite{Bartolo:2016ami} discuss three broad categories of mechanisms leading to enhanced GW emission during inflation: the presence of extra fields that are amplified in the later stages of inflation (and therefore affect only scales much smaller than the CMB ones); GW production in the effective field theory framework of broken spatial reparametrizations, and GWs sourced by (large) scalar perturbations.
In the following we will briefly summarize these three cases.

\paragraph{Extra Particle Species}

Several mechanisms of particle production during inflation have recently been considered in the context of GW amplification. Here, for definiteness, we discuss a specific mechanism in which a pseudo-scalar inflaton $\phi$ produces gauge fields via an axion-like coupling of the form $(\phi / (4 f_a)) F \tilde{F}$,  where $F_{\mu\nu}$ is the gauge field strength tensor,  ${\tilde F}_{\mu \nu}$ is its dual, and $f_a$ is the decay constant of $\phi$. The motion of the inflaton results in a large amplification of one of the two gauge field helicities due to a tachyonic instability. The produced gauge quanta in turn generate inflaton perturbations and GW via $2 \to 1$ processes \citep{Barnaby:2010vf,Sorbo:2011rz}. The spectrum of the sourced GWs is \citep{Barnaby:2010vf}
\begin{align}
    \Omega_{\rm GW, 0}(f) \simeq 3.6 \cdot 10^{-9} \, \Omega_{\text{rad},0}
                              \frac{H^4}{M_{\rm Pl}^4} \, \frac{e^{4 \pi \xi}}{\xi^6} \,,
    \qquad
    \text{with}\ \xi \equiv \frac{\dot{\phi}}{2 f_a H} \,.
    \label{PGW-frequency}
\end{align}
In this relation, $H$ and $\dot{\phi}$ are evaluated when a given mode exits the horizon, and therefore the spectrum in \cref{PGW-frequency} is in general scale-dependent. In particular, in the $\xi \gg 1$ regime, the GW amplitude grows exponentially with  the speed of the inflaton, which in turn typically increases over the course of inflation in single-field inflation models. As a consequence, the spectrum in \cref{PGW-frequency} is naturally blue-tilted. The growth of $\xi$ is limited by the backreaction of the gauge fields on the inflaton. Within the limits of a perturbative description, $\xi \lesssim 4.7$ \citep{Peloso:2016gqs}, and GW amplitudes of $\Omega_{\text{GW},0} \simeq 10^{-10}$ can be obtained. \cite{Domcke:2016bkh, Garcia-Bellido:2016dkw} explored the resulting spectrum for several inflaton potentials. In particular hill-top potentials are characterized by a very small speed close to the top (that is mapped to the early stages of observable inflation), and by a sudden increase of $\dot\phi$ at the very end of inflation. Interestingly, hill-top type potentials are naturally present \citep{Peloso:2015dsa} in models of multiple axions such as aligned axion inflation \citep{Kim:2004rp}.

The axionic coupling to gauge fields discussed above can also lead to gravitational wave production in contexts that go beyond inflation. If the pseudoscalar $\phi$ does not play the role of the inflaton, as long as it remains light it is generally expected to remain displaced from its minimum-energy configuration during inflation due to the combined effect of vacuum fluctuations and Hubble friction. The field will only start rolling towards the minimum at a temperature $T_{\rm osc} \sim \sqrt{m_\phi M_{\rm Pl}}$, activating the tachyonic instability of the gauge field, which results again in gravitational waves~\citep{Machado:2018nqk, Machado:2019xuc, Ratzinger:2020oct, Kitajima:2018zco, Chatrchyan:2020pzh, Kitajima:2020rpm}. The signal in this case is strongly peaked and chiral, with the peak amplitude scaling as $\Omega h^2 \sim 10^{-7}(f_\phi/M_{\rm Pl})^4$, which can saturate the dark radiation bound of \cref{eq:CMBbound}. The peak frequency is determined by $T_{\rm osc}$ and thus by the scalar field's mass. Other scenarios that produce similar signals include models of axion kinetic misalignment~\citep{Co:2021rhi, Madge:2021abk} and models with spectator fields that oscillate in the early Universe~\citep{Cui:2023fbg}.

\paragraph{Effective Field Theory of Broken Spatial Reparametrization Symmetry}

Inflationary scenarios based on modifications of general relativity can give rise to enhanced GW production and to a blue-tilted GW spectrum, rendering this emission relevant to high-frequency GW detectors. From the theoretical point of view, the effective field theory (EFT) approach \citep{Cheung:2007st} represents a powerful tool to describe the relevant degrees of freedom at the energy scales of interest and to make predictions for observable quantities.

In the standard single-field effective field theory of inflation \citep{Cheung:2007st}, only time-translation symmetry ($t \to t + \xi_0$) is broken by cosmological expansion. However when space-reparameterization symmetry ($x_i \to x_i + \xi_i$) is also broken \citep{Bartolo:2015qvr, Graef:2015ova}, scalar and tensor perturbations -- the latter corresponding to GWs -- acquire interesting features. In particular tensor perturbations can acquire a mass $m_h$ and sound speed $c_T$, making them potential targets for high-frequency detectors since in this case the spectrum gets enhanced on small scales. At quadratic order, the EFT Lagrangian for graviton fluctuations $h_{ij}$ around a conformally flat Friedmann--Lema\^{i}tre--Robertson--Walker background can be expressed as in \citep{Cannone:2014uqa, Bartolo:2015qvr, Ricciardone:2016lym}:
\begin{align}
    \mathcal{L}_h = \frac{M_{\rm Pl}^2}{8} \, \bigg[
                        \dot h_{ij}^2
                      - \frac{c_T^2(t)}{a^2} \, (\partial_l h_{ij})^2
                      - m_h^2(t) \, h_{ij}^2 \bigg] \,.
    \label{sol-qac}
\end{align}
The corresponding tensor power spectrum and its spectral tilt are
\begin{align}
    {\cal P}_T = \frac{2 H^2}{\pi^2 M_{\rm Pl}^2 c_T^3} \bigg( \frac{k}{k_*} \bigg)^{n_T} \,,\qquad
    n_T = -2 \epsilon + \frac{2}{3} \frac{m_{h}^2}{H^2} \,.
    \label{eq:nteft}
\end{align}
The GW energy density is given by $\Omega_{\text{GW},0} \sim \Omega_{\text{rad},0} {\cal P}_T$.
We see that, if the quantity $m_h/H$ is sufficiently large, the tensor spectrum is blue-tilted with no need to violate the null energy condition in the early Universe. The spectrum is bounded at high frequencies by the observational BBN and CMB bounds, see \cref{eq:BBNbound}.

This scenario shows how GW detectors at high frequency might be useful to test modification of gravity at very high-energy scales.

\paragraph{Second-Order GW Production from Primordial Scalar Fluctuations}

In homogeneous and isotropic backgrounds, scalar, vector and tensor fluctuation modes decouple from each other at first order in perturbation theory. These modes can nevertheless source each other through non-linear effects, starting from second order. In particular, density perturbations (scalar modes) can produce  `induced' (or `secondary') GWs (tensor modes) through a $\zeta + \zeta \to h$ process, where $\zeta$ represents a scalar fluctuation and $h$ the tensor mode \citep{Tomita:1975kj, Matarrese:1993zf, Mollerach:2003nq, Ananda:2006af, Baumann:2007zm} (see also \cite{Kohri:2018awv, Espinosa:2018eve, Braglia:2020eai}). This production, which involves only gravity, is mostly effective when the modes re-enter the horizon after inflation. (Second-order GWs would also be produced in an early matter-dominated era, see \cite{Inomata:2019ivs, Inomata:2019zqy}.) The amplitude of this signal is quadratic in the scalar perturbations.

Scale-invariant $\mathcal{O}(10^{-5})$  perturbations, as measured on large scales in the CMB, result in GWs with unobservably small amplitude. On the other hand, if the spectrum of scalar perturbations produced during inflation has a localized bump at some scale (significantly smaller than the scales probed by the CMB and by large scale structure), a larger GW signal could be generated \citep{Inomata:2016rbd, Garcia-Bellido:2017aan, Bartolo:2018rku}. Such bumps in the spectrum of inflationary perturbations are also interesting in other contexts, for instance they can lead to the production of a sizable primordial BH abundance at some specific mass scale. Conversely, the non-detection of a stochastic GW background can also be used to constrain fluctuations \citep{Byrnes:2018txb, Inomata:2018epa}. The induced GWs have a frequency $f_*$ parametrically equal to the wave number $k_*$ of the modes from which they are produced and can hence be related to the number of $e$-folds, $N$, at which the scalar perturbation exits the horizon through \cref{eq:fN}.

The precise GW yield depends on the statistics of the scalar perturbations \citep{Nakama:2016gzw, Garcia-Bellido:2017aan, Cai:2018dig, Unal:2018yaa}. A reasonable estimate is obtained by simply looking at the scalar two-point function,
\begin{align}
    P_h^{\rm ind} \propto \langle h^2 \rangle 
                  \propto \langle \zeta^4 \rangle
                  \propto P_\zeta^2 \,,
\end{align}
where $P_h^{\rm ind}$ is the power spectrum (two-point function) of the induced GW background and $P_\zeta$ is the power spectrum of the gauge invariant scalar density fluctuations such that $\langle \zeta_{\bf k} \, \zeta_{\bf k'} \rangle \propto  \frac{\delta ({\bf k+k'})}{k^3} \, P_\zeta (k)$. From this relation, the present-day energy density of the induced stochastic GW background is given by
\begin{align}
    \Omega_{\text{GW},0} \sim \Omega_{\text{rad},0} \, P_\zeta^2 \,.
\end{align}
At the largest scales of our observable Universe, $P_\zeta \simeq \num{2e-9}$, resulting in $\Omega_{\text{GW},0} \sim  {\cal O}(10^{-22})$. Primordial BH limits are compatible with $P_\zeta$ as large as $\lesssim 10^{-2.5}$ at some (momentum) scales $k_*$. Scalar perturbations saturating this bound would lead to $\Omega_{\text{GW},0} \sim  {\cal O}(10^{-9})$.

\subsubsection{(P)reheating}
\label{sec:Preheating}

Preheating is an out-of-equilibrium particle production process driven by non-perturbative effects \citep{Traschen:1990sw, Kofman:1994rk, Shtanov:1994ce, Kaiser:1995fb, Khlebnikov:1996mc, Prokopec:1996rr, Kaiser:1997mp, Kofman:1997yn, Greene:1997fu, Kaiser:1997hg}, which takes place after inflation in many models of particle physics (see \cite{Allahverdi:2010xz, Amin:2014eta, Lozanov:2019jxc} for reviews).  After inflation, interactions between the different fields may generate non-adiabatic time-dependent terms in the field equations of motion, which can give rise to an exponential growth of the field modes within certain momentum ranges. The field gradients generated during this stage can be an important source of primordial GWs, with the specific features of the GW spectra depending strongly on the considered scenario, see e.g.\ \cite{Khlebnikov:1997di, GarciaBellido:1998gm, Easther:2006gt, Easther:2006vd, GarciaBellido:2007dg, GarciaBellido:2007af, Dufaux:2007pt, Dufaux:2008dn, Figueroa:2011ye, Ringwald:2022xif}. If instabilities are caused by the inflaton field's own self-interactions, we refer to it as \emph{self-resonance}, a scenario which will be discussed in more detail below. Here we consider instead a multi-field preheating scenario, in which a significant fraction of energy is successfully transferred from the inflationary sector to other fields.

For illustrative process, let us focus on a two-field scenario, in which the post-inflationary oscillations of the inflaton excite a second, massless, field. More specifically, let us consider an inflaton with power-law potential $V(\phi) = \frac{1}{p} \lambda \mu^{4-p} |\phi|^p$, where $\lambda$ is a dimensionless coefficient, $\mu$ is a mass scale, and $p \geq 2$. Let us also define $t_{\star}$ as the time when inflation ends. For $t \gtrsim t_{\star}$, the inflaton field oscillates with a time-dependent frequency $\Omega_{\rm osc} \equiv \omega_{\star} (t /t_{\star})^{1- 2/p}$, where $\omega_{\star} \equiv \sqrt{\lambda} \mu^{(2 - p/2)} \phi_{\star}^{(p/2 - 1)}$ and $\phi_{\star} \equiv \phi(t_{\star})$ \citep{Turner:1983he}. Let us now include a quadratic interaction term $g^2 \phi^2 \chi^2$ between the inflaton and a secondary massless scalar field $\chi$, where $g$ is a dimensionless coupling constant. In this case, the driving post-inflationary particle production mechanism is a parametric resonance \citep{Kofman:1994rk, Kofman:1997yn, Greene:1997fu}. In particular, if the so-called \emph{resonance parameter} $q_{\star} \equiv g^2 \phi_{\star}^2 /\omega_{\star}^2$ obeys $q_{\star} \gtrsim 1$, the secondary field gets excited through a process of broad resonance, and the amplitude of the field modes grows exponentially inside a Bose-sphere of radius $k \lesssim k_{\star} \sim q_*^{1/4} \omega_{\star}$. The GW spectrum produced during this process has a peak at approximately the frequency and amplitude \citep{Figueroa:2017vfa},
\begin{align}
    f \simeq \SI{8e-9}{Hz} \times \bigg( \frac{\omega_\star}{\rho_\star^{1/4}} \bigg) \,
             \epsilon_\star^{\frac{1}{4}} q_\star^{\frac{1}{4} + \eta}
                                                    \label{eq:preheating-1} \,, \\
    \Omega_{\text{GW},0}(f) \simeq \mathcal{O}(10^{-9}) \times \epsilon_\star \, \mathcal{C}
             \frac{\omega_\star^{6}}{\rho_\star M_{\rm Pl}^2} \, q_\star^{-\frac{1}{2} + \delta}
                                                    \label{eq:preheating-2} \,,
\end{align}
where $\rho_\star$ is the total energy density at time $t=t_{\star}$, $\eta \sim 0.3$--0.4 and $\delta \sim 0.1$ are two parameters that account for non-linear effects while $\mathcal{C}$ is a constant that characterizes the strength of the resonance with ${\cal C} \omega_\star^{6}/(\rho_\star M_{\rm Pl}^2) \sim 0.01$--0.001, depending on the model details.
The factor $\epsilon_\star \equiv (a_{\star} /a_{\rm RD})^{1 - 3 w}$ parametrizes the period between the end of inflation and the onset of the radiation dominated era with a transitory effective equation of state parameter $w$. If non-linear effects are ignored, the frequency and amplitude scale as $f \sim q_\star^{1/4}$ and $\Omega_{\text{GW},0} \sim q_{\star}^{-1/2}$, respectively.

The values for $\mathcal{C}$, $\eta$, and $\delta$, can be determined for specific preheating models with classical lattice simulations. For chaotic inflation with quadratic potential $V(\phi) \propto \phi^2$, one finds a frequency in the range $f \simeq (10^8 - 10^9) \, \si{Hz}$ and $\Omega_{\text{GW},0} \simeq (10^{-12} - 10^{-11})$ for resonance parameters $q_{\star} \in (10^4,10^6)$ (assuming $\epsilon_{\star} = 1$). Similarly, for a quartic potential $V(\phi) \propto \phi^4$, one obtains $f \simeq (10^7 - 10^8)\,\si{Hz}$ and $\Omega_{\text{GW},0} \simeq (10^{-13} - 10^{-11})$ if $q_\star \in (1,10^4)$. The GW spectrum in the quartic case also features additional peaks \citep{Figueroa:2017vfa, Ringwald:2022xif}.

GWs be efficiently produced also by fields that carry spin, or when the resonant phenomena driving preheating are different from a parametric resonance. For example, GWs can be produced during the out-of-equilibrium production of fermions after inflation, for both spin-1/2 \citep{Enqvist:2012im, Figueroa:2013vif, Figueroa:2014aya} and spin-3/2 \citep{Benakli:2018xdi} fields. Similarly, GWs can be generated when the produced particles are Abelian or non-Abelian gauge fields. These gauge fields can for example be coupled to a complex scalar field via a covariant derivative \citep{Dufaux:2010cf, Figueroa:2016ojl, Tranberg:2017lrx}, or to a pseudo-scalar field via an axial coupling \citep{Adshead:2018doq,Adshead:2019igv,Adshead:2019lbr}. Preheating can be remarkably efficient in the second case, and the resulting GW energy density can be as larger as $\Omega_{\rm GW} \sim \mathcal{O}(10^{-6} - 10^{-7})$ for certain coupling strengths, see \cite{Adshead:2019igv, Adshead:2019lbr} for more details. Production of GWs during preheating with non-minimal couplings to the curvature scalar $R$ has also been explored in \cite{Fu:2017ero}. Finally, the stochastic background of GWs from preheating may develop anisotropies if the inflaton is coupled to a secondary light scalar field, see \cite{Bethke:2013aba, Bethke:2013vca}.

\paragraph{Oscillon Production.}

Oscillons are long-lived compact scalar field configurations \citep{Gleiser:1993pt} that can be formed in the early Universe in a variety of post-inflationary scenarios which involve a preheating-like phase \citep{Amin:2010jq, Amin:2010dc, Amin:2011hj, Zhou:2013tsa, Amin:2013ika, Lozanov:2014zfa, Antusch:2015vna, Antusch:2015ziz, Antusch:2016con, Antusch:2017flz, Antusch:2017vga, Lozanov:2017hjm, Amin:2018xfe, Antusch:2019qrr, Sang:2019ndv, Lozanov:2019ylm, Fodor:2019ftc, Hiramatsu:2020obh}. Their dynamics is a possible source of GW production. Oscillons are pseudo-solitonic solutions of real scalar field theories: their existence is due to attractive self-interactions of the scalar field that balance the outward pressure.\footnote{If the scalar field is complex and the potential features a global $U(1)$ symmetry, non-topological solitons like Q-balls \citep{Coleman:1985ki} can be formed during the post-inflationary stage, giving rise to similar GW signatures \citep{Chiba:2009zu}.} The real scalar field self-interactions are attractive if the scalar potential is shallower than quadratic at least on one side with respect to the minimum. Oscillons can be thought of as bubbles in which the scalar field is undergoing large oscillations that probe the non-linear part of the potential, while outside the scalar field is oscillating with a very small amplitude around the minimum of the potential.

As discussed in the previous section, during preheating the quantum fluctuations of the scalar field that may ultimately form oscillons are amplified due to a resonant process. The Universe ends up in a very inhomogeneous phase in which the inflaton (or any other scalar field that drives preheating) is fragmented and there are large fluctuations in the energy density. At this point, if the field is subject to attractive self-interactions, the inhomogeneities can clump and form oscillons. The geometric shape of the oscillons initially deviates significantly from being spherically symmetric, therefore their dynamics produce GWs. After many oscillations of the scalar field, oscillons tend to become spherically symmetric and GW production stops. However, during their entire lifetime oscillons can produce GWs also due to interactions and collisions among each other \citep{Helfer:2018vtq}. Oscillons are very long-lived: their lifetime is model-dependent but typically $\gtrsim 10^4/m$ \citep{Gleiser:2008ty, Amin:2010jq, Amin:2010dc, Amin:2011hj, Salmi:2012ta, Saffin:2014yka, Antusch:2019qrr, Gleiser:2019rvw, Zhang:2020bec}, where $m$ is the mass of the scalar field. Oscillons eventually decay through classical \citep{Segur:1987mg} or quantum radiation \citep{Hertzberg:2010yz}.

The peak of the GW spectrum at production is centered at a frequency slightly below the mass of the field, which typically lies well above the LIGO range today \citep{Zhou:2013tsa, Antusch:2017flz, Lozanov:2019ylm}.\footnote{See however \cite{Antusch:2016con, Liu:2017hua, Kitajima:2018zco} for models that lead to a GW peak at lower frequencies.} Typically, an oscillating massive scalar field forming oscillons quickly comes to dominate the energy density of the Universe until the perturbative decay of the field itself. For the simplest case of a gravitationally coupled massive field that starts oscillating at $H \simeq m$ and decays at $H \sim m^3/M_{\rm Pl}^2$) the frequency today can be estimated as
\begin{align}
    f \simeq \SI{e6}{Hz} \times X \times \bigg( \frac{m}{\SI{e12}{GeV}} \bigg)^{5/6} \,.
    \label{eq:OscillonFrequency}
\end{align}
Here the factor $X$, which is typically in the range $X \simeq (10-10^3)$, is due to the uncertainty in the precise time at which GWs are produced. $X$ can be obtained in concrete models from lattice simulations: $X \simeq 10$ would hold if GWs were produced immediately when the scalar field starts oscillating.\footnote{This rough estimate assumes that the field starts oscillating when $H \simeq m$. Since the potential contains self-interactions, assuming that the field starts at rest, the actual requirement for the start of the oscillations is $V''(\phi_{\text{in}}) \sim H$, where $\phi_{\rm in}$ is the initial value of the field. Note also that if the field is the inflaton itself, the initial conditions are different from those assumed in \cref{eq:OscillonFrequency}, and therefore this estimate does not necessarily hold, see e.g., \cite{Antusch:2016con}.} On the other hand, the later GWs are produced, the less the frequency is red-shifted and the larger is $X$. The maximum value of the GW energy density today for these processes, inferred from numerical simulations, is in the range $\Omega_{\text{GW},0} \simeq (10^{-13} - 10^{-10})$ \citep{Antusch:2016con, Antusch:2017flz, Amin:2018xfe}, see \cite{Dufaux:2007pt} for a discussion on how to compute the GW amplitude.

Depending on the model, gravitational effects can become important and play a crucial role for the existence/stability of the oscillon solution \citep{Seidel:1991zh}. In particular, the requirement that the potential must be shallower than quadratic is no longer necessary, as the attractive force is provided by gravity \citep{UrenaLopez:2002gx}. In this case oscillons are equivalent to oscillatons, see \cref{sec:ECOs}, and can give rise to interesting additional effects, such as the collapse to BHs \citep{Muia:2019coe, Giblin:2019nuv, Kou:2019bbc, Nazari:2020fmk}.

\subsubsection{The Cosmic Gravitational Microwave Background}
\label{sec:CGMB}

The hot thermal plasma of the early Universe acts as a source of GWs, which, similarly to the relic photons of the CMB, peak in the $\sim \SI{100}{GHz}$ range today. This makes this range of frequencies particularly interesting to target, since the source is the well-established Standard Model and the prediction is based on standard cosmology. The spectrum of this signal is determined by the particle content and the maximum temperature $T_{\rm max}$ reached by the thermal plasma in the history of the Universe \citep{Ghiglieri:2015nfa, Ghiglieri:2020mhm, Ringwald:2020ist}. The energy density in GWs per logarithmic frequency interval can then be written as follows,
\begin{multline}
    \Omega_{\rm GW,0}(f) \simeq
        \frac{1440 \sqrt{10}}{2 \pi^2} \,
        \frac{g_{\star s}(T_{\rm end})^{1/3}}{g_{\star s}(T_{\rm max})^{5/6}} \,
        \Omega_{{\rm rad}, 0} \\
    \times \frac{f^3}{T_0^3} \frac{T_{\rm max}}{M_{\rm Pl}} \, \hat{\eta} \Big(
        T_{\rm max}, 2 \pi \big( \tfrac{g_{\star s}(T)}{g_{\star s}(T_{\rm end})} \big)^{1/3}
                                 \tfrac{f}{T_0} \Big) \,.
    \label{eq:CGMB}
\end{multline}
In the above expression, $T_0$ is the temperature of the CMB today, while $T_{\rm end}$ denotes the temperature at which thermal production of gravitational waves stopped. $T_{\rm end}$ is generally taken as the temperature of the electroweak phase transition. Given that GW production is ultraviolet-dominated this is a reasonable approximation. The function $\hat\eta$ encodes the sources of GW production in the thermal plasma, which is dominated by long range hydrodynamic fluctuations at $2 \pi f < T_0$ and by quasi-particle excitations in the plasma at $2 \pi f \sim T_0$, see \cite{Ghiglieri:2015nfa, Ghiglieri:2020mhm, Ringwald:2020ist} for more details. For frequencies corresponding to modes that were superhorizon at the time when $T = T_{\rm max}$, and thus would be prevented from evolving until horizon entry, $T_{\rm max}$ in \cref{eq:CGMB} should be replaced with the horizon crossing temperature $T_{\rm hc}(f) \sim M_{\rm Pl} f / (\SI{6e10}{Hz})$ \citep{Drewes:2023oxg}. Corrections to $\hat\eta$ from two-graviton emission have been computed in \citep{Ghiglieri:2022rfp, Ghiglieri:2024ghm}.

The peak frequency of the spectrum in \cref{eq:CGMB} is 
\begin{align}
    f_{\rm peak}^{\Omega_{\rm CGMB}}\approx 79.8\, {\rm GHz}\left(\frac{106.75}{g_{\star s}(T_{\rm max})}\right)^{1/3} .
\end{align}
where $g_{*s}(T = T_{\rm max})$ is the number of entropic relativistic degrees of freedom at $T_{\rm max}$. The peak amplitude of $\Omega_{\rm GW, 0}(f)$ approaches the dark radiation bound, \cref{eq:CMBbound}, if $T_\text{max} \sim {\cal O}({\rm few})\times \SI{e19}{GeV}$, and thus close to $M_{\rm Pl}$. The CMB constraints on the tensor-to-scalar ratio, however, impose a tighter constraint, namely $T_\text{max} < \SI{6.6e15}{GeV} \times (106.75/ g_{\star s}(T_{\rm max}))^{1/4}$ \citep{Akrami:2018odb} under the assumption of slow-roll inflation and instantaneous reheating. Therefore the detection of the cosmic gravitational microwave background with a spectrum pointing to $T_{\rm max} > \SI{e16}{GeV}$ would rule out slow-roll inflation as a viable pre hot Big Bang scenario. Note that since at leading order $\Omega_{\rm GW,0}(f)$ scales linearly with $T_{\rm max}$ and the peak frequency depends on $g_{*s}(T_{\rm max})$, the detection of the peak of the cosmic gravitational microwave background would determine both $T_{\rm max}$ and $g_{*s}(T_{\rm max})$, see \cite{Ringwald:2020ist} for more details.

Going beyond standard scenarios, the possibility of nonstandard cosmological histories has been considered in \cite{Muia:2023wru}. These authors, as well as \cite{Drewes:2023oxg}, have also considered the possible existence of several decoupled hidden sectors with different temperatures. The impact of strong coupling on the emission rate has been analyzed in \cite{Castells-Tiestos:2022qgu}, and graviton emission from high-temperature fundamental strings has been considered in \cite{Frey:2024jqy}. The resulting spectrum from the latter process has robust characteristics: it peaks at frequencies of order 50--\SI{100}{GHz}, and contrary the predictions of other scenarios beyond the Standard Model the amplitude is hierarchically larger than the one in the Standard Model. Notably, it is directly proportional to the string scale, indicating that a potential signal may also determine the string scale.

\subsubsection{Phase Transitions}
\label{sec:PhaseTransitions}

A first order phase transition in the early Universe proceeds by the nucleation of bubbles of the low-temperature phase as the Universe cools below the critical temperature \citep{Steinhardt:1981ct, Hogan:1984hx}.\footnote{The \emph{critical} temperature, $T_c$, denotes the temperatures at which the low-temperature vacuum state becomes energetically favorable compared to the high-temperature state that the Universe is in before the phase transition. The \emph{nucleation} temperature, $T_*$, denotes the temperature at which the first bubbles form. It is usually similar to $T_c$, but can be much lower in the case of supercooled phase transitions.}  Due to the higher pressure inside, the bubbles expand and collide until the stable phase fills the whole Universe. The process disturbs the fluid, generating shear stress and hence GWs \citep{Witten:1984rs, Hogan:1986qda}. As the perturbations are mostly compression waves, they can be described as sound waves, and their collisions are often the main source of GWs \citep{Hindmarsh:2013xza, Hindmarsh:2015qta, Hindmarsh:2017gnf}. The peak frequency of an acoustic contribution to the relic GW background from a strong first order transition is controlled by the temperature of the transition, $T_*$, and by the mean bubble separation $R_*$.\footnote{The subscript $*$ denotes quantities evaluated at the bubble nucleation time.} Numerical simulations show for bubble wall speeds well above the speed of sound that \citep{Hindmarsh:2017gnf}
\begin{align}
    f_{\rm peak} \simeq \SI{260}{MHz} \times \bigg( \frac{1}{H_\ast R_\ast} \bigg)
                                            \bigg( \frac{T_\ast}{\SI{e15}{GeV}} \bigg)
                                            \bigg( \frac{g_\ast (T_\ast)}{100 } \bigg)^{1/6} \,,
    \label{eq:PTf}
\end{align}
where $H_\ast$ is the Hubble rate at nucleation. The theoretical expectation is that $1 \lesssim (H_\ast R_\ast)^{-1} \lesssim 10^4$.
Remarkably, phase transitions in the very early universe, possibly associated with grand unification or the breaking of $B$$-$$L$, a Peccei-Quinn symmetry or flavour symmetries are natural candidates for high frequency gravitational waves.
The intensity of the GW emission depends on $H_\ast R_\ast$, on the fraction $K$ of the energy density of the Universe which is converted into kinetic energy during the phase transition, and on the lifetime of the source, which can last for up to a Hubble time.
With the lifetime of the velocity perturbations given by $\tau_{\rm v}\approx4 R_\ast/(3{K})$, the GW spectrum can be estimated as \citep{Hindmarsh:2015qta, Guo:2020grp}
\begin{align}
    \Omega_{\text{GW},0}(f) \simeq & 2 H_\ast R_\ast
        \left(1 - (1+2 H_\ast \tau_{\rm v})^{-1/2} \right)
        \left( \frac{100}{g_\ast (T_\ast)} \right)^{1/3} K^2
        \tilde\Omega_\text{GW} \nonumber \\
        & \times \Omega_{\text{rad},0} \, S\left(\frac{f}{f_{\rm peak}}\right) ,
    \label{eq:PTOmega2}
\end{align}
where $\tilde\Omega_\text{GW} \approx 0.058$ is an efficiency factor obtained from simulations. The frequency dependence of $\Omega_{\text{GW},0}(f)$ is determined by the function $S(f/f_{\rm peak})$, with
\begin{align}
   S(x) = x^3 \bigg(\frac{7}{4 + 3 x^2} \bigg)^{7/2} ,
\end{align}
which takes its maximal value of 1 at $x=1$. Numerical simulations indicate $\tilde\Omega_\text{GW} = \mathcal{O}(10^{-2})$.
Hence, $\Omega_{\text{GW},0} \lesssim 10^{-7}$ today, with the upper bound reached only if most of the energy available in the phase transition is turned into kinetic energy. This is only possible if there is significant supercooling. 

The calculation of the kinetic energy fraction and the mean bubble separation requires knowledge of the free energy density ${\cal F}(T,\phi)$, a function of the temperature and the order parameter of the phase transition. If the underlying quantum theory is weakly coupled, and the scalar particle corresponding to $\phi$ is light compared to the masses gained by gauge bosons in the phase transition, this is easily calculated, and shows that first order transitions are generic in gauge theories in this limit \citep{Kirzhnits:1972iw, Kirzhnits:1976ts}, meaning that there is a temperature range in which there are two minima of the free energy as a function of $\phi$.  The critical temperature is defined as the temperature at which the two minima are degenerate, separated by a local maximum.

The key parameters to be extracted from the underlying theory, besides the critical temperature $T_c$, are the nucleation rate $\beta$, the strength parameter $\alpha$ and the bubble wall speed $v_w$. The nucleation rate parameter $\beta = d \log p/dt$, where $p$ is the bubble nucleation rate per unit volume, is calculable from ${\cal F}(T,\phi)$ by applying homogeneous nucleation theory \citep{Langer:1969bc} to fields at high temperature \citep{Linde:1981zj}.
This calculation also gives $T_\ast $ as the temperature at which the volume-averaged bubble nucleation rate peaks.
The strength parameter is roughly, but not precisely, one quarter of the latent heat divided by the thermal energy (see \cite{Hindmarsh:2019phv} for a more precise definition) at the nucleation temperature and also follows from knowing ${\cal F}(T,\phi)$.
The wall speed is a non-equilibrium quantity, which cannot be extracted from the free energy alone, and is rather difficult to calculate accurately (see \cite{Dorsch:2018pat, Laurent:2020gpg, Ai:2023see, Ekstedt:2024fyq} and references therein). In terms of these parameters, it can be shown that $R_\ast \approx (8\pi)^{1/3} v_w/\beta$ \citep{Enqvist:1991xw}.
The kinetic energy fraction $K$ can be estimated from the self-similar hydrodynamic flow set up around an isolated expanding bubble, whose solution can be found as a function of the latent heat and bubble wall velocity by  a simple one-dimensional integration \citep{Turner:1992tz, Espinosa:2010hh, Hindmarsh:2019phv}. $K$ is usually parameterized in terms of an efficiency factor $\kappa$ and the phase transition strength $\alpha$, with $K = \kappa \alpha / (1 + \alpha)$. Approximate fits for $\kappa$ in terms of $\alpha$, $v_w$ can be found in \cite{Espinosa:2010hh}; for example, in the limit of near-luminal velocity, one has $\kappa\approx\alpha/(0.73+0.083\sqrt{\alpha}+\alpha)$.  Typically, $K$ falls in the range between $K = 1$--$10^{-6}$.

Current projected sensitivities for Einstein Telescope and Cosmic Explorer indicate that these observatories will be able to probe cosmological first order transition occurring at temperatures of at most $\text{few} \times \SI{100}{TeV}$ assuming a modest amount of supercooling \citep{Evans:2016mbw, Punturo:2010zz, Hild:2010id} (i.e., when $T_\ast \sim T_c$ and $(R_\ast H_\ast)^{-1} \gtrsim 100$).
When considering high-scale transitions, it should be kept in mind that if the transition happens immediately after inflation, the gravitational wave signal could be substantially diluted by an early matter-dominated epoch that typically follows inflation. Additionally, it should be noted that the frequency has an upper bound of $\sim \SI{e2}{GHz}$, since the maximal temperature of the Universe is bounded by CMB observations and the distance scale $R_*$ cannot be smaller than the mean free path associated with thermal fluctuations, $\sim 1 /T_*$~\citep{Ghiglieri:2024ghm}.

\subsubsection{Topological Defects}
\label{sec:TopologicalDefects}

Cosmic strings are one-dimensional topological defect solutions to a field equation which may have formed after a phase transitions in the early Universe if the first homotopy group of the vacuum manifold associated with the symmetry breaking is non-trivial \citep{Kibble:1976sj, Jeannerot:2003qv}. They can also be fundamental strings from string theory, formed for instance at the end of brane inflation \citep{Dvali:2003zj, Copeland:2003bj}, and stretched to cosmological scales. The energy per unit length of a string is $\mu \sim \eta^2$, with $\eta$ the characteristic energy scale. (In the case of topological strings, $\eta$ is the energy scale of the phase transition that generated the strings.) Typically, the tension of the strings is characterized by the dimensionless combination $G \mu \sim (\eta/M_{\rm Pl})^2$. The current upper bound from the CMB is $G \mu \lesssim 10^{-7}$, whereas GW searches in pulsar timing arrays constrain the string tension to $G \mu \lesssim 10^{-11}$. Cosmic strings are energetic objects that move at relativistic speeds. The combination of these two factors immediately suggests that strings should be a powerful source of GWs.

When cosmic strings are formed in the early Universe, their dynamics rapidly drive them into an attractor solution, characterized by their fractional energy density relative to the background energy density of the Universe remaining constant. This is known as the `scaling' regime. During this regime, strings will collide and possibly intercommutate. For topological strings the intercommutation probability is $\mathcal{P}=1$, whereas $\mathcal{P}<1$ is characteristic for cosmic superstrings networks. Closed string configurations -- loops -- are consequently formed when a string self-intersects, or when two strings cross. Loops smaller than the horizon decouple from the string network and oscillate under their own tension, which results in the emission of gravitational radiation (eventually leading to the decay of the loop). The relativistic nature of strings typically leads to the formation of \emph{cusps}, corresponding to points where the string momentarily moves at the speed of light \citep{Turok:1984cn}. Furthermore, the intersections of strings generates discontinuities on their tangent vector known as \emph{kinks}. All loops are typically expected to contain cusps and kinks, both of which generate GW bursts \citep{Damour:2000wa,Damour:2001bk}. Hence, a network of cosmic (super-)strings formed in the early Universe is expected to radiate GWs throughout the entire cosmological history, producing a stochastic background of GWs from the superposition of many uncorrelated bursts. While searches for cosmic string are normally searched for this stochastic background, an alternative strategy is to search for individual strong bursts, which could manifest as transient GW signals \citep{Aasi:2013vna, Abbott:2017mem}.

A network of cosmic strings in the scaling regime contains, at every moment of its evolution, sub-horizon loops and long strings that stretch across a Hubble volume. The latter are either infinite strings or they form super-horizon loops, and they are also expected to emit GWs. However, the dominant contribution is generically the one produced by the superposition of radiation from many sub-horizon loops along each line of sight.

The power emitted into gravitational radiation by an isolated string loop of length $l$ can be calculated using the standard formalism in the weak gravity regime, see \cite{Weinberg:1972kfs}. More explicitly, we can assume that, on average, the total power emitted by a loop is given by $P_{\rm 1Loop} = \Gamma \times (G \mu) \times \mu$, where $\Gamma$ is a dimensionless constant independent of the size and shape of the loops. Estimates from simple loops \citep{Vachaspati:1984gt, Burden:1985md, Garfinkle:1987yw}, as well as results from Nambu--Goto simulations \citep{Blanco-Pillado:2017oxo}, suggest that $\Gamma \simeq 50$. The GW radiation is only emitted at discrete frequencies by each loop, $\omega_n = 2 \pi n/T$, where $T=l/2$ is the oscillation period of the loop, and $n$ is an integer $\geq 1$. We can write $P_{\rm 1Loop} = \mathrm{G} \mu^2 \sum_n P_n$, with $P_n$ characterizing the power emitted at each frequency $\omega_n$ for a particular loop, depending on whether the loop contains cusps or kinks, and whether kink--kink collisions occur \citep{Burden:1985md, Allen:1991bk}. It can be shown that for large $n$, $P_{ n} = (\Gamma / \zeta(q)) n^{-q}$, where $\zeta(q)$ is the Riemann zeta function, which appears here as a normalization factor to ensure that the total power of the loop is equal to $\Gamma = \sum_n P_n$. The parameter $q$ takes the values $4/3$, $5/3$, or $2$ depending on whether the emission is dominated by cusps, kinks or kink--kink collisions, respectively, see for e.g.\ \cite{Vachaspati:1984gt, Binetruy:2009vt, Auclair:2019wcv}.

The stochastic GW background emitted by loops generated during the radiation domination period is characterized by a scale-invariant energy spectrum, spanning many decades in frequency. The high-frequency cut-off of this spectrum is determined by the temperature of the thermal bath at formation of the string network, with the CMB bound on the reheating temperature, $T_\text{max} \lesssim \SI{e16}{GeV}$, implying a cut-off frequency of $f_\Delta \lesssim \SI{e9}{GeV}$ \citep{Gouttenoire:2019kij}.
The amplitude of the plateau is given by \citep{Auclair:2019wcv}
\begin{align}
    \Omega^{\rm plateau}_{\text{GW},0}(f) \approx 8.04 \, \Omega_{\text{rad},0}
        \sqrt{\frac{G \mu}{\Gamma}} \,.
    \label{eq:plateauStringsNG}
\end{align}
Note that this estimate does not depend on the exact form of the loops' individual power spectra, nor on whether the GW emission is dominated by cusps or kinks. Rather, it depends only on the total GW radiation emitted by the loops. \Cref{eq:plateauStringsNG} indicates that the stochastic GW background from cosmic strings can be rather strong.\footnote{{\it Important remark}: as the characteristic width $\delta \sim 1/\eta$ of a cosmic string is generally much smaller than the horizon scale, it is commonly assumed that strings can be described by the Nambu--Goto action, which is the leading-order approximation when the curvature scale of the strings is much larger than their thickness. The plateau in \cref{eq:plateauStringsNG} applies only for the case of Nambu--Goto strings. For these strings to reach the scaling regime, GW emission from loops is actually crucial as it is the loss of loops from the network that guarantees scaling, and GW emission provides a mechanisms for loops to decay. However, in field theory simulations of string networks \citep{Vincent:1997cx, Hindmarsh:2008dw, Daverio:2015nva, Hindmarsh:2017qff}, the network of infinite strings reaches a scaling regime thanks to energy loss into classical radiation of the fields involved in the simulations. The simulations show the presence of extensive radiation of massive particles being emitted, and the loops that are formed decay within a Hubble time. This intriguing discrepancy has been under debate for the last $\sim 20$~years, \reply{and has very significant impact on the amplitude of the predicted GW spectrum.}}

Moreover, if the phase transition responsible for cosmic string formation is originating from symmetry breaking in a grand unified theory (GUT), then, depending on the structure of the GUT symmetry group, cosmic strings may be metastable, decaying via the (exponentially suppressed) production of monopoles \citep{Vilenkin:1982hm, Monin:2008mp, Monin:2009ch, Leblond:2009fq}. In this case, the low-frequency end of the spectrum, corresponding to GW emission at later times, is suppressed and the signal may \emph{only} be detectable at high frequencies \citep{Leblond:2009fq, Dror:2019syi, Buchmuller:2019gfy, Buchmuller:2020lbh}. In this case, the string tension is only constrained by the BBN bound on $N_{\rm eff}$, $G \mu \lesssim 10^{-4}$, and the scale-invariant part of the spectrum may extend from \SI{e3}{Hz} (LIGO constraint) up to \SI{e9}{Hz} (network formation).

Finally, let us recall that long strings (infinite and super-horizon loops) also radiate GWs. One contribution to this signal is given by the GWs emitted around the horizon scale at each moment of cosmic history, as the network's energy--momentum tensor adapts itself to the scaling regime \citep{Krauss:1991qu,JonesSmith:2007ne,Fenu:2009qf,Figueroa:2012kw,Figueroa:2020lvo}. This emission is expected from any network of cosmic defects in the scaling regime, independent of the topology and origin of the defects \citep{Figueroa:2012kw}. In the case of cosmic string networks modeled by the Nambu--Goto approximation (where the thickness of the string is taken to be zero), this irreducable background represents a very sub-dominant signal compared to the GW background emitted from sub-horizon loops. In the case of field theory strings (for which simulations to date indicate an absence of `stable' loops), it is instead the only GW signal emitted by the network.

The GW energy density spectrum of this irreducible background from long strings is predicted to be exactly scale-invariant for the modes emitted during radiation domination \citep{Figueroa:2012kw}. The power spectrum from long strings therefore mimics the spectral shape of the dominant signal from loop decay, but with a smaller amplitude. The amplitude depends on the fine details of the unequal-time correlator of the network's energy-momentum tensor. This correlator can be obtained accurately only from sufficiently large scale lattice simulations. For strings based on a global symmetry (global strings), the scale-invariant GW power spectrum has been obtained numerically from massively parallel lattice field theory simulations, with a predicted energy density of \citep{Figueroa:2012kw}
\begin{align}
    \mathrm{\Omega}_{\text{GW},0}(f) \simeq 4\times 10^4 \, (G \mu)^2 \, \Omega_{\text{rad},0} \,.
    \label{eq:GWSOSFplateau}
\end{align}
The irreducible background from the more interesting case of an Abelian Higgs model has unfortunately not been studied yet. Despite the large numerical prefactor in \cref{eq:GWSOSFplateau}, the quadratic dependence on $(G\mu)^2$ suppresses the energy density significantly, see e.g., \cite{Buchmuller:2013lra} for a comparison among GW signals emitted from the same string network. This amplitude is clearly subdominant when compared to the amplitude of the GW signal from loops, which scales as $(G\mu)^{1/2}$ according to \cref{eq:plateauStringsNG}.

Finally, we point out that, since the irreducible GW emission described above is expected from any network of defects in the scaling regime, global texture networks also emit a GW background due to their self-ordering during scaling \citep{JonesSmith:2007ne, Fenu:2009qf, Giblin:2011yh, Figueroa:2012kw,  Figueroa:2020lvo}. Textures are formed when the second (or higher) homotopy group of the vacuum manifold is non-trivial \citep{Vilenkin:2000jqa}. Such conditions can be realized in case of the breaking of a global or gauge symmetry. In the case of a global symmetry, the GW spectrum is scale invariant for radiation domination \citep{JonesSmith:2007ne, Fenu:2009qf, Figueroa:2012kw}, and exhibits a peak at the horizon today for matter domination \citep{Figueroa:2020lvo}. In the case of gauged textures one instead expects a peaked spectrum, with the peak frequency and amplitude of the GW background set by the symmetry breaking scale $v$ \citep{Dror:2019syi},
\begin{align}
    f \sim \SI{e11}{Hz} \times \frac{v}{M_{\rm Pl}} \,,
    \qquad
    \Omega_{\text{GW},0} \sim 2 \times 10^{-4} \, \Big( \frac{v}{M_{\rm Pl}} \Big)^4 \,.
    \label{eq:gaugetextures}
\end{align}
Given that the frequency and amplitude both increase with $v$, it is not unlikely that such signals will be most easily detectable by high-frequency detectors.

\subsubsection{Evaporating Primordial Black Holes}
\label{sec:EvaporatingPBHs}

In \cref{sec:PBHmergers}, we have discussed GW signals emitted by primordial BHs merging in the late Universe. Very light primordial BHs (with masses smaller than \SI{e11}{kg}), which  evaporate before BBN, could produce an $\mathcal{O}(\si{GHz})$ stochastic spectrum of GWs by merging and scattering \citep{Dolgov:2011cq}.  Here we consider yet another source of GWs tied to primordial BHs, namely the emission of gravitons as part of their Hawking radiation. This is particularly relevant for light primordial BHs evaporating either before BBN, or between BBN and the present day. 

The graviton emission from a population of primordial BHs induces a stochastic background of GWs \citep{Anantua:2008am, Dong:2015yjs, Ireland:2023avg} that peaks at very high frequencies, between $f \sim \SI{e13}{Hz}$ and \SI{e22}{Hz}. The shape and amplitude of the resulting GW frequency spectrum depends on multiple factors, such as the primordial BHs' abundance at formation, their mass spectrum, their spin, and the number of degrees of freedom in the particle physics theory. Due to the redshift of the GW amplitude and frequency, the observable GW spectrum today is dominated by the latest stages of primordial BH evolution, and the frequency is hence set by the evaporation time (which in turn depends on the initial mass) of the primordial BH.

Taking into account the limits on the primordial BH abundance from BBN and from extra-galactic background radiation, the maximum amplitude can be up to $\Omega_{\text{GW},0} \approx 10^{-7.5}$ for primordial BHs evaporating just before BBN, corresponding to an initial mass $m_{i\,\rm PBH} \lesssim \SI{e9}{g}$ at formation.  For heavier BHs that might have not fully evaporated yet today, $\SI{e9}{g} \lesssim m_{i\,\rm PBH} \lesssim \SI{e17}{g}$, it can be up to $\Omega_{\text{GW},0} \approx 10^{-6.5}$ \citep{Dong:2015yjs}, with a spectrum peaked at frequencies between \SI{e18}{Hz} and \SI{e22}{Hz}. See also \cite{Ireland:2023avg} for more details. Finally, much lighter primordial BHs that would have completely evaporated long before BBN are of interest as well. Because the primordial BH density decreases $\propto 1/a^3$ \reply{(with $a$ denoting the scale factor of the expanding Universe)}, while the radiation density is $\propto 1/a^4$, such early decaying primordial BHs can be very abundant in the early Universe, leading to an early matter dominated phase. GWs produced in their decay could then constitute a sizable fraction of the energy density during the subsequent radiation dominated epoch, limited only by the BBN and CMB constraints (see \cref{eq:CMBbound}). For primordial BHs produced close to the grand unification scale, $E \sim \SI{e15}{GeV}$, the GW frequency spectrum has a peak around \SI{e15}{Hz} and can reach an amplitude $\Omega_{\text{GW},0}(f) \sim 10^{-8}$ for a Universe with $\sim 10^3$ degrees of freedom \citep{Anantua:2008am}.

For primordial BHs in theories with large extra dimensions, the peak frequency can be lowered substantially, since the true bulk Planck scale $M_*$ can be much smaller than the effective 4d Planck scale. For an optimal choice of parameters, the peak frequency may then be $< \si{MHz}$ \citep{Ireland:2023zrd}.


\subsubsection{Miscellaneous}
\label{sec:Misc}

In the following we summarize a few additional sources of high-frequency GWs that require more exotic setups.

\paragraph{Brane-world Scenarios.}

In a brane-world scenario \citep{Rubakov:1983bb}, the very weak force of gravity in our $(3+1)$-dimensional Universe arises from a stronger gravitational force that is felt in a fifth dimension at a level commensurate with the other forces. This scenario suggests that two $(3+1)$-dimensional branes -- one of which represents our four-dimensional Universe, while the other is a `shadow' brane -- are separated in a fifth dimension by a small distance \citep{Randall:1998uk, Maartens:2010ar}. If violent gravitational events -- such as BH mergers -- take place on the shadow brane, they would excite oscillations not only in the shadow brane but also in the five-dimensional space separating the branes. This leads to GW production on our visible brane as well \citep{Seahra1, Seahra2}.

\paragraph{Pre-Big Bang Cosmology.}

The pre-Big Bang scenario provides an alternative to cosmological inflation as a mechanism for setting the initial conditions for the hot Big Bang. The scenario exploits the fundamental symmetries of string theory to build a model in which the Universe starts in a cold and empty state in the infinite past and moves towards a state of high curvature through accelerated expansion \citep{Gasperini:2002bn, Gasperini:2007vw}. The state of high curvature corresponds to a region in the parameter space in which the theory is strongly coupled. It is then assumed that the strongly coupled theory is able to match this initial accelerated expansion to the usual hot Big Bang cosmology. Interestingly, this scenario predicts a blue spectrum of GWs, with a peak at high frequency \citep{Brustein:1995ki}.

\paragraph{Quintessential Inflation.}

If the inflationary epoch is followed by a phase in which the equation of state is stiffer than radiation ($w > 1/3$), the stochastic spectrum of GWs features a growth at high frequency, followed by a sharp cutoff \citep{Giovannini:1999bh}. Such behavior is expected in quintessential models of inflation such as the one investigated in \cite{Peebles:1998qn}. The position of the peak depends very weakly on the number of minimally coupled scalar fields of the model, but it is independent of the final curvature at the end of inflation. Therefore, it is always located at $\sim \SI{100}{GHz}$. The amplitude of the GW spectrum can become very large: in \cite{Giovannini:1999bh} the authors present a choice of the parameters such that $\Omega_{\rm GW, 0} \simeq 10^{-6}$ at the peak.

\paragraph{Magnetars.}

Magnetars are neutron stars with extremely large surface magnetic fields $\sim 10^9$--\SI{e11}{Te}. Ref.~\cite{Wen:2017itr} suggests that gamma-ray bursts produced by the magnetar or by a companion object in a binary system, and interacting with the surface magnetic field of the magnetar could be a source of high-frequency GW, with frequency around \SI{e20}{Hz} and energy density at Earth up to $\Omega_{\rm GW, 0} \sim 10^{-6}$.

\paragraph{Reheating.}

The oscillations of the inflaton (or another scalar field in the inflationary sector) around the minimum of its potential at the end of inflation constitutes a model-independent source of stochastic GWs \citep{Ema:2020ggo}. The oscillations act as a driving force in the equation of motion for the tensor modes, leading to GW production at high frequency $\gtrsim \SI{e5}{Hz}$. The amplitude of this signal is bound to be quite small: in \cite{Ema:2020ggo} the authors present a choice of parameters such that $\Omega_{\rm GW, 0} \lesssim 10^{-21}$.

\paragraph{Plasma Instabilities.}

Ref.~\cite{Servin:2003cf} studied interactions of electromagnetic waves and GWs in a magnetized plasma. In the high-frequency regime, a circularly polarized electromagnetic wave traveling parallel to the background magnetic field present in a plasma generates GWs with the same frequency as the electromagnetic wave. However, no specific estimates for the amplitude and spectrum of the resulting GW background at Earth have been derived yet.


\subsection{Gravitational Wave Generation in Laboratory Setups}
\label{sec:GWgeneration}

The possibility of laboratory control of gravitational fields was considered in the early 1960s in \cite{PhysRev117306} and \cite{Gertsenshtein}. The power radiated into gravitational waves at \SI{300}{MHz} by electrically-induced stresses in a piezoelectric crystal with a size of fifty centimeters on a side was calculated to be up to \SI{e-20}{W}, seventeen orders of magnitude above the maximal power generated by a spinning rod having the same length as the crystal. Assuming isotropy, the corresponding strain is $h \approx 10^{-38}$ ten meters away from the source.

In 1973, gravitational radiation generated by alternating electromagnetic fields inside resonant cavities has been investigated \citep{GrishchukSazhin1973}. Assuming a hypothetical rectangular cavity of dimensions $\SI{e-2}{m} \times \SI{1}{m} \times \SI{1}{m}$ with an average energy density of $\SI{e4}{J/m^3}$, the gravitational energy flux at a distance of $r = \SI{10}{m}$ is \SI{e-30}{W/m^2}. The emitted gravitational waves would have a frequency of \SI{e10}{Hz} and a strain of $h \approx 10^{-43} \times (\SI{10}{m}/r)$.

In the following decade, gravitational waves radiated by circulating bunches of charged particles in high-energy accelerators were considered. In \cite{1987PhLB197302D} the radiated power was calculated to be \SI{5.5e-25}{W} for the LEP-2 collider at CERN. The frequency of the generated gravitational waves in this case is $f \approx \SI{e4}{Hz}$ and, assuming isotropic emission, the associated strain is $h \approx 10^{-43}$ a hundred kilometers away from the source.
Updates on gravitational radiation emitted by particles circulating in storage rings or by conversion of electromagnetic radiation into GWs can be found in \cite{Berlin:2021uov}.

Since the turn of the millennium, advances in high-power and high-energy lasers have provided appealing platforms to study gravitational aspects of light under laboratory conditions.  Laser-accelerated ions are potential sources of gravitational waves in the THz band \citep{Gelfer:2015fbj}. The generated gravitational strain depends on the total kinetic energy of the accelerated ions, with a maximal value $h \approx 10^{-43}$ at a distance $r = \SI{10}{m}$. Two counter-propagating laser beams are also expected to generate gravitational waves at twice the laser frequency $1/\lambda_{\rm las}$, with a strain given by \citep{GianlucaVacalis}:
\begin{align}
    h \approx \num{5.2e-38} \times
              \bigg( \frac{\lambda_\text{las}}{\SI{10}{\mu m}} \bigg)^2
              \bigg( \frac{\tau}{\SI{e-12}{sec}} \bigg)
              \bigg(\frac{I}{\SI{e23}{W\,cm^{-2}}} \bigg)
              \bigg( \frac{\SI{10}{cm}}{r} \bigg) \,.
    \label{eq:laserstrain}
\end{align}
Here, $\tau$ is the duration of the laser pulse and $I$ is the laser intensity.
Refinements with the use of twisted laser beams carrying orbital angular
momentum have been proposed in \citep{Atonga:2023psc}. Estimations of the produced gravitational strain are compatible with the above expressions. In addition, properties of the emitted gravitational waves, such as polarization, direction of emission, or beaming are all highly controllable by the experimental setup.

Beyond the generation of classical waves, controlled emission of gravitons has also been considered. 
Notably, the rate at which gravitons are spontaneously emitted by the quadrupolar transition $\text{3d (m=2)} \rightarrow \text{1s}$ in a hydrogen atom has been performed in \cite{Weinberg:1972kfs}. This was later improved upon in \cite{Boughn:2006st}, resulting in $\Gamma(3d \rightarrow 1s) \approx 10^{-40} ~\text{Hz}$.
This rate can be substantially increased by considering spontaneous or stimulated emission of gravitons in macroscopic quantum systems (see e.g.~\cite{Tobar:2023ksi}). However, the achievable event rates remain orders of magnitude too small for conceivable applications.



\begin{landscape}

\subsection{Summary of Sources}
\label{sec:SummaryTable}

\vspace*{-0.3 cm} 
\begin{center}
    \captionsetup{type=table} 
    \caption[Summary of late universe sources.]
    {Summary of late universe sources. We distinguish between coherent and stochastic sources by reporting the strain $h(f)$ or characteristic strain $h_{c,{\rm sto}}$, respectively. See \cref{sec:lateU} for details on these expressions and the assumptions made.}
    \label{tab:summary-coherent}

    \newlength{\usableWidth}
    \setlength{\usableWidth}{\dimexpr\linewidth-\headheight-\footskip\relax}
    \newcolumntype{P}[1]{>{\raggedright\arraybackslash}p{#1}}
    \newcolumntype{C}[1]{>{\centering\arraybackslash}p{#1}}
    \renewcommand{\arraystretch}{1.3}
    \begin{footnotesize}
    \begin{longtable}{P{.3\usableWidth}C{.25\usableWidth}C{.4\usableWidth}}
        \toprule
        \textbf{Source} & \textbf{Typical frequency} & \textbf{Amplitude} \\
        \midrule
        \endhead
        Neutron star mergers
            & $\lesssim (1-5) \, \text{kHz}$
            & $h(f) \lesssim \SI{e-26}{sec}$ \\
        \midrule

        Phase transitions in neutron star mergers 
            & $\displaystyle \simeq \SI{0.6}{MHz} \times
              \bigg( \frac{0.1}{v_w} \bigg)
              \bigg( \frac{\SI{1}{ms}}{\tau} \bigg)$
            & $\displaystyle h_{c,{\rm sto}} \simeq \num{1.5e-24} v_f^2
              \bigg(\frac{\SI{100}{Mpc}}{D} \bigg)$ \\
        \midrule
        Disk around supermassive BHs
            & $\displaystyle \simeq \SI{3.3e19}{Hz}$
            & $\displaystyle h_{c,{\rm sto}} \lesssim \num{3e-44}$ \\
        \midrule
        Sun
            & $\displaystyle \simeq \SI{e14}{Hz}$
            & $\displaystyle h_{c,{\rm sto}} \lesssim \num{3e-42}$ \\
        \midrule
        Primordial BH mergers
            & $\displaystyle \lesssim \frac{4400}{(m_1 + m_2)}{\rm Hz}$
            & $\displaystyle h(f) \approx \SI{e-31}{sec}
                \bigg( \frac{\rm kpc}{D} \bigg)
                \bigg( \frac{m_{\rm PBH}}{10^{-5} M_\odot} \bigg)^{5/6}
                \bigg( \frac{f}{\rm GHz} \bigg)^{-7/6}$ \\
        Primordial BH mergers: SGWB
            & $\displaystyle \lesssim \frac{4400} {(m_1 + m_2)} {\rm Hz}$
            & $\displaystyle h_{c,{\rm sto}} \approx 5 \times 10^{-31}
                \bigg(\frac{f_{\rm ISCO}}{\rm GHz} \bigg )^{-1.07}$ \\

        Primordial BH mergers: memory
            & $\displaystyle \lesssim \frac{4400}{(m_1 + m_2)}{\rm Hz}$
            & $\displaystyle h(f) \approx \SI{5e-25}{sec}
                \bigg( \frac{f_{\rm ISCO}}{f} \bigg)
                \bigg( \frac{m_{\rm PBH}}{10^{-5} M_\odot} \bigg)
                \bigg( \frac{\si{kpc}}{D} \bigg)$ \\
%
        Primordial BH hyperbolic encounters
            & $\displaystyle \simeq \SI{0.5}{GHz} \bigg (\frac{10^{-5} M_\odot}{m_{\rm PBH}} \bigg)
                \bigg( \frac{R_S}{r_p} \bigg)^{3/2}$
            & $\displaystyle h(f) \approx \SI{e-24}{sec}
                \bigg( \frac{f}{\si{GHz}} \bigg)^{2/3} 
                \bigg( \frac{m_{\rm PBH}}{10^{-5} M_\odot} \bigg)^{5/3} 
                \bigg( \frac{\si{Mpc}}{D} \bigg)$ \\
        \midrule
        Exotic compact objects
            & $\displaystyle \lesssim C^{3/2} \bigg(\frac{\SI{6e-3}{M_{\odot}}}{M}\bigg) \, \si{MHz}$
            & $\displaystyle h(f) \approx \SI{e-31}{sec}
                \bigg( \frac{f}{\si{GHz}} \bigg )^{-7/6}
                \bigg( \frac{m_{\rm PBH}}{10^{-5} M_\odot} \bigg)^{5/6}
                \bigg( \frac{\si{kpc}}{D} \bigg)$ \\
        \midrule
        Superradiance: annihilation
            & $\displaystyle \simeq \SI{5}{MHz}
                \bigg( \frac{\mu}{\SI{e-8}{eV}} \bigg)$
            & $\displaystyle h_S(f) \approx \SI{5e-30}{sec}
                \bigg( \frac{m_{\rm PBH}}{10^{-5} M_\odot} \bigg)
                \bigg( \frac{\si{kpc}}{D} \bigg)$ \newline
              $\displaystyle h_{V,T}(f) \approx \SI{e-26}{sec}
                \bigg( \frac{m_{\rm PBH}}{10^{-5} M_\odot} \bigg)
                \bigg( \frac{\si{kpc}}{D} \bigg)$ \\

        Superradiance: nonlinear effects
            & $\displaystyle \simeq \SI{5}{MHz}
                \bigg( \frac{\mu}{\SI{e-8}{eV} } \bigg)$
            & $\displaystyle h(f) \approx \SI{e-27}{sec}
                \bigg( \frac{{m_{\rm PBH}}}{10^{-5} M_\odot} \bigg)
                \bigg( \frac{\si{kpc}}{D} \bigg)$ \\
        \bottomrule    
    \end{longtable}
    \addtocounter{table}{-1}
    \end{footnotesize}
    \renewcommand{\arraystretch}{1.0}
\end{center}

\begin{center}
    \captionsetup{type=table} 
    \caption[Summary of stochastic sources.]
    {Summary of stochastic sources. For the conversion between energy density $\Omega_\text{GW}$ and characteristic strain, see \cref{eq:OmegaShRelation,eq:hcStoch}. The amplitudes reported are maximum values: for all the details on how to obtain these expressions, the dependence on the parameters of the models and the assumptions behind them, see the corresponding sections above.}
        \label{tab:summary-stochastic}

    \setlength{\usableWidth}{\dimexpr\linewidth-\headheight-\footskip\relax}
    \newcolumntype{P}[1]{>{\raggedright\arraybackslash}p{#1}}
    \newcolumntype{C}[1]{>{\centering\arraybackslash}p{#1}}
    \renewcommand{\arraystretch}{1.5}
    \begin{footnotesize}
    \begin{longtable}{P{.22\usableWidth}C{.2\usableWidth}C{.2\usableWidth}C{.2\usableWidth}}
        \toprule
        \textbf{Source} & \textbf{Frequency Range} & \textbf{Amplitude $\Omega_{\rm GW}(f)$} & \textbf{Characteristic Strain $h_{c,{\rm sto}}$} \\
        \midrule
        \endhead
        Inflation: vacuum amplitude 
            & flat in the range \mbox{$(10^{-16} - 10^{8})\,\si{Hz}$}
            & $\lesssim 10^{-16}$
            & $\lesssim 10^{-32} \,
                \Big( \frac{\si{MHz}}{f}\Big)$ \\
        Inflation: extra-species 
            & $(10^5 - 10^{8}) \, \si{\rm Hz}$
            & $\lesssim 10^{-10}$
            & $\lesssim 10^{-29} \, \Big(\frac{\si{MHz}}{f} \Big)$ \\
        Inflation: broken spatial reparametrization
            & Blue in the range \mbox{$(10^{-16} - 10^{8}) \, \si{Hz}$} 
            & $\lesssim 10^{-10}$ 
            & $\lesssim 10^{-29} \,
                \Big( \frac{\si{MHz}}{f} \Big)$ \\
        Inflation: secondary GW production
            & Flat or bump 
            & $\lesssim 10^{-8}$ 
            & $\lesssim 10^{-28} \,
                \Big( \frac{\si{MHz}}{f} \Big)$ \\
        Preheating
            & $(10^{6}-10^9)$ \, \si{Hz}
            & $\lesssim 10^{-10}$
            & $\lesssim 10^{-29} \,
                \Big( \frac{\si{MHz}}{f}\Big)$ \\
        Oscillons
            & $(10^{6}-10^9)$ \, \si{Hz}
            & $\lesssim 10^{-10}$
            & $\lesssim 10^{-29} \,
                \Big(\frac{\si{MHz}}{f} \Big)$ \\
        Cosmic gravitational microwave background
            & $f_{\rm peak} \sim (10-100)$ \, \si{GHz}
            & $\Omega_{\rm GW}(f_{\rm peak}) \lesssim 10^{-6}$
            & $h_c(f_{\rm peak}) \lesssim 10^{-31} \,
                \Big( \frac{\si{MHz}}{f} \Big)$ \\

        Phase transitions 
            & $\lesssim \SI{e9}{Hz}$
            & $\lesssim 10^{-8}$
            & $\lesssim 10^{-28} \,
                \Big(\frac{\text{MHz}}{f}\Big)$ \\
        Defects
            & Scale invariant
            & $\Omega_{\text{rad,0}} \, \frac{v^4}{M_{\rm Pl}^4} \, F_U$
            & $10^{-26} \frac{v^4}{M_{\rm Pl}^4} F_U 
                            \Big(\frac{\text{MHz}}{f}\Big)$ \\
        Gauge textures
            & $\sim 10^{11} \, \frac{v}{M_{\rm Pl}} \, \si{Hz}$
            & $\lesssim 10^{-4} \frac{v^4}{{M_{\rm Pl}}^4}$
            & $\lesssim 10^{-26} \frac{v^4}{M_{\rm Pl}^4}
                             \Big(\frac{\text{MHz}}{f}\Big)$ \\
        Grand unification \newline primordial BH evaporation
            & $(10^{18}-10^{15})$ \, \si{Hz}
            & $\lesssim 10^{-8}$
            & $\lesssim 10^{-28} \,
                \Big( \frac{\si{MHz}}{f} \Big)$ \\
        \bottomrule    
    \end{longtable}
    \addtocounter{table}{-1}
    \end{footnotesize}
    \renewcommand{\arraystretch}{1.0}
\end{center}

\end{landscape}


\section{Detection of Gravitational Waves at High Frequencies}
\label{sec:exp}

After the first detection of GWs at frequencies in the range (0.1--2.0)\,kHz \citep{PhysRevX.9.031040} and indications of a stochastic GW signal at pulsar timing arrays \citep{NANOGrav:2023gor, Antoniadis:2023ott, Reardon:2023gzh, Xu:2023wog}, expanding the frequency coverage of the worldwide gravitational wave program is a natural next step -- as it was for electromagnetic observations in the 1950s when radio, X-ray and UV astronomy became possible with new technology. As detailed in the previous section, many exciting questions in astrophysics, cosmology, and fundamental physics are tied to GW signals with frequencies well above the capabilities of current detectors or their upgrades. Even GW upper limits in regions of parameter space with no known Standard Model sources may be valuable in restricting current or future physical theories.

The detection of gravitational waves at LIGO, Virgo, and KAGRA requires measuring minuscule space-time deformations, smaller than the size of a proton (see \cref{sec:existing_detectors}). Achieving this has required the development of highly efficient mechanical-to-electromagnetic transducers. Similarly, a large class of high-frequency GW detector concepts relies on detecting mechanical deformations (see \cref{sec:ModernResonantMass}), with the main differences between different detector designs being the method used to engineer these transducers. A second large class of high-frequency GW detectors relies instead on the direct coupling between gravity and electromagnetism (see \cref{sec:EMoscillators,sec:PhotonRegen,sec:EMother,sec:AstroDetectors}). In electromagnetism in curved spacetime, the effect of a GW is to alter the vacuum’s dielectric properties, to generate effective currents that source induced electromagnetic fields, and to allow for GW--photon mixing. Relying on the coupling of GW to electromagnetism removes the need for a mechanical-to-electromagnetic transducer, though at the cost of working with a stiff system: Maxwell's equation in vacuum, governed by the speed of light, are more difficult to deform by a GW-induced force than typical materials of, e.g., resonant bars, which are characterized by the speed of sound $v_s/c \sim 10^{-5}$ rather than the speed of light.
While the technological challenges differ between observational methods, the core concepts are often closely linked to specific ranges of GW frequency, which explains the use of very different technologies across the frequency spectrum, also in the high frequency range. 
For a heuristic approach to estimate the sensitivity for a range of these concepts, see \citep{TitoDAgnolo:2024uku}.

In the frequency range from kHz to GHz, the GW frequency can be matched to the mechanical or electromagnetic resonant modes of a detector. One can thus profit from resonant enhancements, which can significantly boost the sensitivity, provided that the experiment's reaction time (ring-up time) is compatible with the duration of the GW source. For a meter-scale experiment, mechanical resonances lie in kHz regime, while EM resonances lie in the GHz regime.  Weber bars are the most well-known examples of mechanical resonant mass detectors, and modern versions with improved mechanical-to-electromagnetic transducers are being developed particularly for detecting high-frequency gravitational waves (\cref{sec:ModernResonantMass}). Regarding the electromagnetic coupling, GWs can induce an oscillating electromagnetic field within a microwave cavity placed in a static magnetic field, or the oscillating EM field can be read out through a resonant LC circuit (\cref{sec:EMoscillators}).

At frequencies much higher than GHz, it becomes increasingly challenging to design an apparatus that is small enough to match the GW wavelength for resonant enhancement. In this frequency range, photon regeneration experiments offer an alternative (\cref{sec:PhotonRegen}). Although these experiments were initially developed for axion searches, they can be optimized for detecting high-frequency gravitational waves. The detection range typically depends on the type of photon detector used (CCDs, X-ray detectors, etc.). As photon counting detectors these instruments are typically sensitive to the GW intensity (i.e., the square of the GW strain). A lower limit to the frequency range in which such detectors are sensitive often arises due to the detector vessel functioning as an electromagnetic waveguide with limited transmissivity at low frequencies, and due to challenges of implementing single photon detection at infrared frequencies. Instead of GW interactions with a laboratory setup, the magnetic fields of astrophysical or cosmological objects, such as neutron stars or large-scale galactic and cosmological structures, can also be leveraged for GW detection (see \cref{sec:AstroDetectors}). However, backgrounds are more difficult to control in these environments compared to laboratory-based experiments. Other proposals that leverage the direct coupling between gravity and electromagnetism are based on observing modifications in atomic quantum states (see \cref{sec:AlternativeConcepts}). These include detection methods based on the interaction of GWs and fermion spins, or on alterations in electron wave functions.

In this section, we will often use the short-hand notation $S_n = S_h^\text{noise}$ to denote the noise-equivalent strain sensitivity (or strain sensitivity for short) of detectors. We will do this in particular when quoting sensitivities from the literature throughout the text, whereas we will use the more explicit notation $S_h^\text{noise}$ introduced in \cref{sec:Notation} when there is a danger of confusing different power spectral densities in the discussion.


\subsection{Laser Interferometers and Resonant Mass Detectors and their Limitations}
\label{sec:existing_detectors}

The first GWs were detected by the Advanced LIGO \citep{PhysRevLett.116.131103} detectors in the US and the Advanced Virgo detector in Italy \citep{AdvVirgo}. In early 2020, the Japanese KAGRA detector \citep{PhysRevD.88.043007} joined LIGO's third observing run. These detectors are all Fabry--Perot interferometers, using large suspended mirrors several kilometers apart. Several other detectors of this type are in the design phase. These  detectors typically have their peak sensitivity at frequencies of a few hundred Hz.

However, some future detectors are designed specifically to expand the detection band towards either lower or higher frequencies.
To efficiently probe frequencies below 10\,Hz in terrestrial detectors, cryogenically cooled mirrors, large beam diameters, and operation underground are considered \citep{ETdesign, Adhikari_2020}. LISA, also based on laser interferometry, is a planned satellite-based detector to increase the arm length beyond the possibilities on Earth and to reduce environmental noise sources such as seismics \citep{Audley:2017drz}. LISA will have its peak sensitivity in the mHz range. To increase interferometer sensitivity towards higher frequencies, options are an increase of laser power and/or resonant operation. The planned Australian NEMO detector will be targeting frequencies of up to several kHz, see \cref{sec:ozgrav} below.

While increasing the arm length of an interferometer increases the strain signal in some frequency bands, longer arms are only really beneficial as long as the GW wavelength is longer than the interferometer arms. For significantly shorter wavelengths (frequencies $\gtrsim \si{MHz}$), interferometers with arm lengths of order meters are more suitable, but are of course at the same time limited by the smaller strain sensitivity achievable with shorter arms. This constitutes the main limitation of laser interferometers, used as direct strain meters, towards higher GW signal frequencies.

A concept to detect GWs which existed prior to interferometers are resonant bar detectors, initially proposed and built by Joseph Weber in the 1960s. Their modern successors, resonant spheres, have peak sensitivities at several kHz. In \cref{sec:spheres}, we will give a summary of these resonant spheres.

\subsubsection{Neutron Star Extreme Matter Observatory (NEMO)}
\label{sec:ozgrav}

The first detection of a binary neutron star merger in 2017 \citep{abbott2017gw170817} has increased the interest in the development of GW detectors with sensitivity in the few kHz regime, capable of detecting the merger and ringdown part of the waveform \citep{martynov2019exploring}. It is expected that such detectors will need to have strain sensitivities approaching $\sqrt{S_n} \simeq \SI{e-24}{Hz^{-1/2}}$ in the frequency range (1--4)\,kHz to observe several events per year. This sensitivity should be achieved by the third generation terrestrial GW detectors that are anticipated to come online in the later half of the 2030s \citep{Evans:2016mbw, Punturo:2010zz}. The Australian GW community is currently exploring the feasibility of a new detector, `NEMO', dedicated to detecting this merger phase and the following ringdown as well as testing third generation technology on a smaller scale \citep{ozhf, Bailes:2019oma, Adya2020}. The planned sensitivity of this detector would reach $\sqrt{S_n} \simeq \SI{e-24}{Hz^{-1/2}}$ in the range (1--2.5)\,kHz \citep{ozhf}. This detector will work in collaboration with the existing second generation GW detector network that will provide sky localization for electromagnetic follow-up.

The dominant high-frequency noise source for interferometric GW detectors is quantum phase noise, or shot noise as it is otherwise called. The magnitude of this noise source is inversely proportional to the square of the product of the circulating power incident on the test masses and the length of the arms of the detector. This generally necessitates extremely high powers in the arms of the interferometers ($\approx \SI{5}{MW}$ in the case of NEMO). Such high circulating power leads to technical issues such as parametric and tilt instabilities, as well as thermally induced distortions. These issues can be challenging to deal with, but a dedicated high-frequency detector promises to makes their mitigation easier. This is because sacrificing some sensitivity at low frequencies permits larger actuation on the test masses to correct instabilities and distortions. Further, relaxing the low-frequency sensitivity relaxes requirements on seismic isolation and test mass suspension systems, significantly reducing the cost of these systems.

\subsubsection{Interferometers up to 100\,MHz}
\label{sec:100MHzInterferometers}

As was first pointed out in \cite{Mizuno}, in laser interferometers the total stored energy in the form of circulating laser power sets a limit on the achievable sensitivity and bandwidth as a consequence of the quantum Cram\'{e}r--Rao bound. For a given laser power, large bandwidth and good strain sensitivity need to be balanced against each other, as increasing both at the same time is impossible. While opto-mechanical resonances can be introduced in the signal response of interferometers to shape the sensitivity curve for specific frequencies \citep{Somiya:2016pla, Korobko:2018pla}, it appears unlikely that the stored laser power can be further increased by several orders of magnitude. Therefore, broadband interferometric detectors reaching into the MHz range (while maintaining LIGO or Virgo-level strain sensitivity) seem not to be a viable option when taking also the arm-length argument from above into account.

Nevertheless there are three notable efforts (two existing and one under construction) of laser interferometers in the MHz range, which currently set the best experimental upper limits on GWs in their respective frequency bands.

One option is to build interferometers with a bandwidth of order kHz, but centered around much higher frequencies. See~\cite{akutsu} for upper limits from such a system operating at \SI{100}{MHz}. The detector uses a synchronous recycling architecture based on a resonant recycling cavity of dimension \SI{75}{cm} and a Nd:YAG laser with a power output of \SI{0.5}{W}. The limit on stochastic GW signals was reported to be $\sqrt{S_n} \sim \SI{e-16}{Hz^{-1/2}}$, setting a bound on the characteristic strain of $h_{c,\rm sto} \lesssim \num{7e-14}$. A study of the potential of this technique \citep{Nishizawa} showed that a sensitivity of \SI{e-20}{Hz^{-1/2}} is possible at \SI{100}{MHz} with a bandwidth of \SI{2}{kHz}, but the sensitivity decreases with increasing frequency and is not competitive above \SI{1}{GHz}.

The sensitivity of a single instrument can be surpassed by correlating two co-located instruments when searching for stochastic signals. An example is the \emph{Holometer} experiment at Fermilab, which consisted of two co-located power recycled Michelson interferometers with 40-meter long arms. While their primary research target has been signatures of quantization of spacetime, they are also excellent GW detectors, reaching a sensitivity of $\sqrt{S_n} \simeq\SI{e-21}{Hz^{-1/2}}$ in the band (1--13)\,MHz when cross-correlating both detectors \citep{PhysRevD.95.063002} over a \SI{103}{hr} dataset. See~\cite{PhysRevD.95.063002} for both a search for stochastic GW backgrounds and monochromatic GWs.
Using a \SI{704}{hr} dataset from, the authors of \citep{Martinez:2020cdh} concluded that there are no identifiable sources with harmonic frequency patterns (i.e.\ emitting in integer multiples of a fundamental frequency) such as cosmic string loops and eccentric BH binaries emitting in the frequency range (1--25)\,MHz.

Following a similar detection concept is the Quantum-Enhanced Space-Time (QUEST) experiment at Cardiff University. It consists of two wide-band table-top interferometers sensitive in the (1--100)\,MHz band \citep{vermeulen2020experiment}. Cross-correlating these detectors in a coincident observing run of $10^4$~s, upper limits of about $\sqrt{S_n} \simeq \SI{3e-20}{Hz^{-1/2}}$ on a stochastic GW background between 13 and 80~MHz have been achieved~\citep{Patra:2024thm}. The team plans to increase the bandwidth to \SI{200}{MHz} and to increase the sensitivity by another two orders of magnitude.

\subsubsection{Spherical Resonant Masses}
\label{sec:spheres}

The principle of a \emph{resonant mass detector} is that its vibrational eigenmodes can get excited by a GW. These mechanical oscillations are transformed into electromagnetic signals, using electromechanical transducers, and amplified by electrical amplifiers. These resonant detectors have a relatively small bandwidth, usually of less than \SI{100}{Hz}. Thermal noise, Johnson--Nyquist noise, pump phase noise (if the transducer is parametric), back-action noise, and amplifier noise are the internal noise sources in this kind of detector. The resonant mass antenna and transducers must be made of high-quality factor materials in order to decrease thermal (mechanical) and Johnson--Nyquist (electrical) noise.

The idea of a spherical resonant mass antenna for GW detection has a long history and was first proposed in \cite{Forward1971}, followed by several decades of exploration and proposals \citep{wagoner1977multimode, hamilton1990resonant, PhysRevLett.70.2367}. More recently, the Mario Schenberg detector~\citep{Aguiar:2010kn, Da_Silva_Costa_2014} in S\~ao Paulo, Brazil, and Mini-GRAIL~\citep{Gottardi:2007zn}, in Leiden, Netherlands have developed the concept further. At present, both detectors have been decommissioned, but Schenberg is planned to be reassembled at INPE, in S\~ao Jos\'e dos Campos, about \SI{100}{km} from its initial site at the University of S\~ao Paulo.\footnote{These detectors had much smaller masses (1.15 and \SI{1.3}{tonnes}, respectively) and diameters (65 and \SI{68}{cm}, respectively) than originally proposed in the 1990s (up to \SI{120}{tonnes}, \SI{3}{m}, resonant around $\sim \SI{700}{Hz}$).} Such detectors have a bandwidth of 50--\SI{100}{Hz} with peak frequencies around \SI{3}{kHz} for the quadrupole modes. To increase the frequency range, a xylophone configuration of several spheres has been proposed \citep{PhysRevD.54.2409}.

In 2004, Mini-GRAIL operating at a temperature of \SI{5}{K} reached a peak strain sensitivity of $\sqrt{S_n} \simeq \SI{1.5e-20}{Hz^{-1/2}}$ at a frequency of \SI{2942.9}{Hz}. Over a bandwidth of \SI{30}{Hz}, the strain sensitivity was about $\sqrt{S_n} \simeq \SI{5e-20}{Hz^{-1/2}}$ \citep{Gottardi:2007zn}. Schenberg, operating also at \SI{5}{K}, reached strain sensitivities of $\sqrt{S_n} \simeq \SI{1.1e-19}{Hz^{-1/2}}$ for its quadrupolar modes ($\sim \SI{3.2}{kHz}$) and $\sqrt{S_n} \simeq \SI{1.2e-20}{Hz^{-1/2}}$ for its monopolar mode ($\sim \SI{6.5}{kHz}$) in 2015~\citep{Oliveira:2016gds}. Both Schenberg and Mini-GRAIL could reach sensitivities around $\sqrt{S_n} \simeq \SI{e-22}{Hz^{-1/2}}$ when operating at \SI{15}{mK}. Schenberg, because it uses parametric transducers, can reach higher sensitivities if it implements squeezing of the signal.
On a similar time scale the resonant bar detector AURIGA near Padua, Italy, reported reaching strain sensitivities of $\sqrt{S_n} \simeq \SI{e-20}{Hz^{-1/2}}$ at frequencies around \SI{900}{Hz} over a bandwidth of \SI{100}{Hz}~\citep{Vinante:2006uk}.

Spherical antennas provide more information compared to the classical bar antennas because of their quadrupole modes, while also being significantly more sensitive due to their favorable geometry (they offer a larger cross-section at identical mass). From the output of six transducers tuned to the quadrupole modes of a sphere, one can obtain complete information about the polarization and direction of the incoming wave.

The conceptual difficulties in pushing this technology to higher frequencies are similar to the issues faced by laser interferometers: it requires smaller resonating spheres and consequently measuring smaller absolute displacements to achieve the same strain sensitivity. Contrary to laser interferometers, resonant mass detectors have not yet reached the standard quantum limit. It thus seems unlikely that this technology can be pushed significantly beyond the kHz region.

An additional challenge for resonant detectors in general is their small bandwidth, $\Delta f_\text{det} \sim f/Q$, where the quality factor $Q \gg 1$ plays a key role in enhancing the sensitivity on resonance. For transient high-frequency GW signals with $\dot f \sim f^2$ (such as PBH binaries shortly before the merger, see \cref{sec:PBHmergers}) this implies that the signal spends only a very short amount of time, of order $(f Q)^{-1}$, inside the sensitivity band. This time window is often too short to fully ring up the resonance, in which case the high quality factor is not fully brought to bear. This needs to be taken into account when computing the sensitivity of resonant detectors to transient signals.

\begin{figure}
    \centering
    \includegraphics[width=\textwidth]{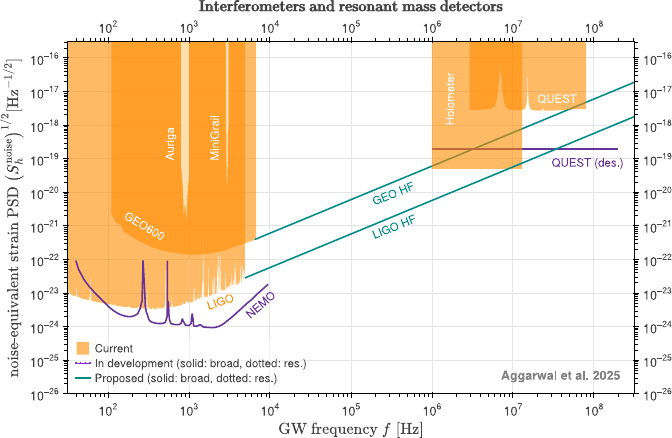}
    \caption[Strain sensitivity of interferometers and resonant mass detectors.]
    {Strain sensitivity of (individual) interferometers (LIGO O4a~\citep{LIGO:2024kkz}, GEO600~\citep{Lough:2020xft}, Holometer~\citep{PhysRevD.95.063002}, QUEST~\citep{Patra:2024thm}) and resonant mass detectors (AURIGA~\citep{Vinante:2006uk}, Mini-GRAIL~\citep{Gottardi:2007zn}), together with the projected sensitivity of the interferometer NEMO~\citep{ozhf} and an extrapolation of the LIGO and GEO sensitivity to higher frequencies (see text). Color coding as in \cref{fig:sens_HF}.}
    \label{fig:41_summary}
\end{figure}
\subsubsection{Summary: Sensitivities of Interferometers and Resonant Mass Detectors}
%
\Cref{fig:41_summary} provides an overview of the typical strain sensitivities achieved by the interferometers and resonant mass detectors described in this section. As in \cref{fig:sens_HF}, instruments which have placed limits on (or detected) GWs are shown in orange, concepts under active R\&D are shown in purple, and other proposals are shown in cyan. Note that  LIGO, the Holometer and QUEST consist of two separate detectors each and can thus increase their sensitivity by cross-correlating the data streams. \Cref{fig:41_summary} shows the strain sensitivity of a single interferometer.
Moreover, we use the cyan color to show a naive extrapolation of the LIGO and GEO sensitivities to higher frequencies (see also \cite{Schnabel:2024hem}). Extending the sensitivity band of these detectors in practice requires overcoming challenges in data acquisition, noise control, and calibration, which requires dedicated R$\&$D. Most likely the sensitivities shown here cannot be reached while simultaneously optimizing the sensitivity around \SI{100}{Hz}.


\subsection{Modern Resonant Mass Detectors}
\label{sec:ModernResonantMass}

Since Joseph Weber’s pioneering developments in the late 1960s, the ability to detect and measure geometrical changes in various systems has progressed significantly. This progress is in particular due to techniques that go beyond the traditional use of large bars or spheres, which relied on monitoring \textit{massive} resonant systems with high quality factors, $Q$. (Modern versions of such systems are discussed in \cref{sec:spheres}.) In particular, using smaller resonators allows for cutting-edge sensing methods, four of which we will explore in this section: optically levitated sensors, bulk acoustic-wave devices, microwave cavities and magnetic Weber bars.

\subsubsection{Optically Levitated Sensors}
\label{sec:OpticallyLevitatedSensors}

Optically levitated dielectric sensors have been identified as a promising technique for resonant GW searches over a wide range of frequencies from a few kHz to $\sim \SI{300}{kHz}$ \citep{arvanitaki:2016gw, aggarwal2020searching}. A dielectric nano-particle suspended at the anti-node of a standing laser wave within an optical cavity will experience a force when a passing GW causes a time-varying strain on the physical length of the cavity. The particle will be displaced from the location of the trapping light anti-node, resulting in periodic kicks on the particle at the frequency of the GW space-time disturbance. The trapping frequency and mechanical resonance linewidth are widely tunable based on the laser intensity and laser cooling parameters chosen. 

When detecting the resulting displacement of the particle at the trapping resonance frequency, the sensitivity is limited by Brownian thermal noise in the particle itself rather than the displacement detection of the particle. This results in improved sensitivity at higher frequency (unlike traditional interferometer style detectors which experience decreased sensitivity at high frequency due to laser shot noise) \citep{arvanitaki:2016gw}.  The low-friction environment made possible by optical levitation in ultra-high vacuum enables extremely sensitive force detection \citep{Ranjit:2016zn}, which becomes ultimately quantum-limited by photon-recoil heating from discrete scattering events of individual trap laser photons \citep{Jain:2016re}.

A 1-meter prototype Michelson-interferometer configuration detector called the `Levitated Sensor Detector' (LSD) is under construction at Northwestern University in the US, with a target sensitivity of better than $\sqrt{S_n}  \sim \SI{e-19}{Hz^{-1/2}}$ at $f \approx \SI{10}{kHz}$ and $\sqrt{S_n} \sim \SI{e-21}{Hz^{-1/2}}$ at $f \approx \SI{100}{kHz}$ \citep{aggarwal2020searching, LSD:2022mpz}. In addition, fiber-based approaches are being investigated to permit longer cavities without the need for expensive optics \citep{Pontin:2018fi}. The ultimate strain sensitivity of a 10-meter room-temperature instrument is estimated to be better than approximately $\sqrt{S_n}  \sim \SI{e-20}{Hz^{-1/2}}$ at $f \approx \SI{10}{kHz}$ and $\sqrt{S_n} \sim \SI{e-22}{Hz^{-1/2}}$ at $f \approx \SI{100}{kHz}$. For a cryogenic 100-meter apparatus, this can be improved by more than an order of magnitude across much of the frequency range \citep{aggarwal2020searching}. A detailed analysis of the search reach for GWs produced by axions via the BH superradiance process is provided in \cite{aggarwal2020searching}.

Another interesting direction is the use of optically-trapped levitated membranes \citep{chang2012ultrahigh}. This idea is based on the use of nano-mechanical resonators which constitute chip-scale implementations of a harmonic oscillator of thin films with high tensile stress, achieving extremely high $Q$-factors ($ > 10^{10}$) \citep{Beccari:2021mwj}. They have a wide range of applications in sensing and cavity optomechanics \citep{RevModPhys.86.1391}. Work towards the design of a corresponding prototype detector is in progress at DESY, together with related R\&D studies \citep{Reinhardt:2023cds}. The realization of this detector, comprising membranes with $Q >  10^{12}$, is a longer term goal.

Among other possible ideas for optomechanical systems to detect GWs, it has also been suggested to use of a volume of superfluid ${}^4$He that responds to mechanical forces. This effect may be read-out by a membrane monitored using interferometric methods. A  sensitivity of $\sqrt{S_n} < \SI{8e-19}{Hz^{-1/2}}$ has been claimed around $f \approx \SI{100}{kHz}$ \citep{Vadakkumbatt:2021fnw}.

The field of optically levitated sensors is rapidly  developing, see, e.g., \cite{Millen:2019bcw, Gonzalez-Ballestero:2021gnu, Winstone:2023whl}. In this regard, it is plausible to assume that the efforts of \citep{arvanitaki:2016gw, aggarwal2020searching, LSD:2022mpz} are only the first steps towards detectors with much better sensitivity to HFGWs in the near future.\footnote{Sensitivities to forces down to yocto-Newtons have been theoretically claimed \citep{Liang:2023plh}. These studies are far from being realistic, but they show that there is ample space for progress.} 

A related approach based on detecting the motion of superconducting spheres levitated in a magnetic field has been proposed in \citep{Carney:2024zzk}, and is discussed in \cref{sec:MWB}.

\subsubsection{Bulk Acoustic Wave Devices}
\label{sec:BAW}

Bulk acoustic wave (BAW) devices are one of the pillars of frequency control and frequency metrology \citep{ScRep}. In the simplest case, a piece of piezoelectric material is sandwiched between two electrodes, converting acoustic waves inside the material into electrical signals. With its relatively compact size and robustness, this technology gives one of the best levels of frequency stability near one second of integration time. More recently, it was demonstrated that quartz bulk acoustic wave devices exhibit extremely high-quality factors (up to $8 \times 10^9$) at cryogenic temperatures for various overtones of the longitudinal mode covering the frequency range (5--700)\,MHz \citep{ScRep,quartzPRL}. For this reason, it was proposed to use the technology for various tests of fundamental physics \citep{ScRep} such as Lorentz invariance tests \citep{PhysRevX.6.011018}, quantum gravity research \citep{PhysRevD.100.066020} and searches for high-frequency GWs \citep{Goryachev:2014yra}. For the latter purpose, a bulk acoustic wave device represents a resonant mass detector whose vibration could be read out through the piezoelectric effect and Superconducting Quantum Interference Devices (SQUIDs). The approach has the following advantages: (i) highest quality factor (high-sensitivity); (ii) internal (piezoelectric) coupling to SQUIDs \citep{Goryachev:2014ab}; (iii) allows parametric detection methods; (iv) a large number of sensitive modes ($> 100$) in a single device; (v) modes scattered over a wide frequency range (1--700)\,MHz; (vi) well-established and relatively inexpensive technology (mass production); (vii) high-precision (insensitive to external influences such as seismic vibration and temperature fluctuations), and (viii) the possibility of building arrays of detectors to extend the frequency range towards lower frequencies and/or to achieve better sensitivity. On the other hand, in practice, identically manufactured devices exhibit significant dispersion in mode frequencies at low temperatures, thus limiting the accuracy of such arrays. The level of sensitivity of bulk acoustic wave detectors is estimated to be at the level of $\sqrt{S_n} \simeq \SI{2e-22}{Hz^{-1/2}}$, depending on the mode geometry \citep{Goryachev:2014yra}. With additional investment into research and development, this sensitivity could be improved and the frequency range extended down to hundreds of kHz.

A search for high-frequency GWs with single bulk acoustic wave devices and two modes, operated at $\sim \SI{4}{K}$, has been running at the University of Western Australia since November 2018. Recently, two interesting events were observed in these searches, at different frequencies around few MHz \citep{Goryachev:2021zzn}. The origin of these events cannot be determined with current data, but given their strength they are not considered to be viable GW candidates (see also  \citep{Lasky:2021naa}). These results have triggered significant interest in further advancing this detection technique. In this context, the possibility of building arrays of BAWs and multimode read-out is being pursued by the Bulk Acoustic Wave Sensors for a High-frequency Antenna (BAUSCIA) program in Milano and by the Multimode Acoustic Gravitational Wave Experiment (MAGE) at the University of Western Australia \citep{Campbell:2023qbf}. The goal is to build networks of $O(10)$ BAWs, accessing $O(100)$ frequency modes.

Further improvements could come from reaching the quantum ground state of the system \citep{Campbell:2022jnq}, or, in general, from counting phonons, and performing quantum state tomography or quantum manipulation and characterization of the states of a BAW resonator \citep{Chu:2018pwc, vonLupke:2021oyu, Bild:2022ues}. Recent theoretical characterization aiming at optimizing the searches of HFGWs with phonons can be found in \cite{Kahn:2023mrj}. Finally, a multi-mode resonant bar concept has been proposed in \cite{Tobar:2024bjr} to absorb GWs with a large mass object, while reading it out with a much lighter one. With the individual components studied in earlier works, see  e.g.~\cite{Tobar:1995,Tobar:2000sy}, a key next step would be the construction of a prototype to understand and verify in more detail their interplay and the achievable sensitivity of this proposal.

\subsubsection{Deformation of Microwave Cavities}
\label{sec:HighQCavities}

An electromagnetic resonator prepared such that it has two nearly degenerate modes, $\omega_1$ and $\omega_2$, may act as a sensitive device to detect GWs. The idea is to inject power into only one of the modes, while an incident GW can resonantly transfer power from this loaded mode 1 into the otherwise quiet mode 2 if the condition $|\omega_2 - \omega_1| - \omega_G \lesssim \Delta \omega_2$ is met, where $\Delta \omega_2$ is the width of mode 2, which is typically wider than the width of mode 1. This process of combining signals with two frequencies is often called \textit{``heterodyning"}, hence the name given to the general approach of \textit{heterodyne detection}. Two mechanisms exist whereby the GW can transfer power from the loaded mode into the quiet mode: directly through the interaction with the electromagnetic energy stored in the cavity in mode 1, or indirectly by deforming the cavity walls in such a way that mode 1 is coupled to mode 2. The latter effect dominates in most of the frequency range of interest (sub-GHz) due to the small speed of sound in materials.

The first studies considering the mechanical coupling of GWs to electromagnetic resonators appeared in the late 70s.  \cite{Pegoraro, Pegoraro:1977uv} proposed and studied a system with a sharp resonance at about \SI{1}{GHz}, while \cite{Caves} contains a theoretical study of a microwave cavity with a high mechanical quality factor. The first experimental efforts based on these ideas were reported  in \cite{Reece:1982sc, Reece}. These schemes offered sensitivity to a range of frequencies from few kHz to GHz, limited by different noise sources, particularly thermal noise at low frequencies. The idea was further developed and eventually started to take shape in the Microwave Apparatus for Gravitational waves Observation (MAGO) \citep{Ballantini:2005am}, which we now describe.

MAGO consists of two microwave resonators (spheres in this case, to maximize the sensitivity), coupled through an \textit{a priori} tunable link. This allows for a control of the frequency split of the ground states of the coupled system and achieve resonance modes with characteristic frequency $\sim \si{GHz}$, but with energy differences as low as $\mathcal{O}(10)$\,kHz. As a result, the device can in principle detect GWs from \SI{10}{kHz} up to MHz and beyond. This detection concept led to the MAGO proposal for a scaled-up experiment with \SI{500}{MHz} cavities as a CERN--INFN collaboration. Although the final project was not funded, three SRF cavities were built during the R\&D activities. The first one (a pill-box cavity) was used as a proof-of-principle experiment, which demonstrated the working principle and the development of an RF system to drive and read out the cavity with the necessary precision~\citep{Ballantini:2005am}. The third cavity was a spherical 2-cell cavity with an optimized geometry, which was never treated and tested (it was placed on display at the University of Genoa after the R\&D efforts stopped). 

The idea was revived in recent work~\citep{Berlin:2023grv} with an improved theoretical treatment and estimate of the various noise sources, as well as the resulting sensitivity of a cavity similar to the third MAGO prototype. The authors found that the noise-equivalent strain PSD could reach $\sqrt{S_n} \sim \SI{e-21}{Hz^{-1/2}}$ in the frequency range $\SI{100}{kHz} \lesssim 2 \pi f \lesssim \si{GHz}$. Furthermore, the authors pointed out that by overcoupling to the signal mode of the cavity, the experiment can be run in broadband mode. (Overcoupling refers to a situation where the energy leaking through the coupling mechanism exceeds the intrinsic loss within the cavity. This effectively reduces the quality factor of the cavity, but increases the bandwidth.) In broadband mode, sensitivities better than $\sqrt{S_n} \lesssim \SI{e-18}{Hz^{-1/2}}$ across two decades in frequency centered around \SI{100}{kHz} could be achieved in a single measurement.

These results lead to a renewed interest in the heterodyne detection with microwave cavities. Currently, with the third MAGO cavity that was on display in Genoa, DESY, the University of Hamburg, and Fermilab continue collaborative R\&D studies \citep{Fischer:2024nte}. The first goal is to obtain a measurement with the MAGO prototype cavity in an existing cryostat at Fermilab, which would lead to a first (albeit weak) bound in the 10--\SI{100}{kHz} range. Long-term goals include developing an improved cavity design, engineering a dedicated low-noise cryostat and suspension system to significantly improve the sensitivity, and ultimately establish coordinated HFGW observatories at DESY and Fermilab.

Further improvements may be possible with larger cavity masses and volumes, as well as with better read-out strategies. Also, the cost of MAGO-like cavities is low enough that operating networks of detectors in different geographic locations may be realistic strategy for enhancing the sensitivity. These efforts are notably pursued within the GravNet collaboration \citep{GravNet} including partners from INFN Frascati (Italy), IFAE/ICREA Barcelona (Spain), as well as the Universities of Bonn and Mainz (Germany).

\subsubsection{Magnetic Weber Bars}
\label{sec:MWB}

A key challenge in resonant mass detectors is the efficient readout of the energy stored in mechanical deformation. In view of this, \cite{Domcke:2024mfu} proposed a superconducting magnet, operated in persistent mode, as a resonant mass detector. A passing gravitational wave leads to a deformation of the current-carrying superconducting coils, modifying the magnetic field. A pickup loop placed close the end caps of a solenoidal magnet and coupled to a SQUID can detect this small, oscillating change in the background magnetic field. The advantage of this setup is that the induced magnetic fields, which are of ${\cal O}(h B_0)$, can profit from the large amount of energy stored in the background magnetic field $B_0$ without any significant transducer loss. In particular, the MRI magnet that is being deployed for the ADMX-EFR experiment would allow for an estimated broadband GW strain sensitivity of $\sqrt{S_n} \sim \SI{e-20}{Hz^{-1/2}}$ for frequencies from a few kHz to about \SI{10}{MHz}, with a peak sensitivity down to $\sqrt{S_n} \sim \SI{e-22}{Hz^{-1/2}}$ at a kHz, exploiting a mechanical resonance.

A related approach was proposed in \cite{Carney:2024zzk}, considering a levitated superconducting sphere in a magnetic field. The superconducting sphere expels the magnetic field, thus leading to a rather inhomogeneous field configuration in its vicinity. A passing gravitational wave results in an oscillation of the superconducting sphere with respect to a pickup loop placed in its vicinity, and consequently to an oscillation of the magnetic flux measured by the pickup loop. Such a system could achieve broadband strain noise sensitivity of $\sqrt{S_n} \sim \SI{e-19}{Hz^{-1/2}}$ for frequencies from \SI{10}{kHz} to \SI{1}{MHz}.

\subsubsection{Summary: Strain Sensitivities of Modern Resonant Mass Detectors}
%
\begin{figure}
    \includegraphics[width = \textwidth]{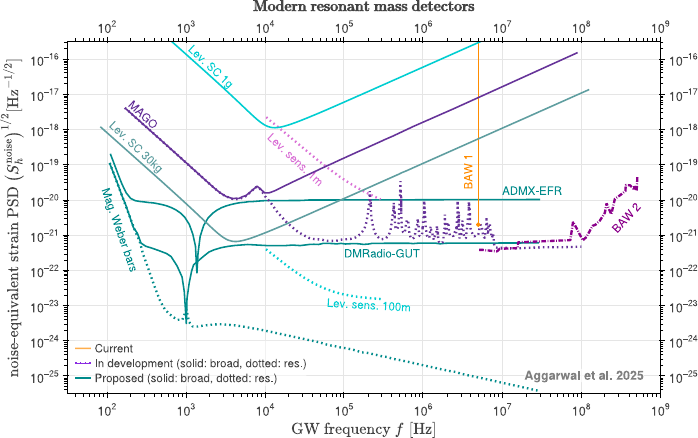}
    \caption[Projected strain sensitivity of modern resonant mass detectors.]
    {(Projected) strain sensitivity of modern resonant mass detectors: levitated sensors (Lev.~sens.)~\citep{aggarwal2020searching}, bulk acoustic wave devices (BAW)~\citep{Goryachev:2014yra}, MAGO~2.0~\citep{Berlin:2023grv}, magnetic Weber Bars~\citep{Domcke:2024mfu} and levitated superconducting spheres (Lev. SC)~\citep{Carney:2024zzk}. Color coding as in \cref{fig:sens_HF}.}
   \label{fig:42_summary}
\end{figure}
\Cref{fig:42_summary} provides an overview of the projected strain sensitivities of a range of modern resonant mass detectors. As in \cref{fig:sens_HF}, instruments which have placed limits on (or detected) GWs are shown in orange, concepts under active R\&D are shown in purple, and other proposals are shown in cyan. Solid lines indicate broadband sensitivities whereas dashed lines indicate a resonant search requiring a scanning strategy. Where available, we show both the projected sensitivity from the initial stage of an experiment as well as possible upgrades. In particular, for levitated sensors we show the \SI{1}{m} disc prototype and a future \SI{100}{m} stack setup~\citep{aggarwal2020searching}, for bulk acoustic wave devices we show estimates for a multimodal cavity cooled to \SI{20}{mK}~\citep{Goryachev:2014yra}, for microwave cavities we show the projected thermal noise limited broadband and resonant sensitivities~\citep{Berlin:2023grv}, for magnetic Weber bars we show the estimated sensitivities using an MRI magnet as well as the larger magnet envisioned for the DMRadioGUT axion experiment, assuming moreover a resonant readout strategy for the latter~\citep{Domcke:2024mfu}, and for levitated superconducting spheres we show the estimated sensitivity for \SI{1}{g} and \SI{30}{kg} spheres \citep{Carney:2024zzk}.


\subsection{Electromagnetic Oscillators}
\label{sec:EMoscillators}

Combining Einstein's theory of general relativity with classical electromagnetism reveals a coupling between gravitational and electromagnetic (EM) waves. This coupling allows for a range of applications for electromagnetic GW detectors, including the conversion of GWs to photons and vice versa~\citep{Gertsenshtein, Boccaletti1970, Fuzfa:2015oaa, Fuzfa:2017ana}.

To see this, let us consider the action of electromagnetism in curved spacetime with the metric $g_{\mu\nu}$~\citep{Landau:1975pou},
\begin{align}
    S = \int \! d^4x \, \sqrt{-g} \, \big(-\tfrac{1}{4} g^{\mu\alpha} g^{\nu\beta}
                                           F_{\mu\nu} F_{\alpha\beta} \big) \,,
\end{align}
with $g \equiv \det g_{\mu\nu}$, and with $F_{\mu\nu} \equiv \partial_\mu A_\nu - \partial_\nu A_\mu$ the electromagnetic field strength tensor. Expressing the metric as $g_{\mu\nu} = \eta_{\mu\nu} + h_{\mu\nu}$ with $\eta_{\mu\nu}$ denoting a flat Minkowski background and $h_{\mu\nu}$ a GW with $\vert h_{\mu\nu} \vert \ll 1$, we obtain 
\begin{align}
    S = \int \! d^4x \, \big(-\tfrac{1}{4} F_{\mu\nu} F^{\mu\nu}
                           + j_{\rm eff}^\mu A_\mu  \big) \,,
\end{align}
where
\begin{align}
    j_{\rm eff}^\mu \equiv \partial_\nu \big(
        - \tfrac{1}{2} h F^{\mu\nu}_0
        + {h^{\nu}}_\alpha F^{\mu\alpha}_0
        - {h^{\mu}}_{\alpha} F^{\nu\alpha}_0 \big) + \mathcal{O}(h^{2}) \,,
    \label{eq:effective_current}
\end{align}
with $h = {h^{\mu}}_{\mu}$, and with $F^{\mu\nu}_0$ denoting the flat space background EM field. Therefore, in the presence of a background magnetic field ($F_0^{\mu\nu}$) a GW generates an effective current oscillating with the GW frequency, which sources induced electromagnetic signals. (See \cref{sec:EMsignals} for other theoretical approaches to GW--EM couplings.)

Expressing the impact of a GW as an effective current highlights possible synergies with axion searches, given that an axion background field $a$ also leads to an effective current. In the axion case, the current is of the form $j_{\rm eff}^a \sim (\partial_\nu a) \Tilde{F}^{\mu\nu}$, with $\Tilde{F}_{\mu\nu} \equiv \tfrac{1}{2} \epsilon^{\mu\nu\rho\sigma} F_{\rho\sigma}$. This has motivated a range of proposals relying on existing or planned axion experiments \citep{Ejlli:2019bqj, Berlin:2021txa, Domcke:2020yzq,Tobar:2022pie}. These experiments typically feature a strong magnetic field and then search for EM signals induced by an axion or axion-like particle. Identical or similar experimental arrangements can also be used to search for GWs, as detailed below.

Many factors enter when estimating the sensitivity of a given experimental setup. The GW couples not only to the electromagnetic fields but also to the mechanical support structure. In the limiting cases of a GW frequency far above or below the mechanical resonance frequencies, this can be treated fairly easily in the free-falling or rigid limit, respectively, whereas the intermediate regime requires a more careful treatment~\citep{Ratzinger:2024spd}. When estimating the signal strength, it is moreover important to account not only for the effective current in the bulk of the magnetic volume but also for effective surface currents on its boundary~\citep{Domcke:2023bat}.

The data analysis of axion searches is optimized for persistent coherent signals, so dedicated searches are necessary to search, e.g., for PBH mergers or stochastic signals, which lead to signals of low coherence and/or very short duration. The sensitivity of techniques relying on relatively long integration times and or high signal coherence (such as the ring-up of cavities) need to be carefully re-evaluated in this regime. Moreover, in most of the proposals outlined below, the coupling factor between the GW and the instrument has been calculated analytically relying on some simplifying assumptions. In a realistic setup, numerical simulations and calibration measurements will be required to determine the relevant order one corrections accurately. Below, all this has been taken into account to the best of our knowledge, unless specified otherwise.

\subsubsection{Microwave Cavities}
\label{sec:SRFcavities}

There are many axion experiments utilizing microwave cavities in strong magnetic fields, such as ADMX \cite{ADMX:2021nhd}, CAPP \cite{CAPP:2020utb}, HAYSTAC \cite{HAYSTAC:2018rwy}, or ORGAN \cite{Quiskamp:2022pks,Quiskamp:2023ehr}. These experiments are designed to detect coherently oscillating axion signals with wavelengths comparable to the detector size of order cm--meters. Thanks to high quality factors $Q \sim 10^{4-5}$, the induced electromagnetic field is resonantly enhanced within the cavity. In a similar fashion, the coupling of coherent GWs to an electromagnetic resonance mode results in an induced EM field which depends on the incoming direction and polarization of the GW. A comparison between the power spectral density expected for such signals and the noise of the instrument provides an estimate of the achievable GW strain sensitivity. One finds $h_0 \sim 10^{-22}$--$10^{-21}$ at $\mathcal{O}(\si{GHz})$ frequencies \citep{Berlin:2021txa}.

In the following we estimate the noise-equivalent strain sensitivity of microwave cavities. The EM field induced by a GW is
\begin{align}
    \tilde E_h(f) \simeq \eta \, Q \, (2 \pi f L) \, B_0 \tilde{h}(f) \,,
\end{align}
where $\eta \simeq 0.1$ indicates the coupling coefficient between the GW and the EM mode, $Q$ is the cavity's quality factor, $L$ its length, and $B_0$ the magnetic field. From this we can estimate the power $P$ delivered to the cavity on resonance as
\begin{align}
    P_\text{sig} &\simeq \frac{1}{2} Q L^5 (2 \pi f)^3 B_0^2
                         \eta^2 S_h(f) \Delta f \,,
\end{align}
with $\Delta f$ being the width of the cavity resonance. We read off corresponding the power spectral density as
\begin{align}
    S_{P,\text{sig}}(f) \simeq \frac{Q}{4} L^5 (2 \pi f)^3 B_{0}^2 \eta^2  S_h(f) \,.
\end{align}
Contrasting this with the power injected by thermal Johnson--Nyquist noise
\begin{align}
    P_{\rm noise} \simeq k_B T_{\rm sys} \Delta f
    \quad \rightarrow \quad
    S_{P,\text{noise}}(f) \simeq k_{B} T_\text{sys} / 2
    \label{eq:thnoise}
\end{align}
yields the noise-equivalent strain sensitivity (see \cref{eq:shnoise})
\begin{align}
    S_h^\text{noise}(f) \simeq \frac{2 k_B T_\text{sys}}{(2\pi f)^3 \eta^2 Q B_0^2 L^5} \,.
\end{align}
\Cref{tab:benchmarks_cavities} lists experimental parameters for a range of experiments. These also serve as reference values for the sensitivities shown in \cref{fig:43_summary} below. These proposals are all based on resonant readout.

\begin{table}[t]
    \centering
    \renewcommand{\arraystretch}{1.1}
    \begin{tabular}{lccccc}
        \toprule
                                 & $f$ [GHz]           & $Q$         & $B_{0}$ [T] & $L$ [m] & $T_{\rm sys}$ [K] \\
        \midrule
        ADMX                     & (0.65, 1.02)        & \num{8e4}   & 7.5         & 0.51    & 0.6 \\
        HAYSTAC                  &  (5.6, 5.8)         & \num{3e4}   & 9           & 0.13    &  0.13 \\
        CAPP                     &  (1.6, 1.65)        & \num{4e4}   & 7.3         &  0.15   & 1.2 \\
        ORGAN                   &  (15,16),~(26,27)
                                                       & \num{e4}
                                                       &  11.5           & 0.023    & 5.3 \\
        \textit{SQMS}            & (1, 2)              & \num{e6}    & 5           & 0.46    &  1 \\
        \textit{Cubic cavity 1}  & 0.1                
                                                       & \num{6.27e5}&  0.6        & 2.1     &  8 \\
        \textit{Cubic cavity 2}  & 1                  
                                                       & \num{1.98e5}& 12          & 0.21    &  1 \\
        \textit{Cubic cavity 3}  & 10                 
                                                       & \num{6.25e4}& 12          & 0.021   &  1 \\
        \textit{RADES-BabyIAXO} & (0.25, 0.33), (2.5-3.4) & \num{e5}  &   2          & (0.5,5) &  4.6 \\
        \hline
        ABRACADABRA              & (\num{e-4}, 0.002)  & 1           & 1           & 0.096   & 0.5 \\
        SHAFT                    & (\num{3e-6}, 0.003) & 1           & 1.51        & 0.046   & 4.2 \\
        ADMX SLIC                & 0.043               & \num{3e3}   & 7           & 0.2     & 20 \\
        BASE                     & \num{4e-4}          & \num{4e4}   & 1.85        & 0.025   & 5.7 \\
        WISPLC                   & (\num{3e-5}, 0.005) & \num{e4}    & 14          & 0.29    & 4 \\
        \textit{DMRadio-\si{m^3}}& (0.005, 0.2)        & \num{e5}    & 4           & 1.3     & 0.02 \\
        \textit{DMRadio-GUT}     & (\num{e-4}, 0.03)   & \num{2e7}   & 16          & 2.2     & 0.01 \\
        \bottomrule
    \end{tabular}
    \renewcommand{\arraystretch}{1.0}
    \caption[Benchmark parameters of microwave cavities and low-mass axion haloscopes.]
    {Benchmark parameters for microwave cavities (\cref{sec:SRFcavities}) and low-mass axion haloscopes (\cref{sec:AxionHaloscopes}) from \citep{Berlin:2021txa, Navarro:2023eii, Valero:2024ncz, Domcke:2023bat}. For more details on the individual setups see \cite{ADMX:2021nhd} (ADMX), \cite{HAYSTAC:2018rwy} (HAYSTAC), \cite{CAPP:2020utb} (CAPP), \cite{Quiskamp:2022pks,Quiskamp:2023ehr} (ORGAN), \cite{Posen-talk} (\textit{SQMS}),
    \cite{Navarro:2023eii} \textit{(cubic cavities)}, \cite{IAXO:2020wwp,Ahyoune:2023gfw} (\textit{RADES-BabyIaxo}),
    \cite{Salemi:2021gck} (ABRACADABRA),
    \cite{Gramolin:2020ict} (SHAFT),
    \cite{Crisosto:2019fcj} (ADMX SLIC),
    \cite{Devlin:2021fpq} (BASE),
    \cite{Zhang:2021bpa} (WISPLC), and
    \cite{DMRadio:2022pkf, DMRadio:2022jfv} \textit{(DMRadio)}.
    Experiments which are proposed or under development are indicated in \textit{italics}.}
    \label{tab:benchmarks_cavities}
\end{table}

The basic idea was further developed in \citep{Navarro:2023eii} using realistic simulations of radio frequency resonant cavities and suggesting a cubic resonator design for the cavity to allow for simultaneous determination of the polarization and the direction of the incoming GW.
The geometry of a quarterly split cavity was proposed in \citep{Gao:2023gph}.
The use of a cavity with tunable resonance frequencies was investigated in \citep{Valero:2024ncz}, studying in particular the RADES-BabyIAXO cavity proposed to search for a dark matter axion background field~\citep{Ahyoune:2023gfw} within the BabyIAXO helioscope~\citep{IAXO:2020wwp} setup.
\cite{Capdevilla:2024cby} estimated that tunable plasma cavities (as being developed by the ALPHA collaboration~\citep{ALPHA:2022rxj} for axion searches) can be sensitive to persistent coherent GWs with amplitudes of $h_0 \sim \num{3e-24} - \num{e-22}$ in the ${\cal O}(10-50)$~GHz frequency range, depending on the choice of an isotropic or anisotropic medium permeating the cavity (see also \citep{Gatti:2024mde} for earlier work).

Moreover, the static external field can be substituted by loading the cavity with a pump mode, as demonstrated in the MAGO prototype designed for GW searches \citep{Ballantini:2003nt,Ballantini:2005am} (see also \cite{Berlin:2023grv}). For the sensitivity of MAGO to the mechanical coupling of the GW, see \cref{sec:HighQCavities}.

It has been suggested in \cite{Herman:2020wao, Herman:2022fau} that the rapidly-oscillating cross-term between the GW-induced EM field and the background magnetic field can lead to improved sensitivity. For stochastic GW backgrounds with $\langle h(f) \rangle = 0$, the term linear in the strain averages to zero, so only a term quadratic in the strain will lead to non-zero signal, since $\langle h(f)^2 \rangle \neq 0$. For signals with $\langle h(f) \rangle \neq 0$, the linear signal can arise, but the sensitivity in this case is independent of the background EM field, contrary to the claims made in \citep{Herman:2020wao,Herman:2022fau}. This can be understood by recalling that a DC magnetic field will not lead to an AC current in an antenna. The appropriate comparison of signal and noise is therefore between the AC signal field and the AC component of the background field sourced by voltage fluctuations in the readout system.

\subsubsection{Low-Mass Axion Haloscopes}
\label{sec:AxionHaloscopes}

Low-mass axion haloscopes based on LC circuits such as ABRACADABRA \cite{Salemi:2021gck}, ADMX SLIC \cite{Crisosto:2019fcj}, BASE \cite{Devlin:2021fpq}, DMRadio \cite{DMRadio:2022pkf, DMRadio:2022jfv}, SHAFT \cite{Gramolin:2020ict}, and WISP-LC \cite{Zhang:2021bpa} target a non-relativistic, wave-like dark matter axion background. They feature a strong static magnetic field, which in the presence of an axion (or gravitational wave) leads to an oscillating effective current which in turn induces small oscillating EM fields. A resonant LC circuit is placed to read out the tiny induced oscillating magnetic flux.

The resulting magnetic flux induced by a coherent GW can be quantified using conventional electromagnetism methods starting from the expression of the effective current in \cref{eq:effective_current}. It can schematically be written as~\citep{Domcke:2022rgu}
\begin{align}
    \tilde \Phi_{h}(f) \simeq \eta B_{0} \, (2 \pi f L)^2
                              L^2 \tilde h(f)
    \label{eq:PhiAbra}
\end{align}
where the coupling constant $\eta \simeq 0.1$ is determined by the detector's geometry~\citep{Domcke:2023bat}.\footnote{This expression assumes a detector sensitive to the leading order term in the GW induced flux (in an $(f L)$-expansion). This can be achieved by suitable detector geometries in which cylindrical symmetry, often employed to maximize the sensitivity to the axion signal, is broken~\citep{Domcke:2023bat}.} By recasting the sensitivities obtained or projected for axion searches, \cite{Domcke:2022rgu, Domcke:2023bat} demonstrated that strain sensitivities of $h_0 \sim 10^{-9}$ (\SI{2}{MHz}, ABRA), $h_0 \sim 10^{-16}$ (\SI{40}{MHz}, ADMX SLIC), $h_0 \sim 10^{-15}$ (\SI{6}{MHz}, WISP-LC) and $h_0 \sim 10^{-21}$ (\SI{100}{MHz}, DMRadio-m$^3$) can be reached for coherent, persistent GW signals.

In terms of power spectral densities, the flux PSD at the readout SQUID is given by
\begin{align}
  S_{\Phi, \text{sig}}^\text{SQUID}(f) = ({\cal T}_1 {\cal T}_2)^2 \, S_h(f)
      \sim \eta^2 (2 \pi f)^4 B_0^2 L^8 Q^2 \alpha^2 \frac{L_{\rm sq}}{4 L_p} S_h(f)\,.
\end{align}
where $B_0$ is the magnetic field strength, $L_p$ denotes the inductance of the pickup loop, and $L_{\rm sq}$ the inductance of the SQUID. A typical value (for the example of DMRadio) is $L_\text{sq} \simeq \SI{1}{nH}$. 

The transfer function $\mathcal{T}_1$ translates from GW strain to flux at the pickup loop (see \cref{eq:PhiAbra}), while $\mathcal{T}_2$ describes the transmission through the LC circuit. The transfer functions are given by
\begin{align}
    {\cal T}_1 = \eta^{2} (2 \pi f L)^2 B_{0} L^2
    \,, \qquad
    {\cal T}_2 = \frac{\alpha \sqrt{L_{\rm sq}}}{2 \sqrt{L_p}} Q \,.
\end{align}
The coupling coefficient between the LC circuit and the SQUID is denoted by $\alpha$, for which $\alpha = 1/\sqrt{2}$ is a typical value \citep{Foster:2017hbq}. Further, $Q$ denotes the quality factor of the LC resonator which we have assumed to operate on resonance in the expression for ${\cal T}_2$.

In resonant readout mode the dominant noise source is the thermal noise of the LC circuit, subject to the same transfer function ${\cal T}_2$ as the signal. This yields the noise-equivalent strain sensitivity
\begin{align}
    S_h^\text{noise}(f) \simeq
        \frac{2 T_\text{sys}}{(2 \pi f)^5 \eta^2 B_0^2 L^7 Q}
    \qquad \text{(resonant)} \,.
\end{align}
In broadband mode the sensitivity is set by SQUID noise  with $S_n^{\rm sq} = (10^{-6} \Phi_0)^2 / \si{Hz}$ (with $\Phi_0 = \pi \hbar/e$). On the signal side, we the enhancement by the quality factor $Q$ is absent in broadband mode, such that we obtain
\begin{align}
    S_h^\text{noise}(f) \simeq 
        \frac{(10^{-6} \, \Phi_0)^2 \, \text{Hz}^{-1}}{2 \pi^4 \eta^{2} f^4 B_{0}^2 L^7 L_{\rm sq}}
    \qquad \text{(broadband)}\,.
\end{align}

Sensitivity curves for several LC circuit-based haloscopes, (namely variations of the DMRadio program) are shown in \cref{fig:43_summary}, based on the benchmark parameters given in \cref{tab:benchmarks_cavities}.


For ABRACADBRA, results of a prototype optimized for GW searches (`figure-8 loop') were recently published~\citep{Pappas:2025zld},  and while the sensitivity is still very far from realistic sources, these studies provide important input for designing detector geometries and optimizing data analysis strategies in the future.

We note that while for axion searches, the advantages of a resonant search are undeniable, a dedicated HFGW search would likely benefit from a broadband search, given the signal expectations discussed in \cref{sec:lateU}.

\subsubsection{Summary: Strain Sensitivities of Electromagnetic Oscillators}
%
\begin{figure}
    \includegraphics[width=\textwidth]{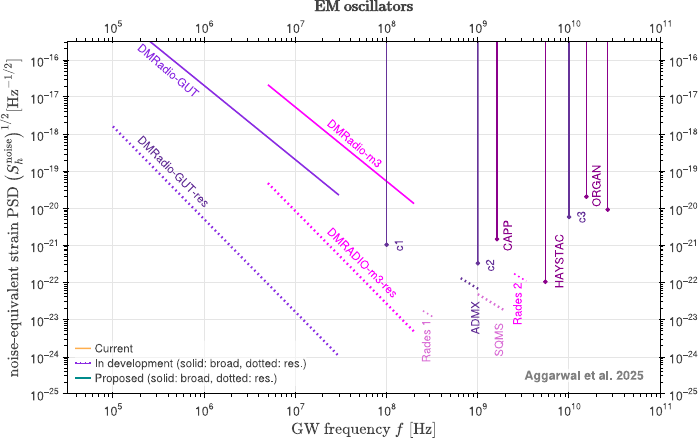}
    \caption[Projected strain sensitivity of electromagnetic oscillators.]
    { Projected strain sensitivity of electromagnetic oscillators employed for gravitational wave detection. The experimental parameters and references used as input for these curves are listed in \cref{tab:benchmarks_cavities}. Color coding as in \cref{fig:sens_HF}.}
    \label{fig:43_summary}
\end{figure}
\Cref{fig:43_summary} provides an overview of the projected strain sensitivities of a range of electromagnetic gravitational wave detectors, in particular low-mass haloscopes and microwave cavities. The experimental parameters on which these sensitivity curves are based are shown in \cref{tab:benchmarks_cavities}. All detector concepts shown are under active experimental development, most of them primarily for axion searches. Solid lines indicate broadband sensitivities whereas dashed lines indicate resonant searches requiring a scanning strategy. Where available, we show both the projected sensitivity at an initial stage as well as the impact of possible upgrades.

\subsection{Photon (Re-)Generation Experiments}
\label{sec:PhotonRegen}

\subsubsection{Light-Shining-through-a-Wall Experiments and Axion Helioscopes}
\label{sec:ALPs_and_axion_helioscopes}

Axion searches based on photon regeneration target relativistic axions originating from the Sun (axion helioscopes) or from powerful lasers in laboratories (light-shining-through-a-wall experiments, LSW). In both cases, axions have to traverse a layer of shielding, which blocks out incoming photons, and are subsequently converted into photons in a conversion region permeated by a strong magnetic field. In a similar manner, GWs can be converted into photons in this conversion region through the \textit{inverse Gertsenshtein effect}, or magnetic conversion effect~\citep{Gertsenshtein, Raffelt:1987im}. An advantage of GW searches in such experiments is that, unlike axions, they are strictly massless, implying that the conversion to photons occurs on resonance. On laboratory scales, the conversion probability can be obtained by solving Maxwell's equations with the GW-induced effective current from \cref{eq:effective_current} (see also \cref{sec:EMsignals}). On astronomical scales, the back-conversion of EM waves into gravitational waves, dictated by the contribution of the EM waves to the energy-momentum tensor in Einstein's equation, becomes relevant and leads to an oscillation between GW and EM waves, see \cref{sec:AstroDetectors}.

Ref.~\cite{Ejlli:2019bqj} set first upper limits on stochastic GW backgrounds at optical and X-ray frequencies (i.e., around \SI{500}{THz} and \SI{e6}{THz}, respectively) using data from light-shining-through-a-wall experiments (ALPS, OSQAR) and axion helioscopes (CAST). As pointed out in \cite{Liu:2023mll}, when interpreted as constraints on an isotropic stochastic GW background, these sensitivities should be reduced by a factor taking into account the field of view of these experiments resulting in sensitivities of of $h_{c, \text{sto}} \simeq \num{4e-23}$ at optical frequencies (OSQAR~II), and  $h_{c, \text{sto}} \simeq \num{8e-26}$ at X-ray frequencies (CAST).

Currently, significant R\&D effort is ongoing in designing more powerful instruments. Notably, ALPS II \citep{Bahre:2013ywa}, featuring a magnetic field of \SI{5.6}{T} and a conversion length of \SI{100}{m}, is currently taking data, and the next generation helioscope (Baby)IAXO~\citep{Armengaud:2014gea} is under active development. Remarkably, these instruments succeed in operating far below the standard quantum limit. Given the strong motivation to search for high-frequency GWs, a dual usage of these detectors could be imagined with dedicated instruments and operational modes to search for GWs.

To estimate the noise-equivalent strain sensitivity of these detectors we start from the GW-induced EM field in frequency space~\citep{Ejlli:2019bqj},
\begin{align}
    \tilde E_h(f),\ \tilde B_h(f) \simeq
        2 \pi f L B_{0} \, \tilde h(f)\,.
\end{align}
Here, $E_h$ and $B_h$ are the GW-induced electric/magnetic fields respectively, $B_0$ is the background magnetic field, and $2 \pi f L \gg 1$ is the enhancement from resonant conversion in vacuum over the length $L$ of the instrument. (See also \cref{sec:EMsignals} for more details.) From this we compute the time-averaged Poynting vector which gives the power per unit area at the receiver as
\begin{align}
    \langle S \rangle
        &= \frac{1}{T} \int_{-T/2}^{T/2} \! dt \, E_h \times B_h
        \simeq \int \! df \, (2 \pi f L B)^2 S_h(f) \,.
\end{align}
The total power is obtained as $P_{\rm sig} = A \langle S \rangle$, with $A$ the area of the receiver. This yields for the power PSD of the signal,
\begin{align}
    P_{\rm sig} = \int \! df \, S_{P,\text{sig}}(f)
    \qquad \rightarrow \qquad
    S_{P,\text{sig}}(f) \simeq A (2 \pi f L B)^2 S_h(f)\,. \label{eq:SP}
\end{align}
Noting that a GW of frequency $f$ will create a photon of the same frequency, we can also compute the number of detected photons,
\begin{align}
    N_{\gamma,\text{sig}} = \frac{\epsilon P_{\rm sig}}{2 \pi f} \Delta t
    \qquad \rightarrow \qquad
    S_{N, \text{sig}} \simeq 2 \pi \epsilon f A L^2 B^2 \Delta t \, S_h(f)\,,
    \label{eq:NsigSPD}
\end{align}
with $\epsilon$ the single photon detection efficiency and $\Delta t$ the observation time.
The use of a resonant regeneration cavity could further increase the number of signal photons at the resonant frequency by the finesse factor ${\cal F}$ of the cavity.
The signal strength degrades at low frequencies due to waveguide effects, which become important when the GW wavelength becomes comparable to the dimensions of the cavity, specifically at frequencies below \citep{Ringwald:2020ist}
\begin{align}
    f \sim \frac{L}{4\pi A} \,,
\end{align}
for an elongated cylindrical cavity of length $L$ and cross-sectional area $A$.

For a single-photon detection scheme, as implemented in current optical and X-ray instruments, we need to compare this with the dark count rate  $\Gamma_D(f)$ in a frequency bin of width $\Delta f$ \citep{Ejlli:2019bqj},
\begin{align}
    N_{\gamma, \text{noise}}
        = \Gamma_D \Delta t
        \simeq \int \! df \, S_{N,\text{noise}} 
        \qquad\rightarrow\qquad
    S_{N,\text{noise}} = \frac{\Gamma_D(f)}{2 \Delta f} \Delta t \,.
\end{align}
This yields the noise-equivalent strain sensitivity (see \cref{eq:shnoise}),
\begin{align}
    S_h^\text{noise}(f) \simeq
        \frac{\Gamma_D(f)/\Delta f}{2 \epsilon A (2 \pi f) L^2 B^2}
    \qquad \text{(single photon)} \,.
\end{align}
We note that that any detection requires $N_{\gamma,\text{sig}} \geq 1$, which for very short signals (as is typically the case, e.g., PBH mergers)  can impose a constraint which is more stringent than overcoming the dark count rate.
Close to this limit, one can moreover not rely on match filtering techniques. Hence, when estimating the sensitivity to PBH mergers, we do not employ \cref{eq:reachPBH}, but instead require directly that $N_{\gamma,\text{sig}}$ as introduced in \cref{eq:NsigSPD}, accumulated over the signal duration during the detector run time, is larger than one.
Similarly, we employ this criterion instead of the PLS curves used commonly in linear detectors to evaluate the sensitivity to stochastic backgrounds.

The inverse Gertsenshtein effect can in principle be exploited over a very broad range of frequencies, and in particular has substantial potential at GHz frequencies where many early Universe signals converge. This would require fitting existing or planned instruments with different electromagnetic receivers.  For example, \cite{Ringwald:2020ist} provides estimates for the sensitivity achievable by a IAXO-type experiment in the GHz region, finding $h_c \lesssim 10^{-22}$ using heterodyne radio receivers (HET) and $h_c \lesssim 10^{-25}$ with single photon detectors (SPD) around a frequency of $f \simeq \SI{4e10}{Hz}$.
This illustrates that single-photon detection at microwave frequencies could be key to unlocking a significant improvement in sensitivity.
While the implementation of these techniques in the GHz range is highly challenging, we note that single-photon detection at microwave frequencies is an area of rapidly advancing experimental development, actively pursued also for dark photon and dark matter searches, see e.g.~\citep{Lescanne:2020awk,Dixit:2020ymh,Chiles:2021gxk,Graham:2023sow,Braggio:2024xed,Pankratov:2024ode}.

In the mean time, current photon (re)generation experiments operating in the GHz regime perform a power measurement with the signal PSD given by \cref{eq:SP}, and the noise PSD given by the thermal noise (see \cref{eq:thnoise}), see also next subsection and heterodyne radio receivers (HET) proposed in~\citep{Ringwald:2020ist}. Combining these yields the noise-equivalent strain sensitivity
\begin{align}
    S_h^\text{noise}(f) \simeq
        \frac{k_{B} T_\text{sys}}{8 \pi^2 f^2 B^2 L^2 A}
    \qquad \text{(thermal)}\,.
   \label{eq:Sn_PG_th}
\end{align}

Importantly, we note that LSW experiments and helioscopes can be designed as broadband instruments, with a bandwidth of about one order of magnitude in frequency. In broadband mode, one cannot employ a resonant cavity and thus does not profit from an enhancement originating from a large finesse.

Benchmark parameters for existing (or decommissioned) experiments as well as upcoming and proposed instruments are shown in \cref{tab:benchmarks_PhotonRegen}, and the resulting sensitivity curves can be found in \cref{fig:43_summary}.

\begin{table}[t]
    \centering
    \renewcommand{\arraystretch}{1.1}
    \begin{tabular}{lcccccc}
        \toprule
                          & $f$ [GHz]            &$B_0$ [T]  &$L$ [m] &$A$ [\si{m^2}]& $\Gamma_D$ [Hz] & $\epsilon$ \\ 
        \midrule
        OSQAR I           & $(0.3, 0.8) \times 10^{6}$ & 9   &  14.3  & \num{5e-4}   & \num{1.76e-3} & 0.5 \\
        OSQAR II          & $(0.3, 1) \times 10^{6}$   & 9   &  14.3  & \num{5e-4}   & \num{1.14e-3} & 0.9 \\
        ALPS I            & $(0.3, 1) \times 10^{6}$   & 5   &   9    & \num{5e-4}   & \num{6.1e-4}  & 0.1 \\
        ALPS II           & $(0.3, 1) \times 10^{6}$   & 5.3 & 106    & \num{2e-3}   & \num{e-6}     & 0.75 \\
        \textit{JURA}     & $(0.3, 1) \times 10^{6}$   & 13  & 960    & \num{8e-3}   & \num{e-6}     & 1 \\
        \midrule
        CAST              & $(0.5, 1.6) \times 10^9$   & 9   &   9.26 & \num{2.9e-3} & \num{1.5e-4}  & 0.7 \\
        \textit{BabyIAXO} & $(0.25, 2) \times 10^{9}$  & 2.5 &  10    &  0.77        & \num{e-3}     & 1 \\
        \textit{IAXO}     & $(0.25, 2) \times 10^{9}$  & 2.5 &  20    &  3.08        & \num{e-4}     & 1 \\
        \bottomrule
    \end{tabular}
    \renewcommand{\arraystretch}{1.0}
    \caption[Benchmark parameters of light-shining-through-wall \& helioscope experiments.]
    {Benchmark parameters of light-shining-through-a-wall and helioscope experiments, see~\citep{Ejlli:2019bqj, Ringwald:2020ist}. For details on the individual setups see  \cite{OSQAR:2015qdv} (OSQAR), \cite{Bahre:2013ywa, Albrecht:2020ntd} (ALPS), \cite{Beacham:2019nyx} (\textit{JURA}), \cite{CAST:2017uph} (CAST),  \cite{IAXO:2020wwp} (\textit{BabyIAXO}) and  \cite{IAXO:2019mpb} (\textit{IAXO}). Experiments which are proposed or under development are indicated in \textit{italics}.}
    \label{tab:benchmarks_PhotonRegen}
\end{table}

\subsubsection{Dielectric Axion Haloscopes}
\label{sec:Dielectric_axion_haloscopes}

At frequencies around \SI{10}{GHz}, dielectric haloscopes are currently being developed for axion searches. Compared to traditional photon regeneration experiments they profit from enhanced axion-to-photon conversion at the surfaces of a stack of dielectric disks. If the disk separation is suitably tuned, the EM waves generated at the surfaces of the different disks interfere constructively. This idea is implemented in MADMAX \citep{MADMAX:2019pub}, which has very recently taken first data with a prototype instrument \citep{MADMAX:2024jnp}.

MADMAX can also be used to search for gravitational waves. Compared to the axion case, the relativistic nature of the GWs enhances conversion in the vacuum region between the disks, but imposes a challenging new requirement to adapt the effective disk thickness to a particular GW frequency. Operated in fully resonant mode, the noise-equivalent strain sensitivity has been estimated as~\citep{Domcke:2024eti}
\begin{align}
    \left( S_h^\text{noise} \right)^{1/2}
        \sim \SI{e-22}{Hz^{-1/2}} \times
              \bigg( \frac{\SI{1}{m^2}}{L^2} \bigg)
              \bigg( \frac{\SI{10}{T}}{B_0} \bigg)
              \bigg( \frac{43}{N_d} \bigg)
              \bigg( \frac{\SI{10}{GHz}}{f} \bigg) ,
    \label{eq:sens-res}
\end{align}
with $N_d \lesssim 50$ the number of dielectric disks inserted. (If the number of disks becomes too large, the sensitivity actually \emph{decreases}.) A sensitivity estimate for a MADMAX-like detector operating in resonant mode and assuming the benchmark parameters from \cref{eq:sens-res} is shown as a dotted purple line in \cref{fig:44_summary}. The sensitivity of the same instrument, but \emph{without} the dielectric disks is shown as a solid line. This corresponds to operating as a standard photon (re)generation experiment, though at radio frequencies. The noise-equivalent strain sensitivity is given by \cref{eq:Sn_PG_th}, and the detector parameters are listed in \cref{tab:benchmarks_dielectric_haloscopes}. As discussed in \citep{Domcke:2024eti}, it seems, however, most beneficial to operate in a hybrid mode, where part of the detector volume is filled with dielectric disks, while the rest is empty. Such a setup still profits from resonant enhancement in a narrow frequency range, but also has broadband sensitivity similar to the photon (re)generation experiments discussed in \cref{sec:ALPs_and_axion_helioscopes} above.

A related approach to search for both axions and gravitational waves is pursued in DALI, which features a superconducting solenoid magnetizing a stack of ceramic wafers, and an array of antennas for readout.  A scaled-down prototype (DALI~PT) is currently under construction, with upgrades to DALI Phase~II (henceforth DALI~II) in planning~\citep{DeMiguel:2020rpn, DeMiguel:2023nmz}.

\begin{table}
    \centering
    \renewcommand{\arraystretch}{1.1}
    \begin{tabular}{lccccc}
        \toprule
                         & $f$ [GHz]          & $B_0$ [T] & $L$ [m] & $A$ [\si{m^2}] & $T_\text{sys}$ [K] \\
        \midrule
        \textit{MADMAX}  & (0.2, 100)         & 10        & 1       & 1.23           & 4.2 \\
        \textit{DALI PT} & (6, 8) \& (29, 37) &  1        & 0.35    & 0.03           & 30 \\
        \textit{DALI II} & (6, 60)            & 11.7      & 1.1     & 1.5            & 1.5 \\
        \bottomrule
    \end{tabular}
    \renewcommand{\arraystretch}{1.0}
    \caption[Benchmark parameters of dielectric haloscopes.]
    {Benchmark parameters of dielectric haloscopes, see \cite{MADMAX:2019pub} (MADMAX) and \cite{DeMiguel:2023nmz} (DALI prototype and phase II).}
    \label{tab:benchmarks_dielectric_haloscopes}
\end{table}

\subsubsection{Summary: Strain Sensitivities of Photon (Re-)Generation Experiments}
%
\begin{figure}
    \centering
    \includegraphics[width=\textwidth]{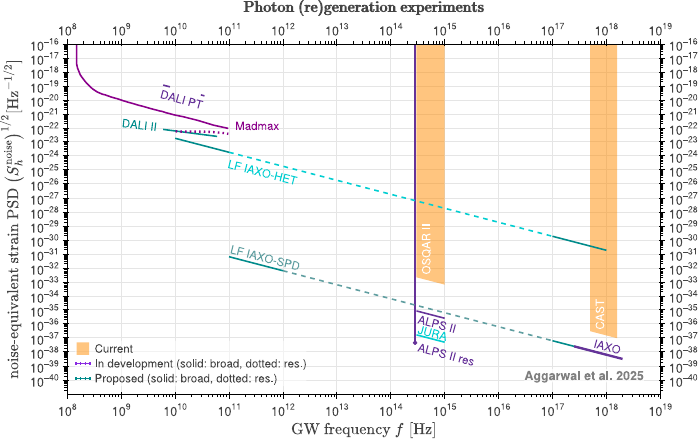}
    \caption[Projected strain sensitivities of GW detectors based on photon regeneration.]
    { Projected strain sensitivities of gravitational wave detectors based on photon regeneration, with parameters and references listed in \cref{tab:benchmarks_PhotonRegen} and \cref{tab:benchmarks_dielectric_haloscopes}. Color coding as in \cref{fig:sens_HF}.}
    \label{fig:44_summary}
\end{figure}

\Cref{fig:44_summary} provides an overview of the projected strain sensitivities of a range of laboratory electromagnetic gravitational wave detectors based on photon regeneration. As in \cref{fig:sens_HF}, instruments which have placed limits on (or detected) GWs are shown in orange, concepts under active R\&D are shown in purple and, other proposals are shown in cyan. Where available, we show both the projected sensitivity at an initial stage, as well as possible upgrades. In particular, we show sensitivity projections for light-shining-through-a-wall experiments (OSQAR~II, ALPS~II, and JURA, with parameters given in \cref{tab:benchmarks_PhotonRegen} and with the dashed curve corresponding to resonant operation of ALPS~II with ${\cal F} = \num{40000}$), helioscopes (CAST and IAXO, with parameters given in \cref{tab:benchmarks_PhotonRegen}), and dielectric haloscopes (DALI, MADMAX, with parameter for broadband operation given in \cref{tab:benchmarks_dielectric_haloscopes}, and with the sensitivity for MADMAX in resonant mode taken from \cite{Domcke:2024eti}). For helioscopes, the upper line refers to a thermal noise limited readout whereas the lower line refers to single photon detection. We also show (in lighter green) the extrapolation to lower frequencies assuming an adapted readout system.

We caution that in the transfer functions used to estimate these sensitivities we have dropped the dependence on the incident angle of the GW, that is, we have not taken into account the antenna pattern. In this sense, these sensitivities should be seen as sensitivities to GWs reaching the detector under an optimal angle. While this is a relatively small effect for many GW detectors, which have rather broad angular response functions, it is a much more important consideration for experiments such as ALPS, which is maximally sensitive only in a very small field of view.

\subsection{Other Electromagnetic Gravitational Wave Detectors}
\label{sec:EMother}

\subsubsection{High Energy Pulsed Lasers}
\label{sec:PulsedLasers}

Ref.~\cite{Vacalis:2023gdz} proposed a method for detecting high-frequency gravitational waves using high-energy pulsed lasers rather than constant magnetic fields. In this approach, GW interactions with the laser field create an electromagnetic signal via the inverse Gertsenshtein effect, and resonance occurs when the frequency of the GW is twice the laser frequency. The method is particularly suitable in the optical frequency range due to the availability of long, high-intensity pulses in this frequency regime. Single-photon counting techniques are used to detect the electromagnetic signal. Targeting the frequency range $(10^{13} - 10^{19})\,\si{Hz}$, this technique can reach strain sensitivities of $h_0 \lesssim 10^{-20}$. With the next generation of optical lasers, strains down to $h_0 \sim 10^{-26}$ may be reachable at specific frequencies.

\subsubsection{GW to Electromagnetic-Wave Conversion in a Static Electric Field}
\label{sec:ConversionEfield}

Ref.~\cite{Lupanov} considered the inverse Gertsenshtein effect in a static electric field rather than a static magnetic field.\footnote{Electric fields are usually not considered in the context of axion searches, as the coupling of non-relativistic axions to electric fields is suppressed compared to their coupling to magnetic fields. For GWs, on the other hand, there is no such suppression.} The physics is essentially the same in the two cases but the intensity of electric fields in laboratory settings is limited due to their tendency to pull electrons from any support structure. Consequently, the energy densities reachable in electric fields are about a million times smaller than those of magnetic fields in the several Tesla range.

This limitation can be overcome by focusing on graviton-to-photon conversion in \emph{atomic} electric fields, which can be much stronger~\citep{Dai:2023rgx}. The conversion happens when the wavelength under consideration is shorter than the atomic radius, making the method sensitive at frequencies of $10^{20}$--$\SI{e24}{Hz}$, or graviton energies between \SI{100}{keV} and \SI{1}{GeV}. \cite{Dai:2023rgx} proposed to search for the generated photons in current and future neutrino detectors, for instance JUNO. A downside of this technique is the limitation to very high frequencies, at which it seems difficult to envisage sufficiently strong GW sources.

\subsubsection{Resonant Polarization Rotation}
\label{sec:ResonantPolarisationRotation}

Ref.~\cite{Cruise1} showed that a GW could induce a rotation of the plane of polarization of electromagnetic waves in certain geometries, some of which might be relevant astronomically. In 2000, the idea of resonant polarization rotation was extended to a situation in which the electromagnetic wave was a circulating wave in a microwave waveguide ring \citep{Cruise2} . The effect is amplified by the (potentially significant) quality factor of the waveguide ring. A proof of concept apparatus was constructed by \citep{Cruise3, Cruise:2006zt}. Such a device would be narrowband, achieving a sensitivity to a stochastic GW background of $\sqrt{S_n} \lesssim \SI{e-14}{Hz^{-1/2}}$ at frequencies around \SI{100}{MHz} by cross-correlating two detectors. It is difficult to see the sensitivity of this GW detection scheme increasing very far beyond this limit though. Recently, this concept has been revisited in the context of optical cavities, emphasizing parallels with axion birefringence searches \citep{Garcia-Cely:2025mgu}. Notably, it has been pointed out that the existing ALPS II infrastructure at DESY can be adapted to measure polarization effects induced by GWs. Utilizing realistic cavity properties and current technology, this approach could, within a few years, enable the exploration of HFGWs in the frequency range of 0.1~\text{MHz} to 0.1~\text{THz} with sensitivities comparable to the aforementioned $\sqrt{S_n}$.

\subsubsection{Heterodyne Enhancement of Magnetic Conversion}
\label{sec:AmplificationMethods}

Refs.~\cite{Li:2004df, Li:2006sx, Baker:2008zzb, Li} suggested enhancing the efficiency of magnetic conversion detectors by seeding the conversion volume with a locally generated electromagnetic wave at the same frequency as the GW being searched for. Concerns were raised in \citep{WOODS201266} and \citep{Eardley:2008}. Furthermore, the claims of outstanding sensitivity rely on technology that does not yet exist, and no experimental results have been produced to suggest it is feasible.


\subsection{Astrophysical and Cosmological Detection Concepts}
\label{sec:AstroDetectors}

The majority of indirect astrophysical and cosmological probes of high-frequency gravitational waves exploit the inverse Gertsenshtein effect, in which gravitational waves convert into photons in cosmological or astrophysical magnetic fields. Schematically, the conversion probability is
\begin{align}
    P_{h \to \gamma} \sim \frac{1}{M_{\rm Pl}^2 } (B L)^2 ,
\end{align}
where $L$ is the characteristic physical length scale over which conversion takes place, and $B$ is the characteristic magnetic field strength. The product $B L$ therefore provides a useful figure of merit for determining the strength of graviton--photon mixing. Typical values are $B L \simeq \SI{e13}{G\,km} \times  (B/\si{nG}) \, (L/\si{Gpc})$ in a cosmological setting, whilst neutron stars can reach $B L \simeq \SI{e13}{G\,km} \times (B/\SI{e12}{G}) \, (L/\SI{10}{km})$ (and even larger values in the case of magnetars) \citep{Domcke:2020yzq}.  This comparison shows how the relative weakness of cosmological magnetic fields can be compensated for by large effective conversion lengths. Beyond this rough figure of merit, the suitability of a given system to search for GWs depends on the details of the environment in question (in particular the effective photon mass, which can be non-vanishing, resulting in suppressed, non-resonant conversion between GWs and photons),
and the flux of background or foreground photons.

Neutron stars have already been used to search for other low-mass particles, notably dark matter axions \citep{Raffelt:1987im, Pshirkov:2007st, Huang:2018lxq, Hook:2018iia}. This initiative has now grown into an established field in its own right, with a wide range of observations and sophisticated modeling. Neutron stars as high-frequency GW detectors have only become an active topic of study recently. \cite{Ito:2023fcr} produced tentative constraints on stochastic gravitational waves in the radio frequency band 0.1--\SI{1}{GHz} and in the range $10^{13}$--\SI{e27}{Hz} spanning the IR, UV, Visible and X-ray regimes. Resulting strain sensitivities range from $h_c \lesssim 10^{-14}$ to $h_c \lesssim 10^{-18}$ in the radio band, and from $h_c \lesssim 10^{-16}$ to $h_c \lesssim 10^{-26}$ in the high-frequency band based on non-resonant conversion of gravitational waves into photons. More recently, \cite{McDonald:2024nxj} explored the role of resonant conversion in setting constraints.

It should be emphasized that the modeling in the pioneering work~\cite{Ito:2023fcr} remained rudimentary, both in terms of the treatment of graviton--photon mixing in 3D magnetized plasmas\footnote{See Ref.~\citep{Macedo:1983wcr} for more systematic attempts in homogeneous plasmas.} and the transport of photons through the magnetosphere. Fortunately, much of the machinery for addressing these issues more accurately has been developed already in the context axion physics, see \cite{McDonald:2023ohd} for improved calculations of the conversion probability, and \cite{McDonald:2023shx} for a discussion of photon transport via ray-tracing techniques, which allow for accurate computation of the photon flux in a non-trivial magnetosphere geometry. Some of these more advanced techniques from axion physics have been applied to graviton--photon conversion in \cite{McDonald:2023shx}. As a cautionary note, in the axion context, the predicted photon signatures from state-of-the-art ray tracing techniques \citep{McDonald:2023shx} differ markedly from more naive early studies \citep{Hook:2018iia, Leroy:2019ghm}.

\begin{figure}
    \centering
    \includegraphics[scale=0.85]{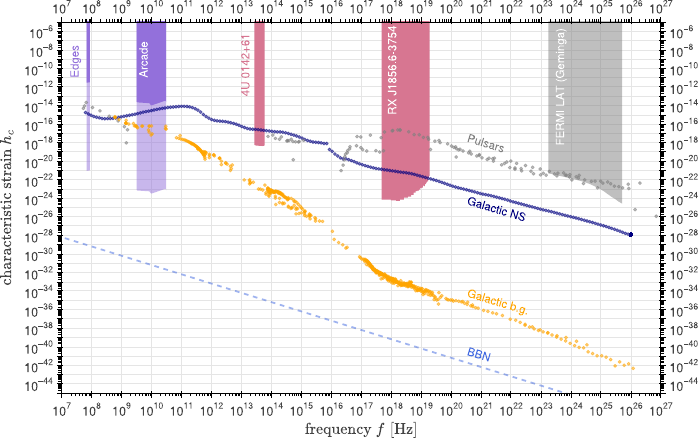}
    \caption[Astrophysical and cosmological constraints on stochastic GW backgrounds.]
    {Tentative astrophysical and cosmological constraints on isotropic stochastic gravitational wave backgrounds. We display constraints from non-resonant conversion \citep{Ito:2023fcr} (gray) and resonant conversion \citep{McDonald:2024nxj} (pink) in individual neutron stars. In purple we show constraints from spectral distortions of the CMB \citep{Domcke:2020yzq}, where the light/dark regions correspond to the range of allowed intergalactic magnetic field values. We also display limits from conversion in galactic magnetic fields \citep{Ito:2023nkq, Lella:2024dus} (orange) and neutron star populations \citep{Dandoy:2024oqg} (dark blue) (taking the conservative decaying magnetic field scenario).  We caution that in the case of galactic, Geminga, Crab and neutron star population constraints, points indicate that the underlying spectral data (see e.g.\ \citep{Hill:2018trh}) may not have continuous frequency coverage, such that there may be gaps in these constraints between observing bins. We refer the reader to the original works for more details. }
    \label{fig:astro}
\end{figure}

More recently, some early studies have been carried out using entire populations of neutron stars to place constraints on gravitational waves \citep{Dandoy:2024oqg}. Results are shown as a dashed blue line in \cref{fig:astro}. Again, we caution that these constraints would benefit from more state-of-the-art approaches to in population modeling, photon production, and photon transport.

Gravitational wave detection using cosmological magnetic fields has been considered in \citep{Pshirkov:2009sf, PhysRevLett.74.634, Dolgov:2012be, Cillis:1996qy,Domcke:2020yzq}. In this case, the magnetic field is weaker and the background is much harder to control, but cosmological magnetic fields can extend coherently over kpc or even Mpc, implying an enormous `detector volume'. Of particular interest is the frequency range from \SI{100}{MHz} to \SI{30}{GHz}, i.e.\ the Rayleigh--Jeans tail of the cosmic microwave background, which is the target of several existing and upcoming radio telescopes. For example, the data of ARCADE~2 \citep{Fixsen_2011} and EDGES \citep{Bowman:2018yin} can be recast into constraints at the level of $h_{c,\rm sto} < 10^{-24}  (10^{-14})$ in the range $\SI{3}{GHz} \lesssim f  \lesssim \SI{30}{GHz}$ (ARCADE~2) and $h_{c,\rm sto}(f \approx \SI{78}{MHz}) < 10^{-12} (10^{-21})$ (EDGES) for the strongest (weakest) cosmic magnetic fields in accordance with current astrophysical data \citep{Domcke:2020yzq}. The large uncertainty in these constraints resides in the unknown power spectrum of the cosmological magnetic fields in the early Universe. Clearly, more accurate modeling of magnetic fields is needed to improve on the 10 orders of magnitude uncertainty in these constraints on $h_c$.

Galactic and planetary magnetic fields have also been used recently to place constraints on stochastic gravitational wave backgrounds \citep{Ito:2023nkq, Lella:2024dus, Liu:2023mll}, though more work is needed to accurately model the conversion and the magnetic fields within the galaxy. Results are displayed in \cref{fig:astro}. Roughly similar sensitivities in the frequency range \SI{100}{TeV} -- \si{PeV} have been found using LHAASO to search for GW-to-photon conversion in the Milky Way \citep{Ramazanov:2023nxz}. Prospects for future radio telescopes and CMB spectrometers are discussed in \cite{He:2023xoh}, though under very optimistic assumptions regarding the magnetic fields and instrumental sensitivities.

Result from astrophysical searches for high-frequency GWs are summarized in \cref{fig:astro}. All of these results apply to isotropic stochastic gravitational wave backgrounds. They are compared with laboratory searches for SGWBs and with possible signals in \cref{fig:PLS}. An important direction of future study is the sensitivity of astrophysical detection techniques to GW signals that are localized in time and/or in space GW, such as typical  signals generated by sources in the late Universe.

\subsection{Other Concepts}
\label{sec:AlternativeConcepts}

In the following, we list several detection concepts not covered in the previous sections.

\subsubsection{Superconductors as GW Detectors}
\label{sec:SuperconductingRings}

GW interactions with matter are typically extremely weak because of ``impedance mismatch'', that is, a mismatch between the way a GW propagates in vacuum and the way deformation waves travel in medium. This impedance mismatch may be significantly reduced in superconductors and superfluids thanks to the macroscopic wave functions of these systems. This has led \cite{Anandan:1982is, Chiao:2002qi} to discuss a detection concept exploiting spin--spin interactions between GWs and vortices in superfluids. The proposed detector consists of a toroidal tube filled with superfluid helium and interrupted by a Josephson junction. An incoming GW leads to a phase difference in the superfluid's wave function across the junction. This phase difference, in turn, leads to a mass current, against which the apparatus recoils. It is proposed to read out this recoil using an electromechanical transducer. But while \cite{Chiao:2002qi} speculates about strain sensitivities at the $h_0 \sim 10^{-30}$ level around \SI{10}{GHz}, it should be kept in mind that this is based on very conceptual and preliminary studies.

An alternative proposal was suggested shortly after in \cite{Anandan:1985pv}, based on the mechanical force exerted by a GW on a superconducting solenoidal magnet. As the magnet coils are infinitesimally deformed by the GW, flux quantization enforces a change in the magnetic field. A suitably constructed and placed pickup loop could read out this change in the magnetic field using a SQUID. \cite{Anandan:1985pv} estimates a possible strain sensitivity of $h \sim 10^{-21}$, under optimistic assumptions regarding the experimental challenges.

A good review of the issues surrounding the interaction of mesoscopic quantum systems such as superfluids and superconductors with gravity can be found in \cite{Kiefer:2004hv}, casting in particular doubt on some of the assumptions made in \cite{Anandan:1982is, Chiao:2002qi}.

\subsubsection{Graviton--Magnon Resonance}
\label{sec:GravitonMagnonResonance}

As pointed out in \cite{Ito:2019wcb}, a GW passing through a ferromagnetic insulator can resonantly excite magnons (collective excitations of particle spins) thanks to an interaction between the GW and the fermion spins~\citep{Ito:2020wxi}.
These collective spin excitation are analogous to the excitation of phonons in resonant bar detectors. 
The readout is achieved by placing the magnetic sample inside a microwave cavity and coupling the magnon to a cavity photon mode. 
This idea builds on the technique of ferromagnetic haloscopes proposed for axion searches~\citep{Crescini:2018qrz, Flower:2018qgb}.
By reinterpreting the data from these axion searches, \cite{Ito:2019wcb, Ito:2020wxi} have shown that the sensitivity of the magnon GW detector reaches strains of $\sqrt{S_n} \lesssim \SI{7.6e-22}{Hz^{-1/2}}$ at \SI{14}{GHz} and $\sqrt{S_n} \simeq \SI{1.2e-20}{Hz^{-1/2}}$ at \SI{8.2}{GHz} \citep{Ito:2019wcb, Ito:2020wxi}. 

Ref.~\cite{Ito:2022rxn} showed that the sensitivity of the magnon GW detector is maximized when  the size of the ferromagnetic insulator is comparable to the wavelength of the GW.
Then, the sensitivity could in principle be improved to $h_0 \sim 10^{-20}$ around GHz by using a bigger sample~\citep{Ito:2022rxn}.
As an another possibility, the sensitivity of this approach can be improved by incorporating single photon counters instead of linear amplifiers. For counters available today, \cite{PhysRevD.88.035020} estimates a sensitivity improvement by several orders of magnitude for axion detection.

\subsubsection{Atomic Precision Measurements}
\label{sec:AtomicMeasurement}

Frequencies of photons in a GW background are modulated, an effect that is exploited for instance in nHz GW measurements using pulsar timing.
\cite{Bringmann:2023gba} extended this concept and proposed to probe high-frequency GWs with optical atomic clock techniques.
These techniques have achieved remarkable precision, allowing for photon frequency measurements with an accuracy of $\lesssim 10^{-20}$~\citep{Bothwell:2021fqe, Zheng:2021rnx}.
They therefore hold promise for probing the tiny frequency modulation of photons caused by GWs.

However, at high frequencies combined with long experimental integration times, this modulation would average to zero. To overcome this challenge, \cite{Bringmann:2023gba} proposed an ``optical rectifier'', which blocks the optical signal during half of each GW period using a shutter. The projected sensitivity to high-frequency GWs under optimistic assumptions for the achievable experimental sensitivity is $h_0 \lesssim 10^{-17}$ to $10^{-21}$ at frequencies from \SI{10}{kHz} to \SI{10}{GHz} for one second of integration time. To estimate the corresponding noise-equivalent strain sensitivity we use \ref{eq:h-sens-persistent-lin} for a linear narrow-band detector, yielding $\sqrt{S_n} = \SI{e-17}{Hz^{-1/2}}$ and $\SI{e-17}{Hz^{-1/2}}$, respectively.

\subsubsection{One-Electron Quantum Cyclotron}
\label{sec:oneelectron}

An electron in a background magnetic field experiences cyclotron motion and spin precession.
By cooling the electron, one can observe the quantization of the energy levels corresponding to the cyclotron motion, that is, the Landau levels.
Such a one-electron quantum cyclotron has been utilized to measure the anomalous magnetic moment of the electron $(g-2)_e$ with a remarkable precision of approximately $10^{-13}$.
To achieve this, an electron in a Penning trap is prepared in the lowest Landau level, which in theory has an infinite degeneracy, with the different degenerate states corresponding to wave functions of different spatial size $R$. 
\cite{Ito:2023bnu} proposed to utilize a similar setup for a gravitational wave search.
The excitation rate from the ground state to the first excited state depends on the size of the electron wave function: an electron with a larger wave function feels the effect of a gravitational wave more strongly (in the limit $2\pi f R \lesssim 1$. 
Interestingly, this enhancement is a particular feature of the excitation by gravitational waves, in contrast to excitations caused by electromagnetic fields (dipole excitation), whose excitation rate is insensitive to the size of the wave function.

The readout in such a setup would be via ``quantum jump spectroscopy'': an additional weak magnetic field is applied to couple the cyclotron motion of the electron to its axial motion (oscillations about the cyclotron orbit). This axial motion can be read out via the currents it induces.

From the dark count rate reported in the context of dark photon searches in~\cite{Fan:2022uwu}, \cite{Ito:2023bnu} estimated that an existing one-electron quantum cyclotron could probe
GW strains down to $h_0 \sim \num{3.8e-20}$ at frequencies around \SI{100}{GHz}.
The sensitivity can be improved by preparing the initial state of the electron even more carefully.
In particular, it is possible to prepare the electron in a state with a particularly large wave function.
This could allow the experiment to achieve a sensitivity of $h_0 \lesssim \num{6.9e-23}$~\citep{Ito:2023bnu}.
Although a GW search with a one-electron quantum cyclotron is a narrow band resonance experiment, it could scan over a frequency range from \SI{20}{GHz} to \SI{200}{GHz} by adjusting the strength of the magnetic field and the frequencies of driving fields, while maintaining the above-mentioned sensitivities~\citep{Fan:2022uwu}.

\subsubsection{Rydberg Atoms}
\label{sec:RydbergAtom}

Rydberg atoms have proven to be a unique type of quantum sensor for numerous applications. A method of exploiting their unusual properties for heterodyne detection of axions \citep{Engelhardt:2023qjf} lends itself also to the detection of gravitational waves with $\mathcal{O}(\si{GHz})$ frequencies \citep{Kanno:2023whr}. The method is based on electromagnetically induced transparency (EIT) \citep{1986JETP...63..945K, Fleischhauer:2005zz}, the starting point for which is a system of three atomic energy levels $|1\rangle$, $|2\rangle$, $|3\rangle$, where $|1\rangle$ and $|2\rangle$ are low-lying states and $|3\rangle$ is a Rydberg state. Two laser beams drive the $|1\rangle \leftrightarrow |2\rangle$ (``probe laser'') and $|2\rangle \leftrightarrow |3\rangle$ (``control laser'') transitions, respectively, leading to two interfering Rabi oscillations. For instance, a transition $|1\rangle \to |2\rangle$ can either happen directly, or via $|1\rangle \to |2\rangle \to |3\rangle \to |2\rangle$, and the two amplitudes interfere. This interference can be destructive, leading to a characteristic narrow absorption feature in the transmission spectrum of the probe laser.

Consider now a second Rydberg level, $|4\rangle$. An incoming gravitational wave can induce an oscillating electric field when coupled to an external magnetic field, and this electric field can drive Rabi oscillations $|3\rangle \leftrightarrow |4\rangle$ if the gravitational wave frequency matches the corresponding energy difference ($\ll \text{eV}$ since both $|3\rangle$ and $|4\rangle$ are Rydberg states). These extra Rabi oscillations split the absorption line that the probe laser experiences into two lines, though the effect is very small (quadratic in the gravitational wave strain). As first proposed in \cite{Jing:2020NatPh..16..911J} in a different context, and applied to the case of high-frequency gravitational wave detection in \cite{Kanno:2023whr}, detection prospects can be significantly enhanced by applying in addition to the probe and control lasers a third laser field (the ``local laser'') tuned to the $|3\rangle \leftrightarrow |4\rangle$ transition. The effect of the local laser is to split the absorption line even in absence of a gravitational wave; the presence of the gravitational wave then changes the separation between the two lines by a small amount. If the splitting induced by the local laser is of order the width of the lines, the change in transmissivity in between the lines becomes \emph{linear} in the strain.

The authors of Ref.~\cite{Kanno:2023whr} estimate that strain sensitivities of $h_0 \lesssim \text{few} \times 10^{-20}$ could be possible at a frequency of order \SI{30}{GHz}. In a hypothetical quantum noise-limited detector, this could be improved to $h_0 \lesssim 10^{-23}$, and with entangled Rydberg atoms, an improvement by a further three orders of magnitude may be possible.



\subsection{Cross-Correlating Multiple Detectors}
\label{sec:crosscorrelation}

The first observations of gravitational waves from coalescing binaries by the LIGO and VIRGO collaborations were performed using template matching techniques.  This was possible because, for a coalescing binary, the expected waveform of the GW can be modeled, so the observed data can be compared to a set of signal templates.

Unfortunately, a similar approach is unsuitable for many of the sources of interest to high-frequency GW searches.  For instance, many cosmological processes produce a GW signal that can be described as a superposition of a very large number of contributions. The waveform in this case is stochastic.  Even for sources that can be modeled deterministically, the number of free parameters is often too large to make template matching practical. The problem is that the set of expected signals does not have the structure of a linear space in the sense that the linear combination of two possible signals does in general \emph{not} produce a possible signal. For this reason the computational cost of a search over a template bank grows very fast with the number of free parameters.

With template matching impractical, high-frequency GW could still be detected as excess noise in the apparatus. A robust result of this kind would, however, require very good understanding of instrumental and environmental noise sources, which is typically not available.  A more promising discovery strategy is therefore the cross-correlation of data from multiple detectors \citep{Allen:1996vm, Michelson}. The basic principle is to compare the signal from two detectors. This means comparing a random signal with another stationary, stochastic, isotropic, Gaussian signal from the same source. Similar to template matching as a means of detecting discrete sources, in this case the template itself is random, and this affects the statistical gain from performing a cross correlation between two detectors.

\subsubsection{The Overlap Reduction Function}

In the cross-correlation approach, the quantity of interest for the detection and parameter estimation of a Gaussian stochastic GW background is the correlation between the strain at two different points $\mathbf{x}$ and $\mathbf{y}$. Focussing on GWs of frequency $f$,
\begin{multline}
    \big\langle h_{ij}^*(\mathbf{x},f) \, h_{k\ell}(\mathbf{y},f)^* \big\rangle \\
        \propto \sum_{a,a'} \int \! d\hat{\mathbf k} \, d\hat{\mathbf k'} \,
        \big\langle \tilde{h}_a^*(\hat{\mathbf{k}},f) \, \tilde{h}_{a'}(\hat{\mathbf{k}}',f) \big\rangle \,
        e_{ij}^a(\hat{\mathbf{k}}) e_{k\ell}^{a'}(\hat{\mathbf{k}}^\prime) \,
        e^{-2 \pi i f (\hat{\mathbf{k}}\mathbf{x}-\hat{\mathbf{k}}^\prime\mathbf{y})}
\end{multline}
We use here the notation from \cref{sec:Notation}, and a hat over the wave vectors $\mathbf{k}$, $\mathbf{k}'$ indicates a unit vector in the corresponding direction.

We see that the correlation is influenced by two effects. First, the detectors will not be in the same position, therefore the phase factor in the integral will oscillate, and the correlation will be reduced. This reduction becomes important when the separation $d$ between the two detectors is larger than the wavelength, $d \gtrsim 1/(2\pi f)$. A further reduction of the correlation can be caused if the two detectors are coupled differently to the GW modes, for example because they are oriented differently.

The reduction of correlation is quantified by the overlap reduction function $\Gamma(f)$ (see \cref{eq:SNR_SB}), a frequency dependent factor with modulus always less than one, which is simply the coherence between the two signal of interest. A derivation of $\Gamma(f)$ can be found in \cite{Michelson}, and \cite{Allen:1996vm} has outlined the process of optimizing the detection efficiency by optimal filtering in the time domain for two detectors with arbitrary separation and orientation.

An interesting possibility with small-scale detectors, like many of the setups envisioned for high-frequency GW detection, is to move detectors relative to one another, hence changing the value of the overlap function $\Gamma$. In this way the correlation of the signal can be modulated, and a detection of this modulation pattern could provide credible evidence of detection.

\subsubsection{Signal Switching}
\label{sec:switch}

As an alternative to cross-correlating signals from different detectors, it may be possible in high-frequency GW searches to turn the sensitivity of a single detector on and off without affecting its other performance properties. If the temporal pattern of switching on and off can be seen in the signal output at a statistically significant level then a credible claim for detection could be made with just a single detector.

Signal switching is possible, for instance, in the case of magnetic conversion detectors, where it is possible to modulate the amplitude of the field and its direction, thereby modulating the instrument's sensitivity. In addition, electromagnetic conversion detectors are sometimes filled with gas to counteract waveguide effects: if the transverse dimensions of the detector are comparable to, or smaller than, the GW wavelength, waveguide effects increase the phase velocity of the generated electromagnetic wave above the vacuum speed of light. This leads to a phase mismatch between the GW and the EM wave, an effect which the introduction of a gas with a sizable refractive index can counteract. By varying the gas pressure, one can then modulate the GW sensitivity of the device.

Note that, statistically, signal switching is a more effective correlation process than cross-correlating two similar detectors because for signal switching, the detector output is compared with an a priori determined template (namely the operational pattern) instead of a random one. The minimum detectable signal in this case is $\propto t_\text{int}^{-1/2}$ allowing a faster gain in sensitivity with time.

\subsubsection{Issues Related to Data Acquisition and Long Term Storage}

To detect correlated periodic events at GHz frequencies at a signal-to-noise ratio of 8, systematic errors related to timing should be of the order of \SI{20}{ps}. This necessitates very accurate timing calibration, a high DAQ sampling rate, and consequently significant data storage capacity of up to several petabytes.\footnote{Note that storage requirements are not proportional to the frequency of interest, but to the bandwidth, as the typical observation frequency can always be scaled down with an appropriate heterodyne technique.}

Timing calibration is challenging as low noise amplifiers, anti-aliasing electronics, and other components of the readout add delay to the data acquisition system. Moreover, quantization errors from the analog-to-digital converters add further bias, and to minimize this effect, the sampling rate would need to be higher than the desired timing resolution. At such sampling rates, making use of super-conducting oversampling ADCs which achieve high dynamic ranges over narrow frequency ranges by pushing quantization noise outside the band-of-interest could turn out to be viable option.

To avoid excessive storage requirements, real-time analysis (as proposed for instance for the Square Kilometer Array (SKA) radio telescope) could be an option. In this approach, the raw data is discarded after the low latency retrieval of relevant information. 

In general, it is reassuring that the combined analysis of time-series data from several sources has been successfully demonstrated by various radio astronomy collaborations. Utilizing cloud storage and grid computing tools, the handling of large datasets seems of no major concern. In addition, data folding techniques based on inherent symmetries, such as the Earth's siderial rotation, have proven effective to decrease data volume in stochastic background searches at audio-band frequencies \citep{PhysRevD.92.022003}. Stacking years of data into a single day while preserving all the statistical properties would even make it possible to carry out the final analysis on personal computing devices.

\subsubsection{Coincidence Counting Experiments}

We have seen in \cref{sec:exp} that single photon detection capabilities in the GHz range could boost the sensitivity of many planned and proposed instruments. Even though in such setups, no continuous time series data is recorded, correlations between multiple detectors can still be exploited. Every detected photon can be timestamped with $\mu s$ resolution or better, allowing for offline coincidence searches involving any number of experiments. This approach is in particular interesting for short transient signals like those from PBH mergers.

For a given coincidence time interval $\tau$, determined by the duration of the signal or by the detector's time resolution, the overall efficiency $\epsilon_{tot}$ can be adjusted via the number of required coincidences $k$ in a system of $N$ detectors:
\begin{align}
    \epsilon_{\rm tot} = \sum_{i \ge k} \binom{N}{i} \epsilon_{\rm det}^i
                         (1 - \epsilon_{\rm det})^{N-i},
\end{align}
where $\epsilon_{\rm det}$ is the probability for each individual detector to see a signal photon. The corresponding rate of accidental coincidences, $R_{\rm acc}$, depends on the dark count rate $R_{\rm dark}$ in an individual detector and is given by:
%
%
\begin{align}
    R_{\rm acc} = \tau^{-1} \sum_{i \ge k} \binom{N}{i} (\tau R_{\rm dark})^i \,
                         \big[1 - (\tau R_{\rm dark})\big]^{N-i},
\end{align}
Note that the coincidence counting approach also allows the combination of information from several narrow band detectors operating at different frequencies if the time evolution of the signal frequency spectrum is known.

\subsection{Summary of Detector Sensitivities}
\label{sec:SummarySensitivities}

Below we summarize the existing and proposed technologies for high-frequency GW detection, referring to the sections above for details.
We also specify the maturity of each technology, that is, whether the experiment has been built, is under active R\&D, or only the physical mechanism has been identified theoretically.
In the frequency column, square brackets indicate a range of frequencies that can be scanned in the case of resonant detectors, whereas round brackets indicate the bandwidths of broadband detectors. Entries marked with a star (*) correspond to setup for which we consider both resonant and broadband operation. Table~\ref{tab:signal_sensing_resonator} gives an overview of the different concepts by technical approach, stating the signal and sensing process used, and what kind of resonant enhancement the setup provides.

\begin{landscape}
\begin{center}
    \captionsetup{type=table} 
    \caption[Summary of existing and proposed detectors and their frequency ranges.]
    {Summary of existing and proposed detectors and the frequency ranges in which they are sensitive. See \cref{sec:SummarySensitivities} for details.}
    \label{tab:SummarySensitivity}
    
    \begin{footnotesize}
    \def\arraystretch{1.1}
    \begin{longtable}{@{}l@{}c@{}}
        \toprule
        \normalsize\textbf{Technology} & \normalsize\textbf{Operational Frequency} \\
        \midrule
        \endhead
        \textbf{Laser Interferometers}, \cref{sec:ozgrav}, \cref{sec:100MHzInterferometers} \\
        NEMO (R\&D), \cite{ozhf, Bailes:2019oma} & $(1-2.5)$\,kHz \\
        \SI{0.75}{m} interferometer (built), \cite{akutsu, Nishizawa:2007tn} & \SI{100}{MHz} \\
        Holometer (built) \cite{PhysRevD.95.063002} & $(1-13)$\,MHz \\
        Twin table-top 3D interferometers (built), \cite{Patra:2024thm} & $(1-250)$\,MHz \\
        \midrule
        \textbf{Spherical Resonant Mass Detectors}, \cref{sec:spheres}, \cite{Forward1971} \\ [0.4ex]
        Mini-GRAIL (built), \cite{Gottardi:2007zn} & \SI{2942.9}{Hz} \\
        Schenberg antenna (built), \cite{Aguiar:2010kn} & \SI{3.2}{kHz} \\
        \midrule
        \textbf{Optically Levitated Sensors}, \cref{sec:OpticallyLevitatedSensors}, \cite{arvanitaki:2016gw} \\
        Levitated Sensor Detector $1$-meter prototype (R\&D), \cite{aggarwal2020searching} & $[10-100]$\,kHz \\
        Levitated Sensor Detector $100$-meter instrument (proposed), \cite{aggarwal2020searching} & $[10-100]$\,kHz \\
        \midrule
        \textbf{Bulk Acoustic Wave Resonators}, \cref{sec:BAW} \\
        Goryachev's detector (built), \cite{Goryachev:2014ab} & $[1-1000]$\,MHz \\
        \midrule
        \textbf{Deformation of Microwave Cavities}, \cref{sec:HighQCavities}, \cite{Pegoraro, Pegoraro:1977uv, Berlin:2023grv} \\
        Caves' detector (proposed), \cite{Caves} & \SI{500}{Hz} \\
        Reece's 1st detector (built), \cite{Reece} & \SI{}{MHz} \\
        Reece's 2nd detector (built), \cite{Reece:1982sc} & \SI{10}{GHz} \\
        Pegoraro's detector (proposed), \cite{Pegoraro} & $[1-10]$\,GHz \\
        DESY/UHH--FNAL collaboration (R\&D), \cite{Fischer:2024nte} & (kHz--GHz) \\
        \midrule
        \textbf{Magnetic Weber Bar}, \cref{sec:MWB}, \cite{Domcke:2024mfu, Carney:2024zzk} & (\SI{10}{kHz}--\SI{1}{MHz}) \\
        \midrule
        \pagebreak
        \textbf{RF Cavities}, \cref{sec:SRFcavities}, \cite{Berlin:2021txa} \\
        ADMX (built), \cite{ADMX:2021nhd} & $[0.65-1.02]$\,GHz \\
        HAYSTAC (built), \cite{HAYSTAC:2018rwy} & $[5.6 -5.8]$\,GHz \\
        CAPP (built), \cite{CAPP:2020utb} & $[1.6 - 1.65]$\,GHz \\
        ORGAN (built), \cite{Quiskamp:2022pks, Quiskamp:2023ehr} & $[15 - 16]$, $[26 - 27]$\,GHz \\
        SQMS (R\&D), \cite{Posen-talk} & $[1-2]$\,GHz \\
        Cubic cavities 1, 2, 3 (R\&D), \cite{Navarro:2023eii} & 0,1, 1, 10\,GHz \\
        \midrule
        \textbf{LC-circuit Axion Haloscopes}, \cref{sec:AxionHaloscopes}, \cite{Domcke:2022rgu, Domcke:2023bat} \\
        ABRACADABRA (built), \cite{Salemi:2021gck} & ($ 0.1 - 2 $)\,MHz \\
        SHAFT (built), \cite{Gramolin:2020ict} & ($\SI{3}{kHz} - \SI{3}{MHz}$) \\
        ADMX SLIC (built), \cite{Crisosto:2019fcj} & \SI{0.043}{GHz} \\
        BASE (built), \cite{Devlin:2021fpq} & \SI{0.4}{MHz} \\
        WISPLC (R\&D), \cite{Zhang:2021bpa} & ($0.03 - 5$)\,MHz \\
        DMRadio-\si{m^3} (R\&D), \cite{DMRadio:2022pkf} & $[5 - 200]^{*}$\,MHz \\
        DMRadio-GUT (R\&D), \cite{DMRadio:2022jfv} & $[0.1 - 30]^{*}$\,MHz \\
        \midrule
        \textbf{Light Shining through a Wall}, \cref{sec:ALPs_and_axion_helioscopes}, \cite{Ejlli:2019bqj, Ringwald:2020ist} \\
        OSQAR I (built), \cite{OSQAR:2015qdv} & $(0.3-0.8) \times 10^6$\,GHz \\
        OSQAR II (built), \cite{OSQAR:2015qdv} & $(0.3-1) \times 10^6$\,GHz \\
        ALPS I (built), \cite{Bahre:2013ywa} & $(0.3-1) \times 10^6$\,GHz \\
        ALPS II (built), \cite{Bahre:2013ywa, Albrecht:2020ntd} & $[0.3-1]^{*} \times 10^{6}$\,GHz \\
        JURA (proposed), \cite{Beacham:2019nyx} & $[0.3-1]^{*} \times 10^6$\,GHz \\
        \midrule
        \textbf{Axion Helioscopes}, \cref{sec:ALPs_and_axion_helioscopes}, \cite{Ejlli:2019bqj, Ringwald:2020ist}  \\
        CAST (built), \cite{CAST:2017uph} & $(0.5-1.6) \times 10^{9}$\,GHz \\
        BabyIAXO (R\&D), \cite{IAXO:2020wwp} & $(0.25-2) \times 10^{9}$\,GHz \\
        IAXO (R\&D), \cite{IAXO:2019mpb} & $(0.25-2) \times 10^{9}$\,GHz \\
        \midrule
        \pagebreak
        \textbf{Dielectric Axion Haloscopes}, \cref{sec:Dielectric_axion_haloscopes}, \cite{Domcke:2024eti} \\
        Madmax (R\&D), \cite{MADMAX:2019pub} & $[\SI{100}{MHz} - \SI{10}{GHz}]^{*}$ \\
        DALI prototype (R\&D), \cite{DeMiguel:2023nmz} & 7, 33\,GHz \\
        DALI phase II (proposed), \cite{DeMiguel:2023nmz} & ($6-60$)\,GHz \\
        \midrule
        \textbf{High Energy Pulsed Lasers}, \cref{sec:PulsedLasers}, \cite{Vacalis:2023gdz} & $[10^4 - 10^{10}]$\,GHz \\
        \midrule
        \textbf{Conversion in a Static Electric Field}, \cref{sec:ConversionEfield} \\
        Atomic electric field, \cite{Dai:2023rgx} & $(10^{11} - 10^{15})$\,GHz \\
        \midrule
        \textbf{Resonant Polarization Rotation}, \cref{sec:ConversionEfield}, \cite{Cruise1} & \\
        Cruise's detector (proposed), \cite{Cruise2} & $[0.1 - 10^5]$\,GHz \\
        Cruise \& Ingley's detector (prototype), \cite{Cruise3, Cruise:2006zt} & \SI{100}{MHz} \\
        Optical cavities of ALPS II (built), \cite{Garcia-Cely:2025mgu} & $[\SI{0.1}{MHz} - \SI{0.1}{THz}]$ \\
        \midrule
        \textbf{Superconducting Rings}, \cref{sec:SuperconductingRings}, \cite{Anandan:1982is, Chiao:2002nv} & \SI{10}{GHz} \\
        \midrule
        \textbf{Graviton--Magnon Resonance}, \cref{sec:GravitonMagnonResonance}, \cite{Ito:2019wcb, Ito:2022rxn} & $[8-14]$\,GHz \\
        \midrule
        \textbf{Atomic Precision Measurement}, \cref{sec:AtomicMeasurement}, \cite{Bringmann:2023gba} & $[\SI{10}{kHz} - \SI{10}{GHz}]$ \\
        \midrule
        \textbf{One-Electron Quantum Cyclotron}, \cref{sec:oneelectron}, \cite{Ito:2022rxn} &
$[20 - 200]$\,GHz \\
        \midrule
        \textbf{Rydberg Atoms}, \cref{sec:RydbergAtom}, \cite{Kanno:2023whr} & $[0.3 - 16]$\,GHz \\
        \bottomrule
    \end{longtable}
    \addtocounter{table}{-1}
    \end{footnotesize}
    \def\arraystretch{1.0}
\end{center}

\clearpage
\begin{center}
    \captionsetup{type=table} 
    \caption{Overview of the different detection concepts for high-frequency gravitational waves by technical approach.}
    \label{tab:signal_sensing_resonator}

    \setlength{\usableWidth}{\dimexpr\linewidth-\headheight-\footskip\relax}
    \newcolumntype{P}[1]{>{\raggedright\arraybackslash}p{#1}}
    \begin{footnotesize}
    \begin{longtable}{P{.24\usableWidth}P{.24\usableWidth}P{.24\usableWidth}P{.24\usableWidth}}
        \toprule
        \normalsize\textbf{Technology} & \normalsize\textbf{Signal}
            & \normalsize\textbf{Sensing} & \normalsize\textbf{Resonator} \\
        \midrule
        \endhead
        \textbf{Laser Interferometers}
            & movement of test masses (mirrors)
            & interferometric monitoring of test mass positions
            & optical cavity \\
        \midrule
        \textbf{Spherical Resonant Mass}
            & deformation of test mass
            & capacitive or superconducting electromechanical transducers
            & vibrational eigenmodes of test mass \\
        \midrule
        \textbf{Optically Levitated Sensors}
            & movement of levitated nanoparticle
            & interferometric measurement of levitated sensor and mirror movement
            & resonance with trapping frequency of levitated sensor \\
        \midrule
        \textbf{Bulk Acoustic Wave Resonators}
            & deformation of test mass
            & electromechanical transducer
            & vibrational eigenmodes of test mass \\
        \midrule
        \textbf{Deformation of Microwave Cavities}
            & electromagnetic mode mixing
            & power in empty cavity mode
            & resonant energy transfer to cavity eigenmode \\
        \midrule
        \textbf{Magnetic Weber Bar}
            & deformation of superconducting coils
            & SQUID-based sensing of oscillating magnetic field
            & vibrational eigenmodes of coils, optionally resonant LC circuit \\
        \midrule
        \textbf{SRF Cavities}
            & induced effective current in magnetic field
            & power in empty cavity mode
            & electromagnetic cavity eigenmodes \\
        \midrule
        \textbf{LC-circuit Axion Haloscopes}
            & induced current
            & SQUID-based low current sensing
            & resonant LC circuit \newline (or none for broadband) \\
        \midrule
        \textbf{Light Shining through a Wall, Axion Helioscopes, Dielectric Haloscopes }
            & magnetic conversion
            & heterodyne, correlation, single photon counting
            & optical cavity eigenmodes \newline
              (or none for broadband) \\
        \midrule
        \textbf{Astrophysical and Cosmological Detection}
            & magnetic conversion
            & radio, IR, optical, UV, X-ray, $\gamma$-ray telescopes
            & none \\
        \midrule
        \textbf{Superconducting Rings}
            & Josephson current induced by GW--spin interaction
            & electromechanical transducer
            & none \\
        \midrule
        \textbf{Graviton–-Magnon Resonance}
            & magnon excitation in ferromagnet
            & coupling the magnon to an eigenmode of a microwave cavity
            & magnon modes \\
        \midrule
        \textbf{Atomic Precision Measurement}
            & modulation of photon frequencies
            & optical atomic clock readout protocol
            & none \\
        \midrule
        \textbf{One-Electron Quantum Cyclotron}
            & excitation of electron in Penning trap
            & quantum jump spectroscopy
            & Penning trap cyclotron modes \\
        \midrule
        \textbf{Rydberg Atoms}
            & electromagnetically induced transparency
            & absorption spectroscopy
            & atomic transition \\
        \bottomrule
    \end{longtable}
    \addtocounter{table}{-1}
    \end{footnotesize}
\end{center}

\end{landscape}


\section{Discussion and Conclusions}
\label{sec:conclusion}

The search for high-frequency gravitational waves is a promising and challenging endeavor. Given the scarcity of astrophysical sources at frequencies $\gtrsim \si{kHz}$, it offers in particular unique opportunities to test theories beyond the Standard Model that could not be tested otherwise.

In fact, numerous models proposed to address open questions in particle physics and cosmology predict gravitational-wave signals in the frequency range $f \simeq (10^3 - 10^{10})$\,Hz. These can be coherent signals, for example from mergers of sub-solar mass compact objects or from axion superradiance around black holes; or they can be stochastic signals, for instance from certain models of cosmic inflation or from first-order phase transition in the very early Universe. In the latter case, physics at higher energies, or equivalently earlier cosmological epochs, corresponds to higher gravitational wave frequencies and correspondingly smaller experimental devices. As can be seen from \cref{fig:PLS}, the ultra-high frequency band, ranging from MHz to GHz, is an exciting window to explore fundamental physics up to the grand unification or string theory scales of order $(10^{16}-10^{17})$\,GeV. It would be remarkable if the experimental test of fundamental physics at the highest energies and of the earliest times in the history of the Universe could eventually be achieved not with huge particle accelerators or with satellite interferometry, but with small table-top experiments.

Many of cosmological gravitational wave sources can lead to relatively large signals corresponding to an ${\cal O}(1)$ fraction of the energy density in the early Universe being converted to gravitational waves. This energy is red-shifted in the expanding Universe, rendering even these strong signals challenging to detect today. Moreover, in many cases the amplitude of the signal depends sensitively on the model parameters and may be significantly lower in large parts of the model parameter space. In \cref{sec:th} of this review, we have given an overview of high-frequency gravitational wave sources, and a concise summary can be found in \cref{tab:summary-coherent,tab:summary-stochastic}. 

The high-frequency band comes with particular challenges and opportunities. High-frequency gravitational waves carry a high energy density, implying that cosmological bounds  on the energy density in relativistic species translate to stringent bounds on the characteristic gravitational-wave strain. This poses a severe challenge for detection, as the magnitude of observable effects is typically governed by the strain and not by the energy density. The detection of cosmological sources of high-frequency gravitational waves is therefore much more challenging than comparable searches at lower frequencies. On the other hand, the lack of known astrophysical gravitational-wave sources in this frequency range presents a unique opportunity for foreground free searches for new physics.

At the moment, there is no general consensus on the most promising detection strategy in this frequency band, though many proposals have been put forward in the past decades. The proposals that we are aware of are summarized in \cref{tab:SummarySensitivity}, and their sensitivities are plotted in \cref{fig:sens_HF,fig:sens_VHF}. We emphasize that a given sensitivity in terms of noise equivalent strain at a higher frequency typically implies a reduced sensitivity to the viable parameter space of a given cosmological source than at lower frequency. Detectors based on magnetic conversion seem to be particularly promising avenues at very high frequencies (above $\sim \si{GHz}$) while relying on mechanical coupling of GWs seems advantageous at lower frequencies. It should be kept in mind, however, that more careful studies of noise levels and of the margin of improvement with foreseeable technology development is needed in many cases. We hope that this document will stimulate the necessary discussion. 

None of the detection concepts listed in this report currently reach the sensitivity needed to probe realistic sources. Even under optimistic assumptions, they fall short by at least several orders of magnitude. However, we recall that, one hundred years ago, the technological gap in strain sensitivity in both the LIGO and LISA frequency ranges was about 15--16 orders of magnitude \citep{Chen:2016isk}. Also, about 50 years ago, Misner, Thorne and Wheeler, declared that `\emph{such detectors have so low sensitivity that they are of little experimental interest}' \citep{Misner:1974qy}, referring to laser interferometers. The first laser interferometer gravitational-wave detector, built at Hughes Research Laboratories in the 1970s \citep{PhysRevD.17.379} had a sensitivity which was eight orders of magnitudes below the design sensitivity of the currently operating LIGO/Virgo/KAGRA detectors. Today, there are clear development paths towards detectors with sensitivities of $\sqrt{S_n} \simeq 10^{-38}\text{Hz}^{-1/2}$ using, e.g., magnetic conversion at optical frequencies (see \cref{sec:PhotonRegen}).

We therefore take the past history of laser interferometry as an encouraging lesson for the development of gravitational-wave detectors in the high-frequency band. The challenges are formidable, but the opportunities and potential rewards are unique.

This white paper sets the stage for the launch of the Ultra-High-Frequency Gravitational Wave (UHF-GW) initiative\footnote{\url{http://www.ctc.cam.ac.uk/activities/UHF-GW.php}.}, a network of researchers with the common goal of further pushing the boundaries of gravitational-wave science in the high-frequency range and to collaboratively work towards the long-term goal of a first detection of a signal in this frequency range.

we strongly encourage feedback regarding additional sources or detection techniques which we may have missed, as well as critical assessments of the ones presented here.

\newpage
\begin{acknowledgements}
\label{sec:acknowledgements}
We thank ICTP Trieste and CERN TH for hosting the first three editions of ultra-high frequency gravitational wave workshops which were key for providing input and organizing this Living Review and its update.
A special thank you to all speakers and participants of these workshops, whose contributions were key in shaping this report.
We moreover thank Masha Baryakhtar, William Campbell, Yiwen Chu, Virgile Dandoy, Aldo Ejlli,  Javier de Miguel, Juan Garcia-Bellido, Ken-Ichi Herada, Alessandro Lella, Axel Lindner, Sotatsu Otabe, Christoph Reinhardt, Seyed Mohammad Sadegh Movahed, Mikel Sanchez, Tommaso Tabarelli, Yutong He for valuable input.
We also thank the referees of \textit{Living Reviews in Relativity} for their supportive and well-thought reports.\\
The Australian High-Frequency Gravitational Wave Effort is supported by the Australian Research Council Centre of Excellence for Gravitational Wave Discovery (OzGrav), Grant number CE170100004. 
N.A. is supported by NSF grant PHY-1806671 and a CIERA Postdoctoral Fellowship from the Center for Interdisciplinary Exploration and Research in Astrophysics at Northwestern University.
A.B. acknowledges support by the European Research Council (ERC) under the European Union's Horizon 2020 research and innovation programme under grant agreement No. 759253 and by Deutsche Forschungsgemeinschaft (DFG, German Research Foundation) - Project-ID 279384907 - SFB 1245 and - Project-ID 138713538 - SFB 881 (`The Milky Way System', subproject A10).
O.D.A. thanks FAPESP / Brazil (grant numbers 1998/13468-9 and 2006/56041-3) and CNPq/Brazil (grants numbers 306467{\textunderscore}2003{\textunderscore}8, 303310{\textunderscore}2009-0, 307176{\textunderscore}2013-4, and 302841/2017-2).
This project has received funding from the Deutsche Forschungsgemeinschaft under Germany's Excellence Strategy\,--\,EXC 2121 `Quantum Universe', 390833306 (V.D., F.M., K.P., A.R.), and
EXC 2118 'Precision Physics, Fundamental Interactions and Structure of Matter' (PRISMA+), 390831469 (J.K, C.T).
D.G.F. (ORCID 0000-0002-4005-8915) is supported by a Ram\'on y Cajal contract by Spanish Ministry MINECO, with Ref. RYC-2017-23493, and by the grant `SOM: Sabor y Origen de la Materia', from Spanish Ministry of Science and Innovation, under no. FPA2017-85985-P.
A.G. is supported in part by NSF grants PHY-1806686 and PHY-1806671, the Heising-Simons Foundation, the W.M. Keck Foundation, the John Templeton Foundation, and ONR Grant N00014-18-1-2370.
M.G. and M.E.T. were funded by the ARC Centre for Excellence for Engineered Quantum Systems, CE170100009, and the ARC Centre for Excellence for Dark Matter Particle Physics, CE200100008, as well as ARC grant DP190100071. F.M. is funded by a UKRI/EPSRC Stephen Hawking fellowship, grant reference EP/T017279/1. This work has been partially supported by STFC consolidated grant ST/P000681/1. A.R. acknowledges funding from Italian Ministry of Education, University and Research (MIUR) through the `Dipartimenti di eccellenza' project Science of the Universe. S.S.  was supported by MIUR in Italy under Contract(No. PRIN 2015P5SBHT) and ERC Ideas Advanced Grant (No. 267985) `DaMeSyFla'. F.T. acknowledges support from the Swiss National Science Foundation (project number 200020/175502). C.U. is supported by European Structural and Investment Funds and the Czech Ministry of Education, Youth and Sports (Project CoGraDS - CZ.02.1.01/0.0/0.0/15\textunderscore003/0000437) and partially supported by ICTP.
D.B. acknowledges the support from the Departament de Recerca i Universitats from Generalitat de Catalunya to the Grup de Recerca 00649 (Codi: 2021 SGR 00649). The research leading to these results has received funding from the Spanish Ministry of Science and Innovation (PID2020-115845GB-I00/AEI/10.13039/501100011033). This publication is part of the grant PID2023-146686NB-C31 funded by MICIU/AEI/10.13039/501100011033/ and by FEDER, UE.
IFAE is partially funded by the CERCA program of the Generalitat de Catalunya.  J.M. acknowledges support form the Science and Technology
Facilities Council (STFC) [Grant No. ST/X00077X/1].  C.G.C. is supported by a Ramón y Cajal contract with Ref.~RYC2020-029248-I, the Spanish National Grant PID2022-137268NA-C55 and Generalitat Valenciana through the grant CIPROM/22/69.
D.B. and V.D.  acknowledge the support by the European Research Area (ERA)
via the UNDARK project (project number 101159929).
\end{acknowledgements}

\newpage
\appendix
\section{Electromagnetic Signals Generated by GWs}
\label{sec:EMsignals}

\normalsize
We have seen in \cref{sec:exp} that many promising detection techniques for high-frequency GWs rely on graviton-to-photon conversion in a magnetic field. Here, we review several calculation methods relevant to such signals.

The starting point are Maxwell's equations in curved spacetime~\citep{Landau:1975pou}
\begin{align}
    \begin{split}
        \nabla_\nu F_{\alpha\beta}
      + \nabla_\alpha F_{\beta\nu}
      + \nabla_\beta F_{\nu\alpha}
        &= \partial_\nu F_{\alpha\beta}
         + \partial_\alpha F_{\beta\nu}
         + \partial_\beta F_{\nu\alpha}
         = 0\,, \\
        \partial_\nu \left(\sqrt{-g} F^{\mu\nu} \right)
        &= \sqrt{-g} \, j^{\mu} \,.
    \end{split}
    \label{eq:covariantinhomN}
\end{align}
Here, as usual, $g_{\mu\nu} = \eta_{\mu\nu} + h_{\mu\nu}$ is the metric, separated into the Minkowski part and the perturbation. The field and the current may be separated accordingly as
\begin{align}
    F_{\mu\nu} = F_{0\mu\nu} + F_{h\mu\nu} + {\cal O}(h^2)
    \qquad\text{and}\qquad
    j^\mu = j_0^\mu + j_h^\mu + {\cal O}(h^2)\,.
 \label{eq:FinFext}
\end{align}
Here subscript $0$ represents quantities in the absence of GWs, and the subscript $h$ indicates terms linear in the metric perturbation. The current $j_h^\mu$ may be attributed to  the effect of the GW on the motion of electric charges. At GW frequencies higher the mechanical eigenfrequencies associated with the experimental apparatus, the system is effectively in free fall \citep{Bringmann:2023gba, Ratzinger:2024spd}, and it is hence convenient to adopt the transverse-traceless (TT) gauge for the GWs. In these coordinates, the system remains at rest while the GW passes, so we can neglect the effect of the GW on external currents, that is, $j_h^\mu = 0$.

For concreteness, let us consider a $+$-polarized GW propagating in the $x$-direction through a region of length $L$ with a uniform magnetic field $B_0$ pointing in the $z$-direction:
\begin{align}
    h_{\mu\nu} &= h_+ \, \big(\delta_{\mu 2} \delta_{\nu 2}
                            - \delta_{\mu 3} \delta_{\nu 3} \big)
                      e^{-i \omega (t-x)} + \text{c.c.} \,, \\
  F_{0\mu\nu} &= \begin{cases}
                     B_0 \, \big( \delta_{\mu 1} \delta_{\nu 2}
                                - \delta_{\mu 2} \delta_{\nu 1} \big)
                                & |x|< \frac{L}{2} \\
                     0          & \text{otherwise}
                 \end{cases} \quad .
    \label{eq:hwave}
\end{align}
This situations corresponds to an external current, $j_{0}^\mu = \partial_\nu F_0^{\mu\nu}$ consisting of two Dirac-$\delta$ peaks at $x=\pm L/2$, thereby sourcing the external field in the region $|x|< L/2$. The solution of \cref{eq:covariantinhomN} for any $x$ with the appropriated boundary conditions can be readily found as
\begin{align}
    F_{h \mu\nu} = \begin{cases}
        \frac{1}{4} \big( -F_{\mu\nu}^{(E)} + F_{\mu\nu}^{(B)} \big)
            \big(e^{i \omega L} - e^{-i \omega L}\big) e^{-i \omega (t + x)}
            + \text{c.c.}
                    & x < -\frac{L}{2} \\[1.5ex]
        \frac{1}{4} \big( -F_{\mu\nu}^{(E)} + F_{\mu\nu}^{(B)} \big)
            e^{i \omega L} e^{-i \omega(t+x)} \\[0.5ex]
            \quad + \frac{1}{4} \big[ F_{\mu\nu}^{(E)}
                    + i \omega L \big(1 + \tfrac{2x}{L} \big)
                                 \big(F_{\mu\nu}^{(E)} + F_{\mu\nu}^{(B)} \big) \\[0.5ex]
            \hspace{4cm}
                             + 3 F_{\mu\nu}^{(B)} \big] e^{-i \omega(t-x)}
                    + \text{c.c.}
                    & |x| < \frac{L}{2} \\[1.5ex]
        \frac{i}{2} \big(F_{\mu\nu}^{(E)} + F_{\mu\nu}^{(B)} \big)
                    \omega L e^{-i \omega (t-x)} + \text{c.c.}
                    & x > \frac{L}{2} \\
    \end{cases}
\end{align}
where $F_{\mu\nu}^{(E)} \equiv h_+ B_0 (\delta_{\mu 2} \delta_{\nu 0} - \delta_{\mu 0} \delta_{\nu 2})$ and $F_{\mu\nu}^{(B)} = h_+ B_0 (\delta_{\mu 1} \delta_{\nu 2} - \delta_{\mu 2} \delta_{\nu 1})$.
It follows that an electromagnetic signal is generated even at $|x| > L/2$, where the external field vanishes. In these regions, the electromagnetic energy-momentum tensor
\begin{align}
    {\cal T}^{\mu\nu} = \Big( g^{\mu\rho} g^{\nu\sigma}
                            - \frac{1}{4} g^{\mu\nu} g^{\rho\sigma} \Big)
                        g^{\alpha\beta} F_{\rho\alpha} F_{\sigma\beta} \,,
    \label{eq:Tem}
\end{align}
averaged over several periods of the signal, is
\begin{align}
    \langle {\cal T}^{\mu\nu} \rangle \big|_{x>\frac{L}{2}}
        &= \frac{1}{2} h_+^2 B_0^2 \omega^2 L^2
          \begin{pmatrix}
              1 & 1 & 0 & 0 \\
              1 & 1 & 0 & 0 \\
              0 & 0 & 0 & 0 \\
              0 & 0 & 0 & 0
          \end{pmatrix} \\
    \langle {\cal T}^{\mu\nu} \rangle \big|_{x<-\frac{L}{2}}
        &= \frac{1}{2} h_+^2 B_0^2 \sin^2(\omega L)
          \begin{pmatrix}
              1 & -1 & 0 & 0 \\
             -1 &  1 & 0 & 0 \\
              0 &  0 & 0 & 0 \\
              0 &  0 & 0 & 0
          \end{pmatrix} \,.
    \label{eq:probabilities0}
\end{align}
The probability of GW conversion into photons can be calculated by taking the ratio of the Poynting vector (the off-diagonal component of $\langle {\cal T}^{\mu\nu} \rangle$) and the flux of gravitational waves, $h_+^2 \, \omega^2/(8\pi G)$. In the limit $\omega L \gg 1$, one finds
\begin{align}
    P \big|_{x>\frac{L}{2}} = 4 \pi G B_0^2 L^2\,,
\label{eq:probabilities}
\end{align}
and $P\big|_{x<-\frac{L}{2}} \simeq 0$. Using the same method, one finds the same conversion probability for $\times$-polarized GWs. This is the inverse Gertsenshtein effect, discussed first in \cite{Gertsenshtein_1962_grav_radiation} (see also \cite{Boccaletti1970}), using an approach very similar to the one just described.

These results can also be derived in other ways without directly solving Maxwell's equations in curved spacetime:

\paragraph{Effective Current Approach}

This method consists of recasting \cref{eq:covariantinhomN} as a standard electrodynamics problem in Minkowski spacetime, with an effective current sourcing the field $F_{\mu\nu}$. Concretely,  in the TT frame, $\sqrt{-g} \simeq {\cal O}(h^2)$  and the expression in parenthesis in the inhomogeneous Maxwell equation (second line of \cref{eq:covariantinhomN}) can be written as
\begin{align}
    \sqrt{-g} \, g^{\alpha\mu} F_{\alpha\beta} \, g^{\beta\nu}
        \simeq F^{\mu\nu} + F_h^{\mu\nu} - h^{\alpha\mu}{F_{0\alpha}}^\nu
             - {{F_0}^\mu}_{\beta } h^{\beta\nu} + {\cal O}(h^2) \,,
\end{align}
where we have raised indices with $\eta$, as we will continue to do. Employing \cref{eq:FinFext} and taking $j_h^\mu = 0$ in the TT frame as explained above, one finds
\begin{align}
    \partial_\nu F_h^{\mu\nu} = j_\text{eff}^\nu \,, 
    \qquad\text{with}\qquad
    j_\text{eff}^\nu \equiv \partial_\nu \big(F_0^{\mu\alpha} {h^\nu}_{\alpha}
                                   - F_0^{\nu\alpha} {h^\mu}_{\alpha} \big) \,.
\end{align}
Note that there are fewer terms here than in \cref{eq:effective_current} in \cref{sec:EMoscillators} because in the TT frame ${h_\mu}^{\mu} = 0$. The effective current can also be expressed as \citep{Domcke:2022rgu}
\begin{align}
    j^\mu_\text{eff} = \big(-\nabla \cdot \mathbf{P}, \,
                       \nabla \times \mathbf{M} + \partial_t \mathbf{P} \big) \,,
\end{align}
introducing the effective polarization, $P_i \equiv -h_{ij} E_{0j}$, and magnetization, $M_i \equiv -h_{ij} B_{0j}$.  As an example, for the specific case of \cref{eq:hwave}, ${\bf P} = 0$ and ${\bf M} = 2 h_+ B_0 \cos(k t - k x) \Theta(L/2 - |x|) \hat{\bf z}$. Imposing the appropriated boundary condition, this oscillating magnetization leads to the induced field $F_{h\mu\nu}$ reported above, and consequently to the conversion probability in \cref{eq:probabilities}. This method readily generalizes to other GW frames beyond the TT gauge, for instance the proper detector frame, and is particularly convenient for studying complicated setups such as resonant cavities, low-mass axion haloscopes or dielectric haloscopes \citep{Berlin:2021txa, Domcke:2022rgu, Domcke:2023bat, Domcke:2024eti}.

\paragraph{$S$-Matrix Approach}

This method exploits the fact that  GWs couple to the energy-momentum tensor of the electromagnetic field. Concretely, this coupling is given by
\begin{align}
    {\cal L} \supset \frac{1}{2} h_{\mu\nu} {\cal T}^{\mu\nu} 
    = h_{\mu\nu} {{F_h}^\nu}_\alpha F_0^{\mu\alpha} + \ldots \,,
    \label{eq:lag}
\end{align}
where we have used \cref{eq:FinFext,eq:Tem}.
This permits an interpretation of the Gertsenshtein effect in terms of Feynman diagrams as conversion of a gravitational perturbation into a photon as it scatters off an external electromagnetic field.

It is easy to see that the $S$-matrix approach yields the same conversion rates as the effective current method by noting that ${\cal L}$ can be rewritten as ${\cal L} \supset A_\mu j_\text{eff}^\mu + \ldots$, which follows from writing $F_{\mu\nu} = \partial_\mu A_\nu - \partial_\nu A_\mu$ in \cref{eq:lag} and integrating by parts. However, the $S$-matrix approach does not provide the induced electromagnetic field, $F_{h\mu\nu}$, at each point in space. Instead, it gives the probability amplitude at large distances, or equivalently, the scattering cross sections.

This method was employed in the seminal paper \cite{Raffelt:1987im} to calculate \cref{eq:probabilities} and to point out the close analogy with axion--photon conversion. The method has been used to calculate graviton-to-photon conversion rates for various external field configurations, including both uniform and dipole electric and magnetic fields, including setup from \cref{eq:probabilities0} above \citep{DeLogi:1977qe}.

\paragraph{Geometric Optics.}

It is known that in a slowly varying background Maxwell's equations admit solutions that correspond to geometric optics in classical electrodynamics. Exploiting the analogy between axions and GWs and allowing for a plasma mass, this has been recently studied for GWs in \cite{McDonald:2024uuh}. This approach yields the conversion probabilities given in \cref{eq:probabilities} in the limit of very high frequencies. Notably, the geometric optics method has been employed recently to calculate conversion rates in the magnetospheres of neutron stars \citep{McDonald:2024nxj}, see also \cref{sec:AstroDetectors}. Its advantage in this context is that it can account for three-dimensional effects that extend beyond the approximations presented in the classical work \cite{Raffelt:1987im}. Moreover, as shown by these authors, the geometric optics limit enables the investigation of polarization effects.

\vspace{1ex}
Finally, it is important to emphasize that while the formal separation of the electromagnetic field and current in \cref{eq:FinFext} is generally straightforward for specific experimental setups, its interpretation requires caution as it is neither coordinate invariant nor gauge invariant \citep{Ratzinger:2024spd}. This difficulty is exacerbated by the existence of multiple methods for calculating a given observable, meaning that only well-defined (gauge-invariant) quantities can be used to compare different calculation methods. Alternatively, one may adopt a coordinate-independent formalism such as the one proposed in \cite{Ratzinger:2024spd}.

\clearpage
\unnumberedtrue
\phantomsection
\addcontentsline{toc}{section}{\listfigurename}
\listoffigures

\phantomsection
\addcontentsline{toc}{section}{\listtablename}
\listoftables


\phantomsection
\renewcommand{\refname}{Bibliography}
\addcontentsline{toc}{section}{\refname}
\bibliographystyle{utphys}
\bibliography{fs_doi}

\end{document}